\documentclass[a4paper,11pt]{article}

\usepackage{geometry}										% [showframe]
\usepackage[english]{babel}									% language
\usepackage[utf8]{inputenc}									% input text encoding
\usepackage[T1]{fontenc}									% output encoding

\usepackage{lmodern}										% Latin Modern

\usepackage{amssymb,amsfonts,amsmath,amsthm,mathtools}

\usepackage{booktabs}
\usepackage{hyphenat}
\usepackage{booktabs}
\usepackage{bbm}
\usepackage{multirow}
\usepackage{graphicx}
\usepackage{algorithm}
\usepackage{algorithmic}
\usepackage{bstnotations}
\usepackage{xcolor}
\usepackage{tikz}

\newcommand{\I}{\mathbbm{1}_{\cd_f}}

\newcommand{\dd}{\textrm{d}}

\usepackage{authblk}										% [auth-sc,affil-it], to be loaded after titling
\title{AL-SPCE -- Reliability analysis for nondeterministic models using stochastic polynomial chaos expansions and active learning}
\author[1]{Anderson V. Pires \thanks{anderson.vinhapires@tno.nl}}
\author[1]{Maliki Moustapha \thanks{moustapha@ibk.baug.ethz.ch}}
\author[1]{Stefano Marelli\thanks{marelli@ibk.baug.ethz.ch}}
\author[1]{Bruno Sudret\thanks{sudret@ethz.ch}}
\affil[1]{Chair of Risk, Safety and Uncertainty Quantification, ETH Z\"{u}rich, Switzerland}

\date{April 29, 2026}

\usepackage{microtype}										% typographical refinements

\usepackage{indentfirst}									% indents first paragraph in new section
\usepackage{caption}
\usepackage{subcaption}

\usepackage{natbib}
\usepackage{hyperref}
\PassOptionsToPackage{unicode}{hyperref}
\hypersetup{
pdftitle = {Active learning reliability method for stochastic emulators},
pdfauthor = {Anderson V. Pires, Maliki Moustapha,  Stefano Marelli, Bruno Sudret},
pdfsubject = {},
pdfkeywords = {Structural reliability, Noisy limit-state functions, Surrogate models, Gaussian process regression, Active learning},
colorlinks = true,            
citecolor = blue,
linkcolor = blue,
urlcolor  = blue                                                                                                                                                              % false: boxed links; true: colored links
}

\usepackage[capitalize]{cleveref}								% Eq., Fig., Table, Section, has to be loaded after hyperref
\creflabelformat{equation}{#2(#1)#3}								% hyperlinks covers the parenthesis around equation
\crefrangelabelformat{equation}{#3(#1)#4 to #5(#2)#6}						% multiple references
\labelcrefformat{subequation}{#2(#1)#3}								% the same for subequations
\labelcrefrangeformat{subequation}{#3(#1)#4 to #5(#2)#6}					% mutiple references

\graphicspath{{./Figures}} 

\begin{document}
\vspace{-.5cm}
\maketitle
\vspace{-.5cm}
\begin{abstract}
Reliability analysis traditionally relies on deterministic simulators, where repeated evaluations with identical inputs produce the same outputs. However, many real-world systems exhibit stochastic behavior, leading to non-repeatable outcomes even under identical conditions. To model such systems, stochastic simulators are employed, where the response is a random variable rather than a deterministic quantity.

The inherent randomness of these models affects their reliability, and must therefore be accounted for in reliability analysis. While Monte Carlo simulation can be used for this purpose, its computational cost is usually prohibitive. To circumvent this issue, stochastic emulators have been introduced as surrogate models capable to reproduce the random response of the simulator at reduced computational cost. Recent contributions have demonstrated that stochastic emulators can be successfully applied to perform reliability analysis for stochastic simulators. However, these approaches still rely on relatively large training sets to achieve accurate reliability estimates, which may become prohibitive for expensive models.

In this work, we propose an active learning framework to further reduce the computational effort required for reliability analysis using stochastic emulators. Focusing on stochastic polynomial chaos expansions (SPCE), we introduce a learning function that identifies relevant regions for reliability estimation in which the emulator exhibits high predictive uncertainty. Additionally, we leverage on the asymptotic normality of the maximum likelihood estimator to quantify the local uncertainty in the predictions of the emulator. The proposed methodology, named active learning stochastic polynomial chaos expansions (AL-SPCE), is validated on three different types of problems. In all cases, the results show that the active learning approach significantly improves efficiency compared to previous surrogate-based approaches and direct Monte Carlo simulation, while maintaining accurate reliability estimates.\\


\end{abstract}

\maketitle

\newpage

%%%%%%%%%%%%%%%%%%%%%%%%%%%%%%%%%%%%%%%%%%%%%%%%%%%%%%%%%%%%%%%%%%%%%%%%%%%%

\section{Introduction}
\label{sec_introduction}

Reliability theory and structural reliability can be viewed as branches of uncertainty quantification focused on estimating the probability that a system will fail due to uncertainties in its environmental conditions. Since physical prototypes and experimental tests are often costly or impractical, computational models have become valuable tools for representing such systems. Advances in numerical methods, modeling techniques, and computational tools have significantly increased the accuracy and realism of these models. However, these improvements come at the cost of increased complexity and computational cost. Despite these challenges, virtual models remain widely adopted because they allow for extensive scenario exploration, providing a practical means to assess the impact of uncertainty on the reliability of a system without relying on physical experiments.

A prevailing paradigm in engineering, aided by the inherently deterministic nature of computers, is to design simulation models as deterministic objects: repeated runs with the same parameters always produce identical outputs. However, many real-world systems are not deterministic and, in such cases, removing all sources of randomness may be either infeasible or undesirable. To realistically reproduce the behavior of such systems, simulations must explicitly account for this inherent variability, leading to the use of so-called \textit{stochastic simulators}.

Nondeterministic behavior may arise from a variety of sources. One common origin is stochastic processes, such as Brownian motion \citep{Heston_1993}, whose evolution is governed by intrinsic randomness. In this case, each realization yields a different trajectory, even under identical initial conditions. A second source is found in physical experiments, where components are subject to unmeasured or uncontrollable effects that make their response inherently random \citep{Tsokanas2021,Hao_2021,Abbiati2022}. A third class involves agent-based models, widely used in fields such as traffic simulation and epidemiology \citep{Cuevas2020, Nguyen_2021}. In these models, the global dynamics are driven by fixed input parameters, but individual agents act according to probabilistic rules, introducing nondeterministic behavior.

In structural reliability analysis, we characterize the state of the system, that is, whether it is in a safe or failed condition, using a \textit{limit-state function}. In this study, the stochastic behavior of the system is accounted for through a \textit{stochastic} limit-state function $g_s$, formally defined as:
\begin{equation} 
\begin{split} 
g_s : D_{\ve{X}} \times \Omega \quad \to \quad & \mathbb{R}, \\
(\ve{x}, \omega) \quad \mapsto \quad & g_s(\ve{x}, \omega),
\end{split}
\label{eq_map_definition_stochastic_models}
\end{equation}
\noindent where $\ve{x} \in D_{\ve{X}} \subset \mathbb{R}^{M_{\ve{X}}}$ denotes the input parameters of the model, which describe the physical conditions under which the system operates. The variable $\omega \in \Omega$ represents a random event defined over the probability space $(\Omega, \mathcal{F}, \mathbb{P})$, and accounts for the nondeterministic nature of the computational model describing the system, and thus the limit-state function.

In practice, however, computers cannot generate true randomness. Instead, stochasticity is introduced into simulators through a set of \emph{latent variables} $\ve{Z}(\omega)$, also defined over the probability space $\prt{\Omega, \cf, \mathbb{P}}$ and with support $D_{\ve{Z}} \subset \mathbb{R}^{M_{\ve{Z}}}$. These variables are internal to the simulator and are not explicitly treated as inputs. Rather, they are used to simulate the nondeterministic behavior of the system. Consequently, the stochastic limit-state function $g_s$ can be expressed in terms of a \textit{deterministic} function $g$ as:
\begin{equation}
g_s(\ve{x}, \omega) = g(\ve{x}, \ve{Z}(\omega)).
\label{eq:sto_to_det_sim}
\end{equation}
Given a fixed input vector $\ve{x}_0$, each evaluation of the simulator corresponds to a different realization of the latent variables $\ve{Z}$, producing a different output. This defines a random variable $Y_{\ve{x}_0} \equiv g(\ve{x}_0, \cdot)$, which is characterized by the \emph{conditional distribution} of the system response given $\ve{x}_0$. In practice, this distribution can be estimated by evaluating the limit-state function multiple times at the same input point $\ve{x}_0$ using different realizations of the latent variables $\{\ve{z}_1, \ve{z}_2, \ldots, \ve{z}_R\}$. This process, known as \emph{replication}, is widely used to characterize the probability density function (PDF) of $Y_{\ve{x}_0}$, although extremely costly.

Given a realization $\ve{z}_0$, the mapping $\ve{x} \mapsto g\prt{\ve{x}, \ve{z}_0}$ is a deterministic limit-state function. Evaluating the latter for different values of $\ve{x}$ can be done by \emph{fixing the random seed} within the stochastic simulator.

In the context of reliability analysis with stochastic simulators, uncertainty arises from two distinct sources: the inherent randomness in the simulator itself, modeled through the latent variables $\ve{Z}$, and the uncertainty in the input parameters, represented by a random vector $\ve{X}$ defined over a probability space with support $D_{\ve{X}} \subset \mathbb{R}^{M_{\ve{X}}}$. The latter accounts for variability in quantities such as loading conditions, material properties, or manufacturing tolerances. Following standard conventions for deterministic models \citep{Ditlevsen_1996,Lemaire_2009,Melchers_2018}, system failure is defined as the event in which the limit-state function takes negative values. Accordingly, the probability of failure is defined as \citep{Choe_2015,Pires_2025b}:
\begin{equation}
P_f = \Prob{g\prt{\ve{X},\ve{Z}} \leqslant 0}.
\label{eq_Pf_stochastic_definition}
\end{equation}
Alternatively, since the failure domain is defined as  \\$\mathcal{D}_f = \left\{(\ve{x}, \ve{z}) \in D_{\ve{X}} \times D_{\ve{Z}}: g(\ve{x}, \ve{z}) \leqslant 0 \right\}
$, and under the assumption that $\ve{X}$ and $\ve{Z}$ are independent, the probability of failure can equivalently be written as:
\begin{equation}
\begin{split}
P_f & = \int \int_{\acc{\prt{\ve{x},\ve{z}}: g\prt{\ve{x}, \ve{z}}\leqslant 0}} f_{\ve{X}}\prt{\ve{x}} f_{\ve{Z}}\prt{\ve{z}} \dd\ve{x} \, \dd\ve{z}\\ 
& = \int_{\mathbb{R}^{M_{\ve{X}}}}\int_{\mathbb{R}^{M_{\ve{Z}}}} \I(\ve{x}, \ve{z}) f_{\ve{X}}\prt{\ve{x}} f_{\ve{Z}}\prt{\ve{z}} \dd\ve{x} \, \dd\ve{z}\\
& = \Esp{\I\prt{\ve{X},\ve{Z}}},
\end{split}
\label{eq_Pf_stochastic_definition_integral}
\end{equation}
where $\I(\cdot, \cdot)$ is the indicator function of the failure domain, equal to $1$ if $g(\ve{x}, \ve{z}) \leqslant 0$, and 0 otherwise.

As in the deterministic case, solving \cref{eq_Pf_stochastic_definition_integral} analytically is generally infeasible, since the failure domain is not known explicitly. A widely used alternative is Monte Carlo simulation (MCS), which provides an unbiased estimate of $P_f$ by repeatedly evaluating the limit-state function. In the stochastic setting, each evaluation of $g$ involves not only a sample from $\ve{X}$, but also a realization of $\ve{Z}$, drawn implicitly by the simulator. Thus, while MCS explicitly draws from the input parameters $\ve{X}$ only, the latent variables $\ve{Z}$ are internally sampled at each run, effectively capturing both sources of variability.

\cref{eq_Pf_stochastic_definition_integral} represents a natural extension of the classical definition of the failure probability.  To better understand the contribution of latent variability, one can define  the \textit{conditional failure probability}, which corresponds to the probability of failure solely due to the latent variability, for a given fixed input $\ve{x}$. Conceptually, this corresponds to running multiple replications at the same $\ve{x}$ with different realizations of $\ve{Z}$. Since it depends on $\ve{x}$, we obtain a deterministic \textit{conditional failure probability function}, denoted by $s\prt{\ve{x}}$ and defined as:
\begin{equation}
s(\boldsymbol{x}) = \Probx{\ve{Z}}{g\prt{\ve{X}, \ve{Z} } \leqslant 0 \mid \ve{X}=\ve{x}} \equiv \int_{\mathbb{R}^{M_{\ve{Z}}}} \I(\ve{x}, \ve{z})  f_{\ve{Z}}\prt{\ve{z}} \, \dd\ve{z}. 
\label{eq_conditional_Pf}
\end{equation}

Substituting \cref{eq_conditional_Pf} into \cref{eq_Pf_stochastic_definition_integral}, the total probability of failure can be written as the expected value of the conditional failure probability, as follows:
\begin{equation}
\begin{split}
P_f & = \int_{\mathbb{R}^{M_{\ve{X}}}}\prt{\int_{\mathbb{R}^{M_{\ve{Z}}}} \I(\ve{x}, \ve{z})  f_{\ve{Z}}\prt{\ve{z}} \dd\ve{z}} f_{\ve{X}}\prt{\ve{x}}\dd\ve{x} \\
& = \int_{\mathbb{R}^{M_{\ve{X}}}}s\prt{\ve{x}} f_{\ve{X}}\prt{\ve{x}}\dd\ve{x} \\
& \equiv \Espe{\ve{X}}{s(\ve{X})}.
\end{split}
\label{eq_Pf_random_variable_POV}
\end{equation}

Similarly to the estimator in \cref{eq_Pf_stochastic_definition_integral}, the probability of failure defined in \cref{eq_Pf_random_variable_POV} can also be estimated via MCS over the distribution of $\ve{X}$. However, as recently shown by \citet{Pires_2025b}, when the conditional failure probability function $s(\ve{x})$ is known, the estimator based on \cref{eq_Pf_random_variable_POV} achieves lower variance compared to that of \cref{eq_Pf_stochastic_definition_integral}.

Monte Carlo Simulation is known to be computationally prohibitive, especially when small failure probabilities are of interest. This challenge naturally extends to stochastic models. However, a key difference from reliability analysis using deterministic models is that few alternative methods have been developed for stochastic simulators, and the literature on this topic remains relatively sparse.

\citet{Choe_2015} were among the first to formalize the reliability analysis problem for stochastic simulators, explicitly accounting for latent variability and introducing an extension of classical importance sampling \citep{Melchers_1989}. Subsequent works addressed different aspects of this method: \citet{Choe_2017} investigated the asymptotic properties of the stochastic importance sampling estimator and proposed an asymptotically valid confidence interval. \citet{Cao_2019} introduced a cross-entropy-based strategy for stochastic importance sampling. \citet{Pan_2020} proposed an adaptive variant of stochastic importance sampling. Finally, \citet{Li_2021} developed a nonparametric approach tailored to wind turbine reliability analysis.

In addition to variance reduction techniques, another commonly used strategy to decrease the cost of performing reliability analysis is the use of surrogate models. These are inexpensive approximations of complex simulators trained on a limited set of full-scale simulations, referred to as the \textit{experimental design}. Surrogate modeling is well-established in the deterministic case, but traditional methods cannot be directly applied to stochastic simulators, as they fail to account for the latent randomness. Due to growing interest in surrogates of stochastic simulators, several \textit{stochastic emulators} have recently been proposed, including stochastic Kriging \citep{Ankenman_2010}, generalized lambda models \citep{Zhu_2020, Zhu_2021_GLaM}, stochastic polynomial chaos expansions \citep{Zhu_2023_SPCE}, or Karhunen-Loève-based spectral surrogates \citep{Azzi2019, Luethen_2023,Mueller_2025}. A more detailed review can be found in \citet{Pires_2025b}.

Regarding the use of stochastic emulators for reliability analysis, only a limited body of literature is currently available. \citet{Gramstad2020} proposed combining Gaussian process models, extreme value theory, and active learning to assess the short- and long-term reliability of wind turbines. \citet{Hao_2021} extended stochastic Kriging \citep{Ankenman_2010} to handle unknown heteroskedastic noise, enabling its application to reliability analysis. However, these methods assume a fixed form for the conditional response distributions, which limits their generality. From a different angle, \citet{Zheng_2022} proposed using feedforward neural networks to perform quantile regression for failure probability estimation. It is worth noting that all these approaches require replicated simulations to characterize the response variability.

Recently, we have demonstrated that it is possible to perform reliability analysis without relying on replication-based surrogate models or assuming Gaussian conditional responses. Specifically, \citet{Pires_2025b} shows that Generalized Lambda Models (GLaM) \citep{Zhu_2020, Zhu_2021_GLaM}, which approximate conditional distributions using flexible generalized lambda distributions, and stochastic polynomial chaos expansions \citep{Zhu_2023_SPCE}, which use polynomial expansions to map latent variables to conditional responses, are viable alternatives. In that study, however, the surrogate models were trained using a \textit{static experimental design}, i.e., a space-filling approach aimed at achieving global accuracy. Reliability analysis was then performed on these globally trained surrogates. While their results showed significant improvement over brute-force Monte Carlo simulation, accurately estimating $P_f$ still required a relatively large number of training samples.

One way to further reduce the number of model evaluations is to adopt active learning strategies, \emph{i.e.}, to adaptively add new experimental design points so that they minimize the uncertainty in the estimated failure probability at each step of an iterative process. The rationale is that global accuracy is not required for reliability analysis. Instead, surrogate models should be refined locally, in regions most relevant to estimating $P_f$. To this end, active learning methods aim to identify training points that maximize information gain with respect to estimating $P_f$, and iteratively enrich the experimental design accordingly. For deterministic models, this strategy has proven highly effective, as shown in the recent review papers by \citet{Teixeira_2021, Moustapha_2022}.

This contribution extends \citet{Pires_2025b} by proposing AL-SPCE, an active learning framework for reliability analysis with stochastic emulators. At the core of our methodology lies a learning function specifically designed to identify regions of high predictive uncertainty that are most relevant for the estimation of the probability of failure. This function guides the enrichment of the experimental design and relies on local error estimates. To compute these estimates efficiently, we leverage the asymptotic properties of the maximum likelihood estimator used for training SPCE, the stochastic emulator of choice in this work. The effectiveness of the proposed method is demonstrated on three case studies: a classical reliability problem modified to include latent variables in the limit-state function, a stochastic agent-based model that simulates an infectious disease outbreak, and a realistic wind turbine application where only operational data is available.

This paper is structured as follows: \cref{sec_stochastic_polynomial_chaos_expansions} introduces the SPCE and its application to reliability analysis. \cref{sec_ALR_SPCE} presents the proposed active learning methodology. Specifically, \cref{sec_learning_function} introduces the learning function, which motivates the need for local error measures, and \cref{sec_local_error_measures_for_SPCE} details how these measures may be computed for SPCE. \cref{sec_enrichment_step} discusses practical aspects of the enrichment strategy. \cref{sec_results} reports the results of three case studies. Finally, \cref{sec_conclusions} concludes the paper, highlighting current limitations and directions for future research.

%%%%%%%%%%%%%%%%%%%%%%%%%%%%%%%%%%%%%%%%%%%%%%%%%%%%%%%%%%%%%%%%%%%%%%%%%%%%
\section{Stochastic polynomial chaos expansions}
\label{sec_stochastic_polynomial_chaos_expansions}

\subsection{Formulation}

Polynomial chaos expansion (PCE) is a widely used surrogate modeling technique for deterministic simulators, which are formally defined by the mapping $g_d: \boldsymbol{x} \in \mathbb{R}^{M_{\ve{x}}} \mapsto g_d(\boldsymbol{x}) \in \mathbb{R}$. In this context, for the random input vector $\boldsymbol{X}$, the random variable $Y = g_d\prt{\ve{X}}$ may be approximated as a weighted sum of orthonormal polynomials in $\ve{X}$, as follows:
\begin{equation}
g_d(\boldsymbol{X}) \approx \sum_{\boldsymbol{\alpha} \in \mathcal{A}} c_{\boldsymbol{\alpha}} \psi_{\boldsymbol{\alpha}}(\boldsymbol{X}),
\end{equation}

\noindent where $\mathcal{A} \subset \mathbb{N}^{M_{\ve{X}}}$ is a finite multi-index set, $c_{\ve{\alpha}}$ are the coefficients to be estimated, and $\{ \psi_{\ve{\alpha}} \}_{\ve{\alpha} \in \mathcal{A}}$ are multivariate polynomials, which are orthonormal with respect to the input PDF $f_{\boldsymbol{X}}$ \citep{Xiu2002}.

Assuming $\boldsymbol{X}$ has independent components, the basis function $\psi_{\boldsymbol{\alpha}}$ is defined by the product of univariate polynomials:
\begin{equation}
\psi_{\boldsymbol{\alpha}}(\boldsymbol{x})=\prod_{j=1}^{M_{\ve{X}}} \phi_{\alpha_j}^{(j)}\left(x_j\right),
\end{equation}
where each $\phi_{\alpha_j}^{(j)}$ is a univariate orthonormal polynomial of degree $\alpha_j$, associated with the marginal distribution of the $j$-th component of $\ve{X}$.

To limit the number of terms in the expansion, truncation schemes are used to define the multi-index set $\mathcal{A}$. A common choice is total-degree truncation, which limits the maximum total polynomial degree. However, this alone often results in a large number of basis functions with negligible contribution. To reduce complexity, hyperbolic truncation, also known as $q$-norm truncation is often used \citep{Blatman2010,Luethen2022}.

Stochastic polynomial chaos expansion (SPCE), originally proposed by \citet{Zhu_2023_SPCE}, is a stochastic extension of PCE that enables the construction of surrogate models for stochastic simulators. Their primary goal is to approximate the probability density function of the conditional response, defined as the random variable \( Y_{\boldsymbol{x}} \equiv Y \mid \boldsymbol{X} = \boldsymbol{x} \). Unlike approaches that require replicated model runs, SPCE does not need replications, allowing the computational budget to be allocated toward more space-filling experimental designs, as well as leveraging existing datasets collected without replications. Furthermore, SPCE does not impose a predefined parametric form on the conditional model response (\emph{e.g.}, it does not assume Gaussianity), providing flexibility to model complex behaviors such as multi-modality.

To accomplish this, SPCE introduces an artificial latent variable $U$ with a known distribution into the polynomial chaos expansion framework, enabling the emulator to represent the stochasticity of the simulator. Such an artificial latent variable encapsulates all sources of randomness from the simulator, regardless of the size of $\ve{Z}$, into a single variable. The isoprobabilistic transform from $U$ to $Y_{\ve{x}}$ reads:
\begin{equation}
Y_{\boldsymbol{x}} \stackrel{\mathrm{d}}{=} F_{Y \mid \boldsymbol{X}}^{-1}\left(F_U(U) \mid \ve{X} = \boldsymbol{x}\right),
\label{eq_PIT}
\end{equation}
where $ F_U$ is the cumulative distribution function (CDF) of $U$, and $\stackrel{\mathrm{d}}{=} $ denotes equality in distribution.

\cref{eq_PIT} implies that the latent variable $U$ can be deterministically mapped to the conditional distribution of $Y_{\boldsymbol{x}}$. Consequently, this deterministic mapping can be captured via a PCE in the augmented space $(\boldsymbol{X}, U)$ of dimensionality $\prt{M_{\ve{X}} + 1}$:
\begin{equation}
F_{Y \mid \boldsymbol{X}}^{-1}\left(F_U(u) \mid \ve{X} = \boldsymbol{x}\right) = \sum_{\mathbb{R}^{M_{\ve{X}}+1}} c_{\boldsymbol{\alpha}} \psi_{\boldsymbol{\alpha}}(\boldsymbol{x}, u).
\label{eq_ill_posed_SPCE}% \label{eq_PIT_SPCE}
\end{equation}

Truncating \cref{eq_ill_posed_SPCE} as-is may lead to singularities in the probability density functions. To address this, \citet{Zhu_2023_SPCE} introduced an additive noise term $\varepsilon \sim \mathcal{N}(0, \sigma_\varepsilon^2) $, which acts as a regularization term. The final emulator becomes:
\begin{equation}
Y_{\ve{x}} \stackrel{\mathrm{d}}{\approx} \hat{Y}_{\ve{x}}=\sum_{\alpha \in \mathcal{A}} c_\alpha \psi_\alpha(\boldsymbol{x}, U)+\varepsilon.
\label{eq_well_posed_SPCE}
\end{equation}

The resulting probability density function of $Y_{\ve{x}}$ comes from the convolution of the PCE output with the Gaussian noise:
\begin{equation}
f_{\hat{Y}_{\boldsymbol{x}}}(y) =\int_{\mathcal{D}_U} \frac{1}{\sigma_\varepsilon} \varphi\left(\frac{y-\sum_{\boldsymbol{\alpha} \in \mathcal{A}} c_{\boldsymbol{\alpha}} \psi_{\boldsymbol{\alpha}}(\boldsymbol{x}, u)}{\sigma_\varepsilon}\right) f_U(u) \mathrm{d} u,
\label{eq_PDF_SPCE}
\end{equation}
where $\varphi $ denotes the standard normal PDF.

\subsection{Parameter estimation}
\label{subsec_parameter_estimation}
The unknown parameters of the model are the PCE coefficients $c_{\boldsymbol{\alpha}} $, for $\alpha \in \mathcal{A} $ and the standard deviation $\sigma_\varepsilon$ of the noise term in Eq.~\eqref{eq_well_posed_SPCE}. Following the methodology of \citet{Zhu_2023_SPCE}, these parameters are estimated using a two-step procedure.

For a fixed value of $\sigma_\varepsilon$, the coefficients are estimated via maximum likelihood. Given an observation $(\boldsymbol{x}, y) $, the likelihood of a set of coefficients reads:
\begin{equation}
l(c, \sigma_\varepsilon; \ve{x}, y) = \int_{D_U} \frac{1}{\sqrt{2\pi\sigma_\varepsilon}}
\exp \left( -\frac{\left(y - \sum_{\alpha \in \mathcal{A}} c_{\alpha} \psi_{\alpha}(\ve{x}, u)\right)^2}{2\sigma_\varepsilon^2} \right)
f_U(u) \dd u.
\label{eq_cond_likelihood}
\end{equation}

Given an experimental design $\cx = \acc{\ve{x}^{\prt{1}}, \cdots, \ve{x}^{\prt{N}}}$ and its associated model responses $\cy = \acc{y^{\prt{1}}, \cdots, y^{\prt{N}}}$, the maximum likelihood optimization is directly cast as a function of the PCE coefficients, as follows:
\begin{equation}
\hat{\boldsymbol{c}}=\arg \max _{\ve{c} \in \mathbb{R}^{\abs{\ca}}} \sum_i^N \log \left(l\left(\boldsymbol{c} ; \boldsymbol{x}^{(i)}, y^{(i)}, \sigma_\varepsilon\right)\right) .
\label{eq_likelihood_optimizations}
\end{equation}

In practice, we assume standard distributions for the latent variable, such as the uniform $\cu\prt{0,1}$ or Gaussian $\cn\prt{0,1}$. Thus, the integral in \eqrefe{eq_cond_likelihood} may be estimated using numerical quadrature \citep{Golub_1969}, as proposed by \citet{Zhu_2023_SPCE}. For details about the solution of the optimization problem in \eqrefe{eq_likelihood_optimizations}, as well as the calibration of the standard deviation $\sigma_\varepsilon$, the reader is referred to the original publication by \citet{Zhu_2023_SPCE}.

\subsection{Reliability analysis with SPCE}
\label{subsec_reliability_analysis_with_SPCE}

When the conditional failure probability $s(\boldsymbol{x})$ is known, the estimator based on \cref{eq_Pf_random_variable_POV} yields lower variance than direct Monte Carlo simulation \citep{Pires_2025b}. In the case of SPCE, we exploit the known structure of the latent space to derive an analytic approximation of \( s(\boldsymbol{x}) \). That is, since the latent variable $U$ is known and follows a standard distribution (\emph{e.g.}, uniform or Gaussian), the conditional probability distribution of the SPCE prediction (in \cref{eq_PDF_SPCE}) may be estimated by numerical quadrature:
\begin{equation}
f_{\hat{Y}_{\ve{x}}}(y) \approx  \sum_{j=1}^{N_Q} \frac{1}{\sqrt{2\pi\sigma_\varepsilon}}
\exp \left( -\frac{\left(y - \sum_{\alpha \in \mathcal{A}} c_{\alpha} \psi_{\alpha}(\ve{x}, u^{(j)}) \right)^2}{2\sigma_\varepsilon^2} \right) w^{(j)},
\label{eq_cond_likelihood_quadrature}
\end{equation}
where $N_Q$ is the number of integration points, $u^{\prt{j}}$ is the $j$-th integration point, and $w^{\prt{j}}$ is the corresponding weight, both associated to the PDF $f_U$.

\eqrefe{eq_cond_likelihood_quadrature} shows that $f_{\hat{Y}_{\ve{x}}}$ is approximated by a Gaussian mixture. Therefore, its associated cumulative distribution function (CDF),  denoted \( F_{\hat{Y}_{\boldsymbol{x}}} \), is given by:
\begin{equation}
F_{\hat{Y}_{\boldsymbol{x}}}(y) \approx \sum_{j=1}^{N_Q} w^{(j)} \, \Phi\left(\frac{y - \sum_{\boldsymbol{\alpha} \in \mathcal{A}} c_{\boldsymbol{\alpha}} \psi_{\boldsymbol{\alpha}}(\boldsymbol{x}, u^{(j)})}{\sigma_\varepsilon}\right),
\end{equation}
where \( \Phi \) is the standard Gaussian CDF.

The conditional failure probability \( s(\boldsymbol{x})=\Prob{\hat{Y}_{\ve{x}} \leqslant 0} \) is obtained by evaluating the above equation at \( y = 0 \), yielding:
\begin{equation}
\hat{s}(\boldsymbol{x}) \approx \sum_{j=1}^{N_Q} w^{(j)} \, \Phi\left(-\frac{\sum_{\boldsymbol{\alpha} \in \mathcal{A}} c_{\boldsymbol{\alpha}} \psi_{\boldsymbol{\alpha}}\prt{\boldsymbol{x}, u^{(j)}}}{\sigma_\varepsilon}\right).
\label{eq_approximation_hat_s}
\end{equation}

Once the SPCE emulator is trained, evaluating the sum \( \sum_{\boldsymbol{\alpha} \in \mathcal{A}} c_{\boldsymbol{\alpha}} \psi_{\boldsymbol{\alpha}}(\boldsymbol{x}, u) \) is computationally inexpensive. The accuracy of this numerical approximation strongly depends on the number of quadrature points used. In our tests, using \( N_Q = 100 \) resulted in a negligible numerical error.

\section{Active learning reliability using SPCE}
\label{sec_ALR_SPCE}

\subsection{Introduction}

The performance of surrogate models strongly depends on the quality of their experimental design. When data are scarce, space-filling designs that aim for uniform coverage often fail to capture critical regions of the input space. To address this, a more effective strategy is to focus the training data in areas that are most informative. This is the core idea behind \emph{active learning reliablity} (ALR), where the experimental design is adaptively constructed to maximize information gain while minimizing computational cost. In the context of reliability analysis, where global accuracy is unnecessary, the focus is on improving $\hat{P}_f$ by refining the surrogate near regions of interest.

Active learning frameworks typically rely on several components, as discussed in \citet{Moustapha_2022}. One of them is the \textit{learning function}, which guides the selection of new points to enrich the experimental design. These functions often depend on local error measures that quantify the predictive uncertainty of the surrogate model. Based on this information, the learning function identifies the additional samples that are expected to have the most significant impact on improving the estimate of $\hat{P}_f$. In this paper, we introduce both a new learning function and a methodology for computing local error measures tailored to SPCE. These components are detailed in the following sections.

\subsection{Learning function}
\label{sec_learning_function}

The learning function typically assigns a score to each point $\ve{x}$ in the input space, which guides an optimization process that selects the most informative points to enrich the experimental design, denoted by $\ve{x}^{\textrm{next}}$. In the case of deterministic models, the reliability problem reduces to a classification task: each input either leads to failure or not. Consequently, learning functions are designed to improve the ability of the surrogate to classify points near the failure boundary, where the limit-state function changes sign.

In contrast, for the SPCE-based framework, our goal is not to directly classify outcomes as failure or not, but to estimate the conditional failure probability function. This turns the problem into a regression task, where the surrogate is trained to approximate the behavior of $s(\boldsymbol{x})$ across the input space, particularly on regions where the input density $f_{\boldsymbol{X}}(\boldsymbol{x})$ is high, according to \eqrefe{eq_Pf_random_variable_POV}. Guided by this heuristics, we propose the following learning function:
\begin{equation}
\boldsymbol{x}^{\text{next}} = \arg\max_{\boldsymbol{x} \in \mathcal{D}_{\boldsymbol{X}}} f_{\boldsymbol{X}}(\boldsymbol{x}) \cdot \Var{\hat{s}(\boldsymbol{x})},
\label{eq_LF}
\end{equation}
\noindent where the estimated conditional failure probability $\hat{s}\prt{\ve{x}}$ is given by \eqrefe{eq_approximation_hat_s}.

This formulation is directly inspired by the estimator shown in \cref{eq_Pf_random_variable_POV}. It combines two key principles: identifying regions where the surrogate-based conditional failure probability $\widehat s\prt{\ve x}$ is most uncertain through the variance term \( \Var{\hat{s}(\boldsymbol{x})} \), and prioritizing areas that are most relevant to the failure probability estimation, through the input density \( f_{\boldsymbol{X}}(\boldsymbol{x}) \).

To apply this learning function in practice, we must evaluate the variance component \( \Var{\hat{s}(\boldsymbol{x})} \), which corresponds to a local error measure for the surrogate prediction of \( s(\boldsymbol{x}) \). Intuitively, this variance is related to the limited information (finite size of the experimental design) used to construct the SPCE and related $\hat{s}\prt{\ve{x}}$.  As this measure is not embedded in the original SPCE emulator, the next section introduces a novel methodology devised to estimate it.

\subsection{Local Error measures for SPCE}
\label{sec_local_error_measures_for_SPCE}
In its current implementation, SPCE does not feature a built-in local measure of uncertainty, which is essential for active learning. A possible workaround for PCE-based techniques is to use bootstrapping \citep{Marelli_2018}, a resampling technique that generates multiple surrogate models from resampled versions of the experimental design. While effective, this approach can be computationally expensive, particularly for large datasets or high-dimensional problems, as it requires re-training the surrogate multiple times (typically $O(10^{2})$) each time the experimental design is updated.

To overcome this limitation, we propose an analytical alternative based on Fisher's fundamental theorem for the maximum likelihood estimator (MLE) \citep{Efron_2016}. Under standard regularity conditions, the MLE $\hat{\theta}$ is asymptotically normally distributed:
\begin{equation}
\hat{\ve{\theta}} \sim \mathcal{N}(\ve{\theta}^*, \mathcal{I}^{-1}(\ve{\theta}^*)),    
\end{equation}

\noindent where $\ve{\theta}^*$ denotes the true parameter and $\mathcal{I}^{-1}(\ve{\theta}^*)$ is the inverse Fisher information matrix.

This implies that if we repeatedly fit the model to independent experimental designs of size $N$, the distribution of the resulting MLEs converges to a normal distribution centered at $\boldsymbol{\theta}^*$ when $N \to \infty$. While bootstrapping could approximate this distribution empirically, the asymptotic normality provides a tractable analytical alternative.

In the context of SPCE, we apply this principle to the coefficient vector $\hat{\ve{c}}$ in Eq.~\eqref{eq_likelihood_optimizations}. Although the true coefficients are unknown, we assume that the MLE provides a sufficiently accurate estimate and approximate its distribution, as follows:
\begin{equation}
    \ve{c} \sim \mathcal{N}\left(\hat{\ve{c}}, \, \mathcal{I}^{-1}(\hat{\ve{c}}) \right).
\end{equation}

In practice, the Fisher information matrix is computed as the Hessian of the log-likelihood function evaluated at the MLE, as follows:
\begin{equation}
  \mathcal{I}\prt{\hat{\ve{c}}} = -\frac{\partial^2 \log \tilde{l}(\ve{c})}{\partial \ve{c} \partial \ve{c}^\top}\Bigg|_{\ve{c} = \hat{\ve{c}}},
\end{equation}
where $\tilde{l}$ corresponds to the Gaussian quadrature-based approximation of the likelihood function in \cref{eq_cond_likelihood}:
\begin{equation}
    l(\boldsymbol{c}; \boldsymbol{x}, y, \sigma) \approx \tilde{l}(\boldsymbol{c}; \boldsymbol{x}, y, \sigma) = \sum_{j=1}^{N_Q} \frac{1}{\sqrt{2\pi}\sigma_\varepsilon} \exp\left(-\frac{\left(y - \sum_{\boldsymbol{\alpha} \in \mathcal{A}} c_{\boldsymbol{\alpha}} \psi_{\boldsymbol{\alpha}}(\boldsymbol{x}, u^{(j)})\right)^2}{2\sigma_{\varepsilon}^2}\right) w^{(j)}.
\end{equation}
For more details on the use of quadrature in this context, including considerations on numerical stability, the reader is referred to \cite{Zhu_2023_SPCE}.

By drawing $m$ realizations of the coefficients from this multivariate Gaussian distribution, we generate multiple surrogate realizations that capture the epistemic uncertainty of the emulator.  Each of these models provides an estimate of the failure probability, and the ensemble of $m$ realizations can be used to quantify the uncertainty on the failure probability estimate, for instance by constructing confidence bounds. Moreover, these realizations can be used to compute local error measures on some quantities of interest, such as the variance of the conditional failure probability used in the learning function (See \cref{eq_LF}).

\begin{figure}[H]
     \centering
     \begin{subfigure}[c]{\textwidth}
         \centering
         \includegraphics[width=\textwidth]{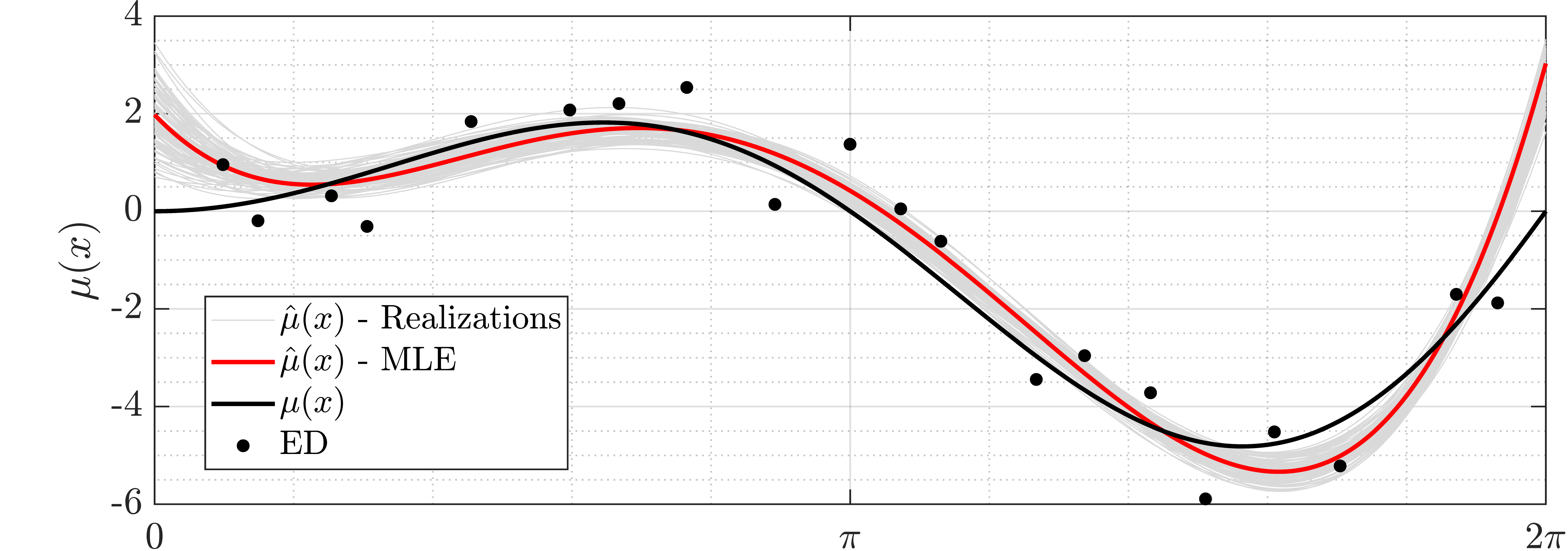}
         \caption{100 realizations of the SPCE mean function (gray), the MLE-based SPCE mean function (red), the true mean function (black), and the experimental design (black dots).}
         \label{fig_realizations_SPCE_mean_functions}
     \end{subfigure}
     \begin{subfigure}[c]{\textwidth}
         \centering
         \includegraphics[width=\textwidth]{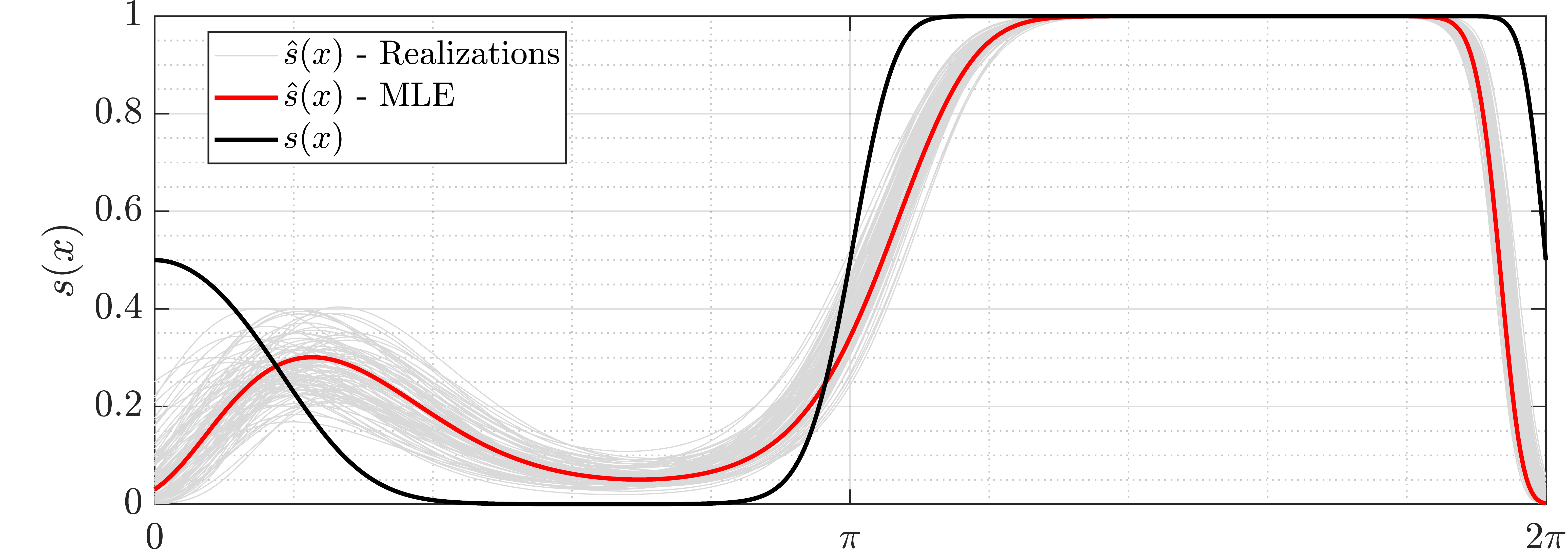}
         \caption{ 100 realizations of the conditional failure probability $\hat{s}(x)$ (gray), the MLE-based estimate (red), and the true conditional failure probability $s\prt{x}$ (black).}
         \label{fig_realizations_SPCE_s_x}
     \end{subfigure}
          \begin{subfigure}[c]{\textwidth}
         \centering
         \includegraphics[width=\textwidth]{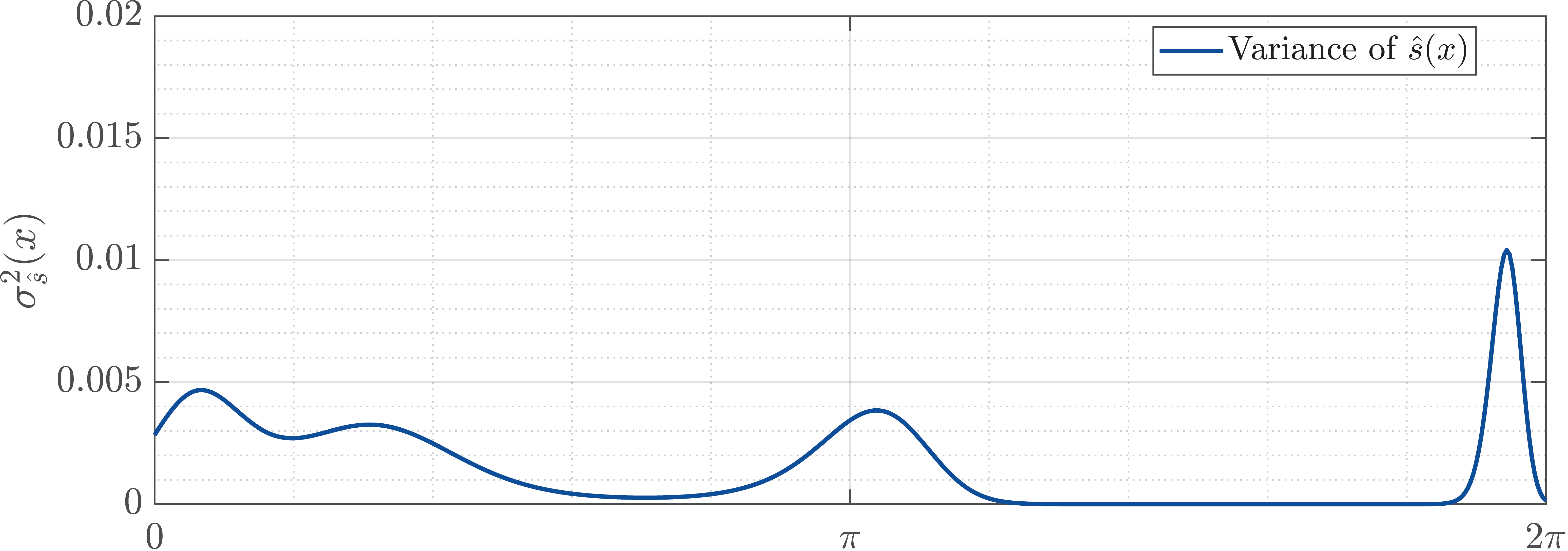}
         \caption{Pointwise predictive uncertainty of $\hat{s}(x)$, computed as the empirical variance of $\hat{s}(x)$ over the 100 realizations.}
         \label{fig_realizations_SPCE_var}
     \end{subfigure}
     \caption{Illustration of the proposed methodology for enabling local error measures for the toy function $h(x, Z) = x \sin(x) + Z$.}
    \label{fig_realizations_SPCE}
\end{figure}

To illustrate the procedure, \cref{fig_realizations_SPCE} presents the results obtained for the following one-dimensional test function:
\begin{equation}
    h(x, Z) = x \sin(x) + Z,
\end{equation}
where the input variable is uniformly distributed $X \sim \mathcal{U}(0, 2\pi)$ and the latent variable is Gaussian, $Z \sim \mathcal{N}(0, 0.5^2)$.

\cref{fig_realizations_SPCE_mean_functions} shows $m=100$ realizations of the SPCE mean function (gray), obtained by resampling the coefficients from the Gaussian approximation. The red curve represents the SPCE mean function, built using the maximum likelihood estimate of the coefficients, while the black curve shows the true mean function, $\mu(x) = x  \sin(x)$. Black dots indicate the experimental design points. \cref{fig_realizations_SPCE_s_x} illustrates the conditional failure probability functions derived from different models. Each gray curve corresponds to a different approximation of $\hat{s}(x)$ associated to one realization of the coefficients. The red curve shows $\hat{s}^{\text{MLE}}(x)$, computed from the SPCE built using the maximum likelihood estimate of the coefficients. The black curve represents the true conditional failure probability, analytically defined as:
\begin{equation}
    s(x) = \Phi\left(-\frac{x  \sin(x)}{0.5}\right),
\end{equation}
where $\Phi$ is the standard normal CDF.

Finally, \cref{fig_realizations_SPCE_var} shows the estimated variance $\Var{\hat{s}^{\prt{m}}(\boldsymbol{x})}$ obtained from the different SPCE realizations, which ultimately serves as a local error measure in our active learning strategy. In this specific example, since the input variable is uniformly distributed, the blue curve in \cref{fig_realizations_SPCE_var} also reflects the shape of the learning function in \cref{eq_LF}, scaled by a factor of $2 \pi$.

It is worth noting that we do not aim to compute precise predictive variances, that is, the exact variability in the predictions of SPCE. Instead, our goal is to identify regions of high epistemic uncertainty, where the SPCE model is not well-calibrated. For this purpose, having measures that merely correlate well with the true local error is sufficient, as they allow us to prioritize regions of the input space where additional training samples are likely to be most beneficial.

\subsection{Practical implementation}
\label{sec_enrichment_step}
We now present practical aspects related to the implementation of the proposed methodology, including details on the optimization procedure, multi-point enrichment, and strategies to mitigate the impact of noise.

\subsubsection{Optimization procedure}
\label{subsec_optimization_procedure}
In active learning for reliability analysis, it is common practice to solve the optimization problem posed in \cref{eq_LF} in a discrete way (see, \emph{e.g.}, \citep{Echard2011,Sun_2017,Marelli_2018,Zhang_2019,Linxiong_2020,Dang_2024}). In this approach, a finite candidate set \( \mathcal{X}_c \) is sampled according to the input probability density function \( f_{\ve{X}}\). The learning function is then evaluated at each candidate point, and the next training point is selected as the one maximizing the learning score. This discrete optimization approach is only practical when the learning function can be evaluated cheaply. In our case, the analytic expression and its vectorized structure make the calculation of \( \hat{s}^{(m)}(\boldsymbol{x}) \), and therefore of its variance, highly efficient, justifying the approach.

Additionally, to monitor the convergence of the active learning process, it is necessary to compute a reliability estimate at each iteration. A common practice when discretizing the optimization is to reuse the candidate set \( \mathcal{X}_c \) as the Monte Carlo sample for estimating the failure probability. In our work, however, we treat these two sets separately: the candidate set \( \mathcal{X}_c \) is resampled at every iteration, whereas the Monte Carlo sample used to compute the failure probability is drawn once at the beginning and kept fixed throughout the process. This choice is motivated by two reasons. First, resampling \( \mathcal{X}_c \) at every iteration allows the candidate set to remain relatively small without quickly exhausting relevant candidate points. Second, fixing the Monte Carlo sample reduces the noise in the convergence curves, ensuring that changes in the estimated probability of failure reflect improvements in the surrogate model rather than fluctuations due to the finite and limited sample size, a strategy widely known as \emph{common random numbers} \citep{Spall2003,Taflanidis_2008}.

\subsubsection{Multi-point enrichment}
\label{subsec_multipoint_enrichment}

Enriching surrogate models with batches of \( K \) points per iteration can significantly reduce the computational time of reliability analysis, especially if parallel computing facilities are available \citep{Schoebi2017}. However, as noted by \citet{Chevalier2014}, such batch enrichment strategies may lead to suboptimal experimental designs compared to sequential approaches that add one point at a time. In the context of stochastic emulators, however, this trade-off becomes less critical. Due to the inherent randomness in the model response, adding single points yields limited informational gain. For this reason, we recommend enriching the surrogate in small batches. The optimal batch size $K$ will generally depend on the available computational resources, ideally matching the number of independent CPUs to maximize parallel efficiency \citep{Schoebi2017}.

We adopt a weighted K-medoids \citep{Zaki2014} clustering approach to prevent the selection of closely spaced points, and to promote more exploratory sampling across the input domain. More specifically, each candidate point in \( \mathcal{X}_c \) is assigned a weight based on its learning function score, which is then used in the clustering process to partition the dataset into \( K \) clusters. From each cluster, the point with the highest learning function score is selected for enrichment. Throughout this paper, we set \( K = 5 \).

\subsubsection{Mitigating noise effects}
\label{subsec_mitigating_effects}

The stochastic nature of the simulators introduces noise into the problem. Optimization problems with noisy objective functions or constraints often exhibit nonsmooth convergence behavior, as discussed in \citet{Hansknecht_2024} and observed in practice \citep{Qiu_2018,Ahmadisoleymani2021,Pires_2025}.

To mitigate the noise introduced during the optimization process, we adopt a damping strategy. In the first iteration, the noise level is selected following the procedure outlined in \cref{subsec_parameter_estimation}. For every successive $i$-th iteration, changes in the noise level are restricted to within $\pm5\%$ of its previous value, with candidate noise levels equally spaced on a logarithmic scale between $0.95\cdot \hat{\sigma_\varepsilon}^{(i-1)}$ and $1.05\cdot \hat{\sigma_\varepsilon}^{(i-1)}$. This strategy prevents abrupt model changes and promotes smoother adaptation as the training set evolves. While the noise level tends to stabilize as the dataset grows, convergence curves still exhibit fluctuations around the true value. To reduce this effect, we apply a moving average smoothing to the failure probability estimates using a window width of three iterations. The final estimate for a given training set size $N$ is therefore defined as:
\begin{equation}
\hat{P}_f = \frac{1}{3} \sum_{j=N-2}^{N} \hat{P}_f^{(j)}.
\label{eq_smoothed_Pf}
\end{equation}

\section{Results}
\label{sec_results}
To demonstrate and validate the proposed methodology, we present three numerical examples. The first is a two-dimensional problem with an analytical solution, used to visualize the learning function and the evolution of the enrichment steps. To account for ~statistical variability due to random initial experimental designs, all experiments are repeated 15 times, and results are visualized using boxplots.

We compare our active learning approach to two benchmarks: a static experimental design using SPCE, and standard Monte Carlo simulation. In both cases, the SPCE models are constructed with an adaptively selected polynomial degree in the range $1$–$10$, and a Gaussian latent variable. For the active learning method, $100$ SPCE realizations are drawn from the Gaussian approximation in \cref{sec_local_error_measures_for_SPCE} to estimate the variance term in the learning function. However, the reported reliability estimates are computed using the SPCE model built from the MLE coefficients. The size of the initial experimental design depends on the problem but is generally set between $10$ and $25$ times the input dimensionality to ensure a reasonable initial surrogate accuracy.

All numerical experiments were carried out using the Stochastic Polynomial Chaos Expansions module \citep{UQdoc_21_121} available in the \textsc{UQLab} uncertainty quantification software \citep{Marelli_2014_UQLab}.

\subsection{\texorpdfstring{Stochastic $R - S$ Function}{Stochastic R - S Function}}

The first benchmark used to assess the proposed methodology is the stochastic $R - S$ function first introduced in  \citet{Pires_2025b}, where we introduce two latent variables into the classical limit-state function to induce stochasticity. The resulting stochastic limit-state function is defined as:
\begin{equation}
g(\boldsymbol{X}, \boldsymbol{Z}) = \frac{R}{Z_1} - S \cdot Z_2,
\label{eq_stochastic_RS}
\end{equation}
where \( \boldsymbol{X} = \{R, S\} \) are the input variables and \( \boldsymbol{Z} = \{Z_1, Z_2\} \) are the latent variables. \( R \) and \( S \) denote resistance and load, respectively.  All variables are assumed to follow lognormal marginal distributions \( \mathcal{LN}(\lambda, \zeta)\), where $\prt{\lambda, \zeta}$ are the mean and standard deviation of the underlying Gaussian variable. \cref{tab_RS} summarizes the mean, standard deviation, and corresponding lognormal parameters for each variable.
\begin{table}[H]
\centering
\caption{Moments, distributions, and parameters of the variables in the stochastic \( R\text{-}S \) problem.}
\label{tab_RS}
\begin{tabular}{@{}cccccc@{}}
\toprule
Variable & Distribution & Mean & Std. Dev. & \( \lambda \) & \( \zeta \) \\ \midrule
\( R \)  & Lognormal    & 5.0  & 0.8       & 1.5968        & 0.1590      \\
\( S \)  & Lognormal    & 2.0  & 0.6       & 0.6501        & 0.2936      \\
\( Z_1 \)& Lognormal    & 1.0  & 0.028     & -0.0004       & 0.0280      \\
\( Z_2 \)& Lognormal    & 1.0  & 0.096     & -0.0046       & 0.0958      \\ \bottomrule
\end{tabular}
\end{table}

This example is particularly suitable as a first test case because both the probability of failure and the conditional failure probability function \( s(\boldsymbol{x}) \) can be computed analytically. Given the structure of \eqrefe{eq_stochastic_RS}, we can define an equivalent limit-state function $\tilde{g}$ as the difference of their logarithms:
\begin{equation}
    \tilde{g}(\boldsymbol{x}, \boldsymbol{z})=\ln{\prt{\frac{R}{Z_1}}} - \ln{\prt{S \cdot Z_2}}= \ln R - \ln Z_1 - \ln S  - \ln Z_2,
\end{equation}

which is a linear combination of independent Gaussian variables. Thus, the failure probability admits the following expression:
\begin{equation}
P_f = \Phi\left( \frac{\lambda_R - \lambda_{Z_1} - \lambda_S - \lambda_{Z_2}}{\sqrt{\zeta_R^2 + \zeta_{Z_1}^2 + \zeta_S^2 + \zeta_{Z_2}^2}} \right) = 3.154 \times 10^{-3}.
\label{eq_true_Pf_RS}
\end{equation}

An analytical expression is also available for the conditional failure probability function \( s(\boldsymbol{x}) \). Conditioning on \( \boldsymbol{x} = (r, s) \) and treating \( \boldsymbol{Z} \) as random yields:
\begin{equation}
s(r, s) = \Phi\left( \frac{\ln r - \lambda_{Z_1} - \ln s - \lambda_{Z_2}}{\sqrt{\zeta_{Z_1}^2 + \zeta_{Z_2}^2}} \right).
\label{eq_sX_RS}
\end{equation}

To evaluate the performance of the proposed active learning strategy, we begin with an initial experimental design of $20$ points generated via Latin Hypercube Sampling (LHS) \citep{Olsson_2003}. At each enrichment step, the failure probability estimate \( \hat{P}_f \) is computed as the average of the conditional failure probabilities over a Monte Carlo sample of size \( N_{\text{MCS}} = 10^6 \), following the approach shown in \cref{eq_Pf_random_variable_POV}. Additionally, to reduce fluctuations in the convergence curves, a moving average smoothing with a window width of three iterations is applied to the sequence of \( \hat{P}_f \)  estimates, as defined in \cref{eq_smoothed_Pf}. The candidate set $\cx_{c}$ consists of \( 10^4 \) points, and the experimental design is enriched by 5 points per iteration until the total reaches 1,000 points. 
%SPCE settings used during training include degree adaptivity in the range \( p \in [1, 4] \), and a Gaussian latent variable.

\begin{figure}[H]
\centering
\includegraphics[width=0.65\linewidth]{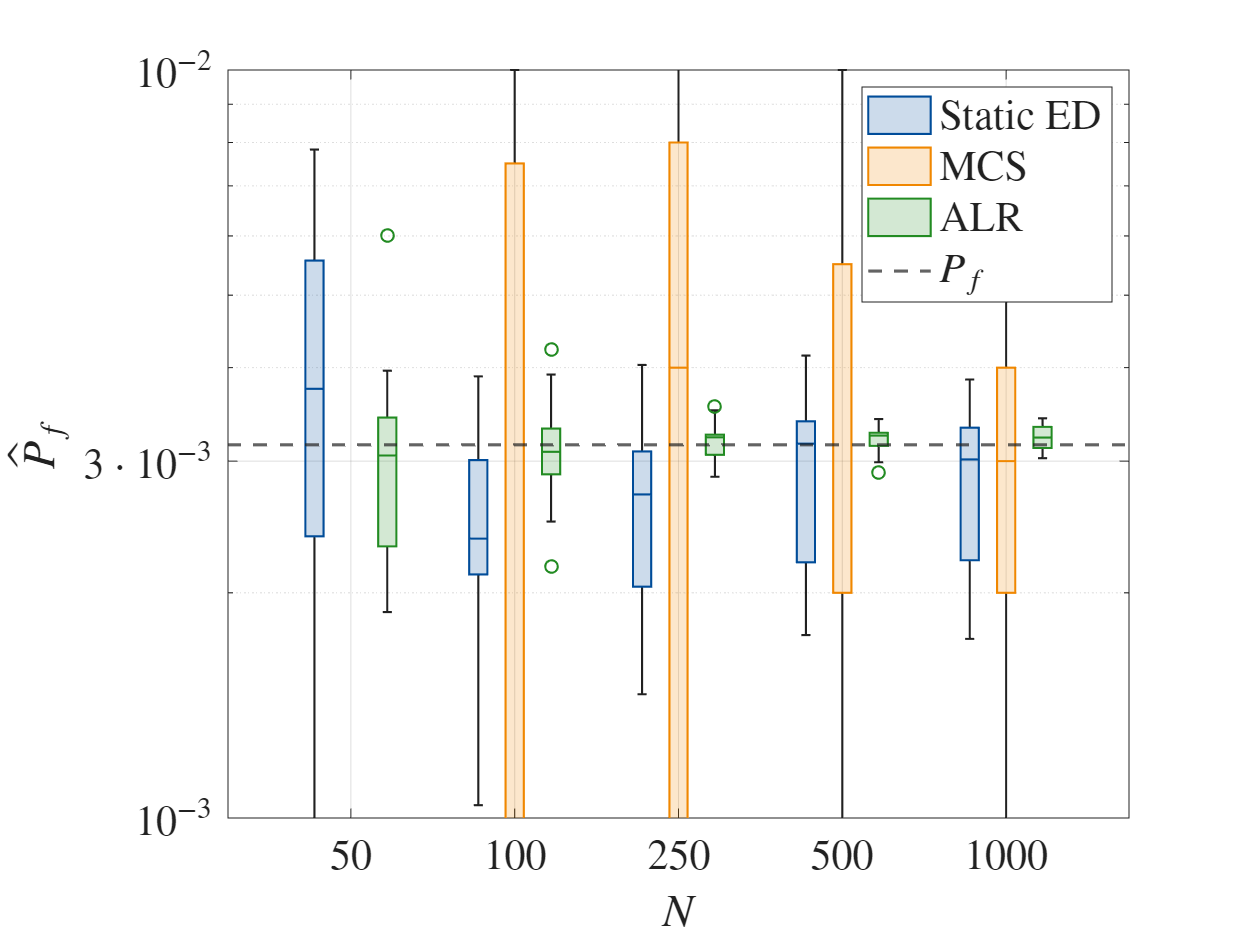}
\caption{R-S example: Boxplots comparing convergence of the ALR method (in green), an SPCE approach using a static ED (in blue), and direct MCS (in orange) across 15 replications. The dashed line indicates the reference $P_f=3.154 \times 10^{-3}$.}
\label{fig_boxplots_RS_ALR_vs_static}
\end{figure}

\cref{fig_boxplots_RS_ALR_vs_static} shows boxplots from 15 runs of the active learning procedure (green), the SPCE approach based on a static experimental design (blue), and direct MCS (orange). In the static approach, the surrogate model is trained on a fixed experimental design of $N$ samples obtained via a space-filling LHS to ensure global accuracy, and the probability of failure is then estimated based on this surrogate. %The globally trained SPCE emulator uses degree adaptivity with polynomial degrees in $p \in \bra{1, 4}$ and hyperbolic truncation with $q$-norms ranging in $q \in \bra{0.7,1}$ with constant increments of $0.1$. 
The analytical solution from Eq.~\eqref{eq_true_Pf_RS} is indicated by the black dashed line. The ALR approach achieves accurate estimates with as few as 50 points and consistently exhibits lower variance across replications. For \( N = 50 \), the MCS results cannot be displayed because no failures were observed in all runs.

Numerical results are presented in \tabref{tab_RS_results}, showing the median and coefficient of variation (CoV) of \( \hat{P}_f \) across the 15 runs for all methods and sample sizes.
\begin{table}[H]
\centering
\caption{R-S function: Comparison of ALR, static ED, and MCS across multiple runs for different sample sizes. Reference $P_f = 3.154 \times 10^{-3}$}
\label{tab_RS_results}
\begin{tabular}{ccccccc}
\toprule
\multirow{2}{*}{\( N \)} & \multicolumn{2}{c}{ALR} & \multicolumn{2}{c}{Static} & \multicolumn{2}{c}{MCS} \\ \cmidrule{2-7}
 & Median \( \hat{P}_f \) & CoV & Median \( \hat{P}_f \) & CoV & Median \( \hat{P}_f \) & CoV \\
\midrule
50   & $3.052 \times 10^{-3}$ & $33.6\%$ & $3.748 \times 10^{-3}$ & $52.0\%$ & — & — \\
100   & $3.086 \times 10^{-3}$ & $16.8\%$ & $2.363 \times 10^{-3}$ & $20.5\%$ & — & — \\
250   & $3.228 \times 10^{-3}$ & $6.4\%$ & $2.708 \times 10^{-3}$ & $28.4\%$ & $4.000\times 10^{-3}$ & $90.2\%$\\
500   & $3.246 \times 10^{-3}$ & $4.1\%$ & $3.165 \times 10^{-3}$ & $26.2\%$ & $2.000\times 10^{-3}$ & $82.9\%$ \\
1,000 & $3.226 \times 10^{-3}$ & $3.7\%$ & $3.016 \times 10^{-3}$ & $25.7\%$ & $3.000\times 10^{-3}$ & $61.1\%$ \\
\bottomrule
\end{tabular}
\end{table}

\begin{figure}[H]
     \centering
     \begin{subfigure}[c]{0.49\textwidth}
         \centering
         \includegraphics[width=0.9\textwidth]{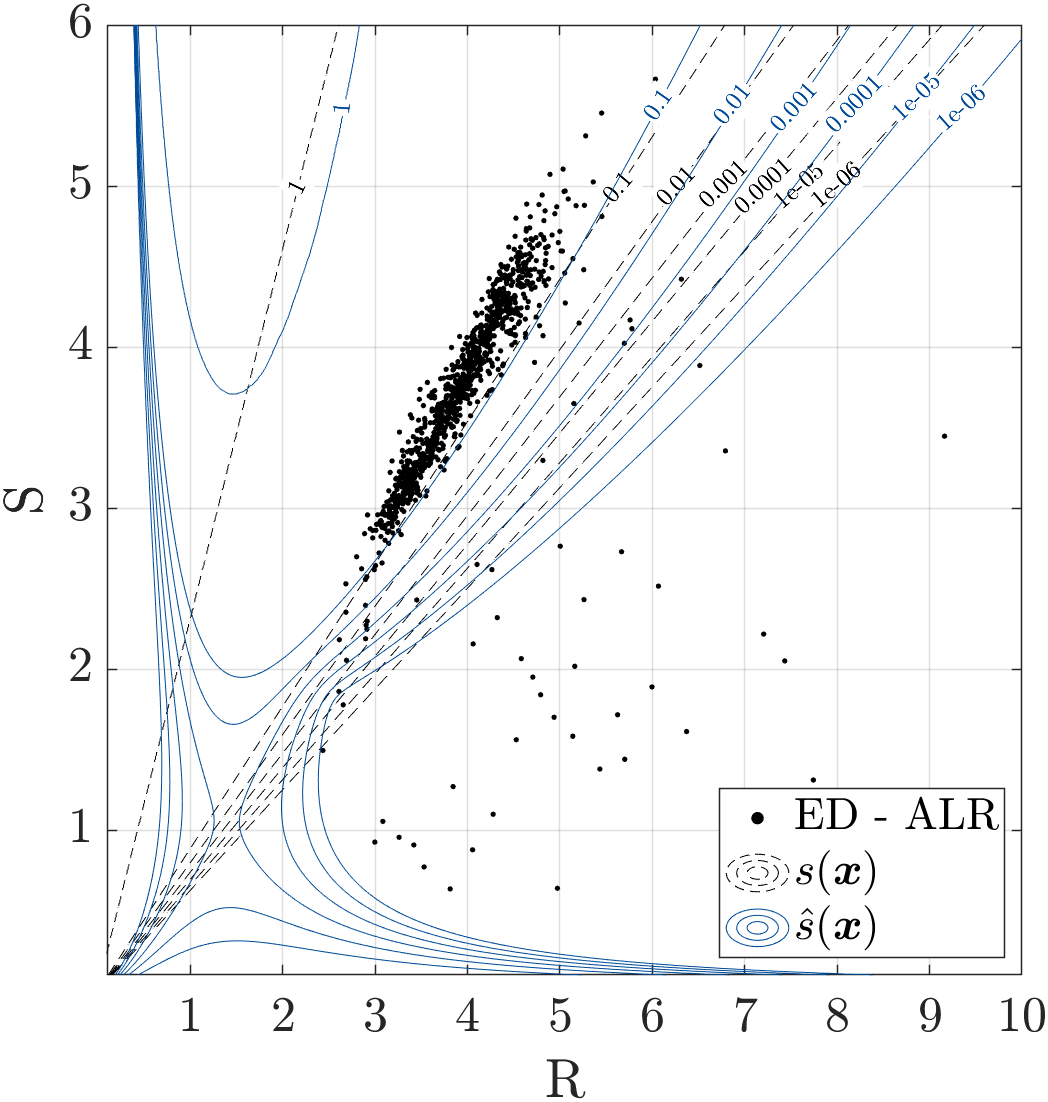}
         \caption{ALR estimate (blue), and experimental design points (black dots).}
         \label{fig_contour_ALR_final}
     \end{subfigure}
     \begin{subfigure}[c]{0.49\textwidth}
         \centering
         \includegraphics[width=0.9\textwidth]{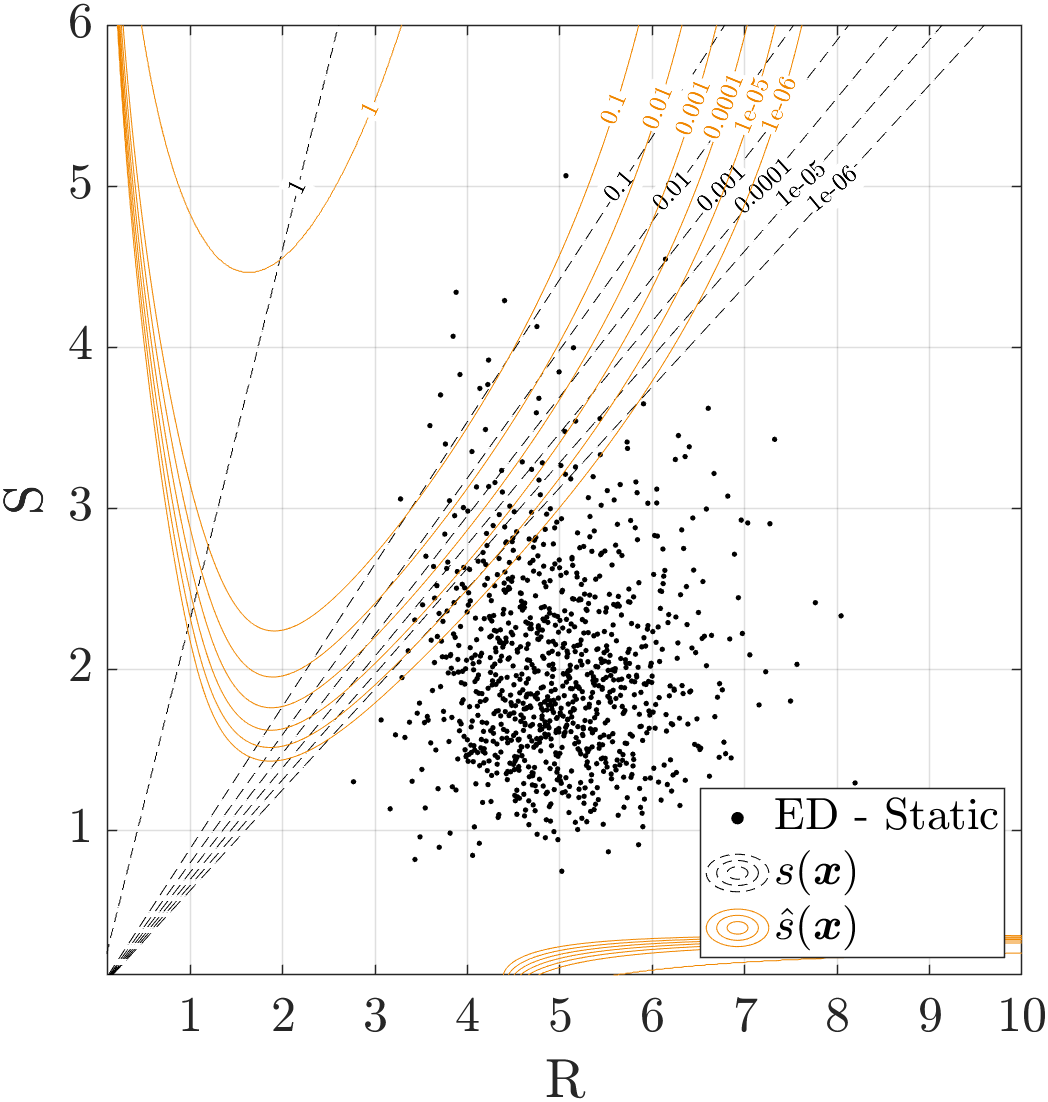}
         \caption{Static ED approach (orange), and experimental design points (black dots).}
         \label{fig_contour_StaticED_final}
     \end{subfigure}
     \caption{R-S example: Comparison between the reference conditional failure probability function (dashed black) and emulator-based estimates (colored) from the run yielding the median $\hat{P}_f$ using $N = 1{,}000$ points.}
     \label{fig_comparison_contours_final}
\end{figure}

In contrast to the deterministic setting, reliability analysis using stochastic simulators does not allow for a direct comparison of the limit-state surfaces. Instead, we compare the conditional failure probability function $s(\ve{x})$, since an accurate estimation of $\hat{s}(\ve{x})$ directly translates into  accurate failure probability estimates $\hat{P}_f$.  In particular, \cref{fig_comparison_contours_final} shows contour plots for both quantities for emulator runs that correspond to the median estimate of $\hat{P}_f$ when $N = 1{,}000$ (in blue for ALR and orange for static ED) and the analytical solution (in black). First, we observe that the ED points cluster around the line $R=S$ in the ALR case (\cref{fig_contour_ALR_final}), as opposed to the scatter plot center around the mean values of $R$ and $S$ for the static design (\cref{fig_contour_StaticED_final}). The contours of the conditional probability of failure $\hat{s}\prt{\ve{x}}$ are also much closer to their analytical counterpart in the ALR case.

\begin{figure}[H]
     \centering
     \includegraphics[scale=0.4]{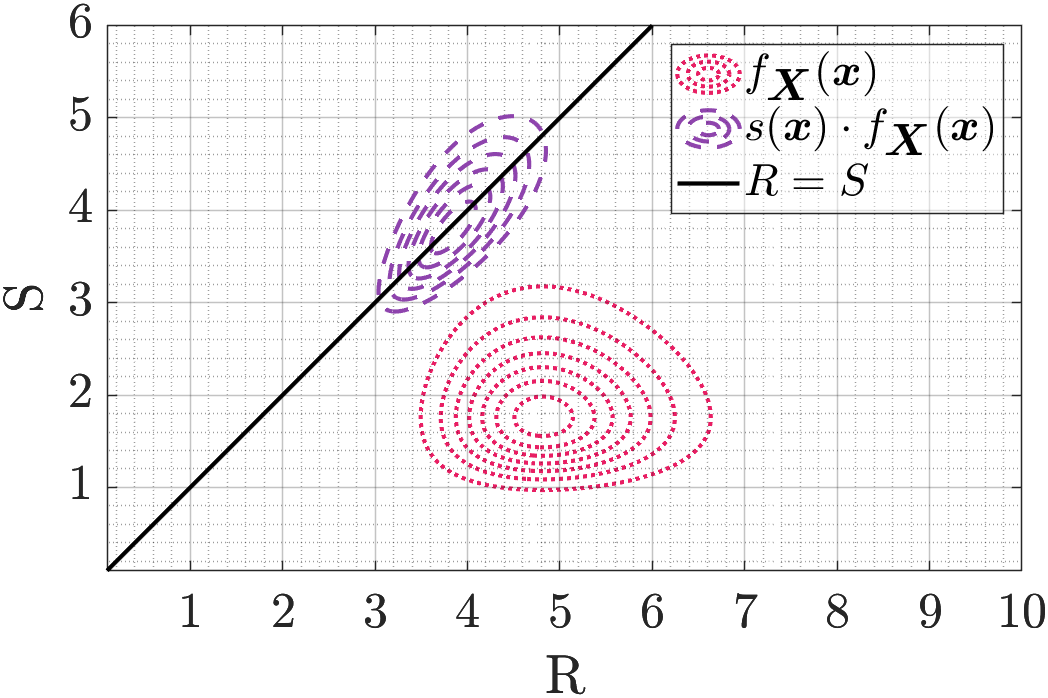}
     \caption{R-S example: Contour plots of the input space and failure region. The red dotted line shows the contour of the joint probability density function $f_{\ve{X}}(\ve{x})$. The purple dashed line shows the contour of the product $s(\ve{x}) \cdot f_{\ve{X}}(\ve{x})$, highlighting the region that contributes most to the failure probability, which is centered around the line $R=S$.}
     \label{fig_contour_PDF_sx_fx_RS}
\end{figure}

To illustrate and better understand this behavior, \cref{fig_contour_PDF_sx_fx_RS} shows the contours of the input density $f_{\ve{X}}(\ve{x})$ and of the product $s(\ve{x}) \cdot f_{\ve{X}}(\ve{x})$. Since $P_f$ is computed as the integral of this product, an accurate approximation of this quantity is essential. Importantly, this does not require $s(\ve{x})$ to be accurate everywhere, but only in regions where the product is significantly large, which are typically limited areas of the input domain, as illustrated in \cref{fig_contour_PDF_sx_fx_RS}.

\begin{figure}[H]
 \vspace{-1cm}
     \centering
     \begin{subfigure}[c]{\textwidth}
         \centering
         \includegraphics[scale=0.3]{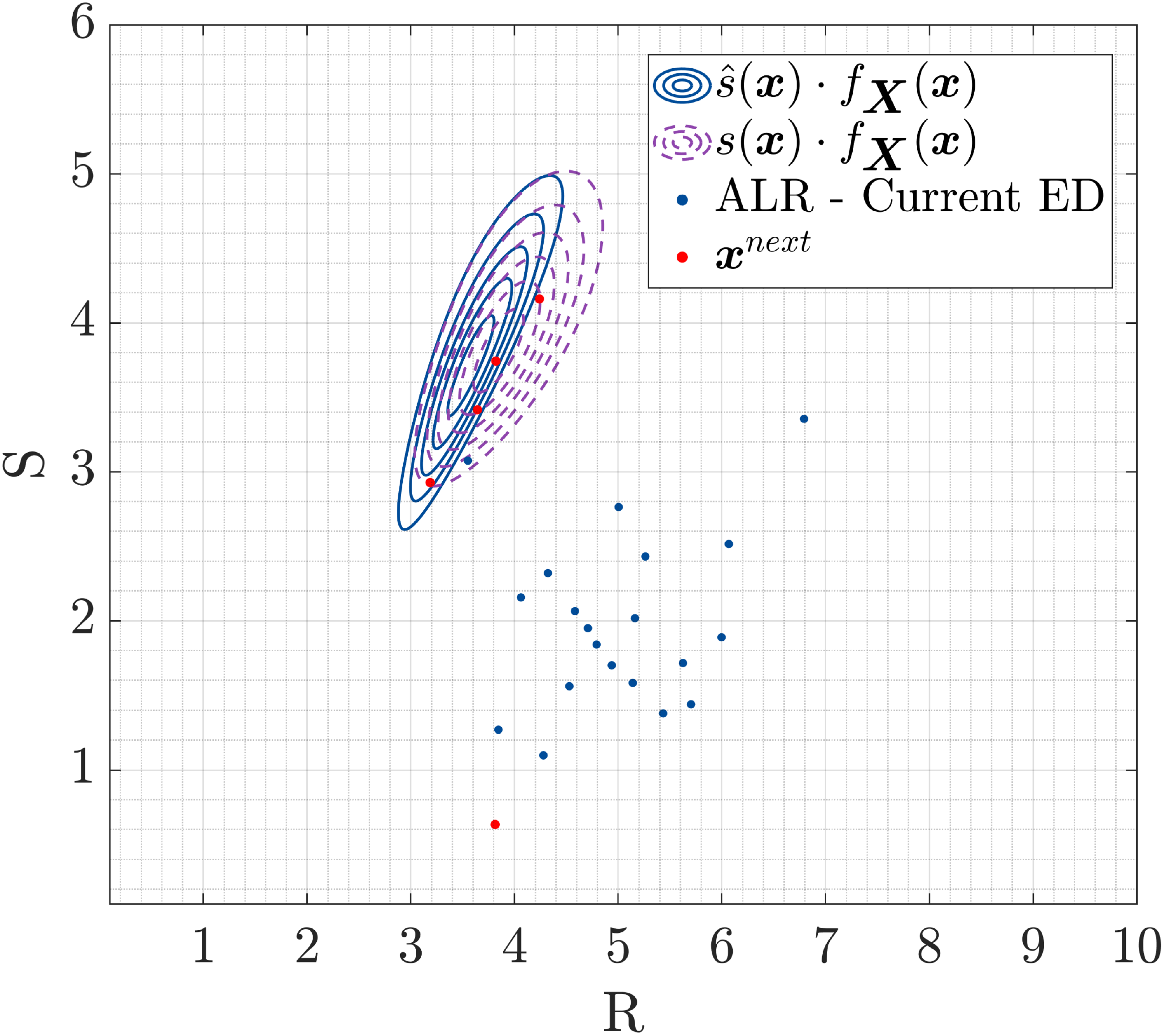}
         \caption{$N=20$}
         \label{fig_comparison_ED20}
     \end{subfigure}
     \begin{subfigure}[c]{\textwidth}
         \centering
         \includegraphics[scale=0.3]{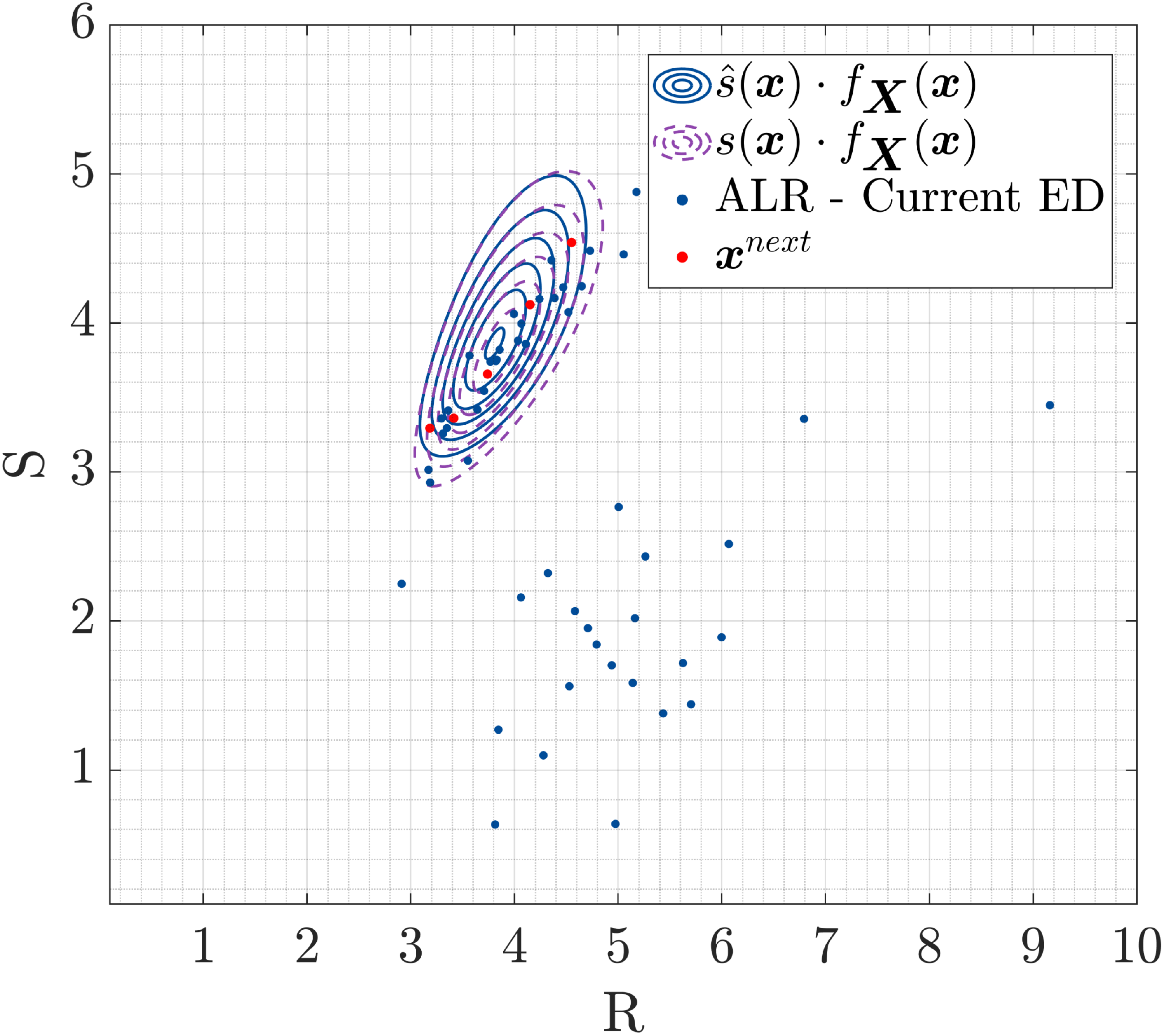}
         \caption{$N=50$}
         \label{fig_comparison_ED50}
     \end{subfigure}
     \begin{subfigure}[c]{\textwidth}
         \centering
         \includegraphics[scale=0.3]{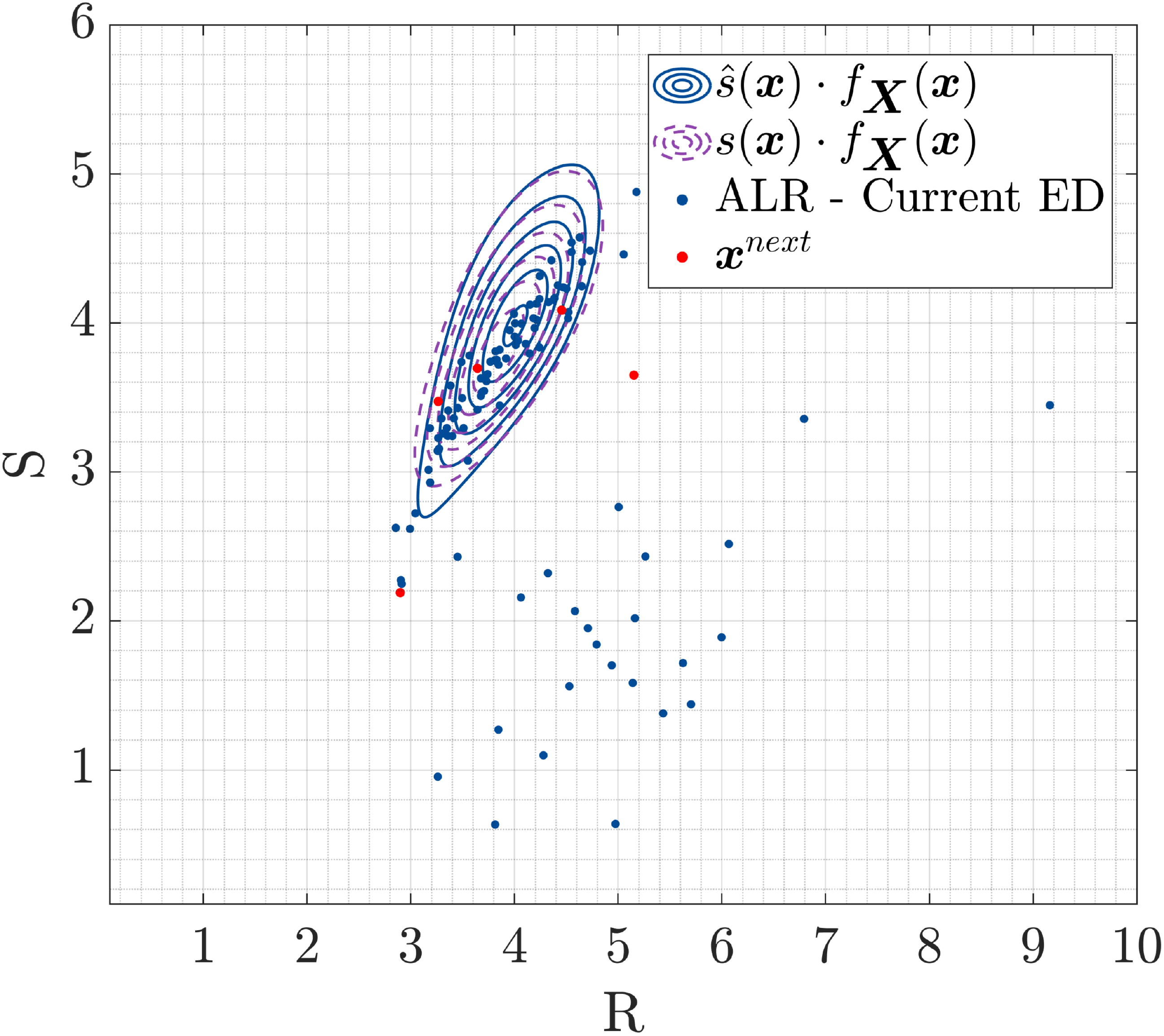}
         \caption{$N=100$}
         \label{fig_comparison_ED100}
     \end{subfigure}
     \caption{R-S example: Comparison of $\hat{s}(\ve{x}) \cdot f_{\ve{X}}(\ve{x})$ (blue) and the reference product $s(\ve{x}) \cdot f_{\ve{X}}(\ve{x})$ (purple) for different $N$. Current ED points are shown in blue and newly added points are marked in red.}
     \label{fig_ALR_steps}
\end{figure}

Additionally, \cref{fig_comparison_ED20,fig_comparison_ED50,fig_comparison_ED100} offer a detailed view of the ALR mechanism by showing intermediate steps of the enrichment phase, namely when $N=20$, $N=50$ and $N=100$. In the left column, the candidate set is colored according to the score obtained by the learning function. Points in blue have higher scores and are more likely to be selected, while gray regions are considered less informative. In the right column, we display $\hat{s}(\ve{x}) \cdot f_{\ve{X}}(\ve{x})$ in blue and the reference product in purple. Existing ED points are shown in blue, and newly added points in red.

\cref{fig_ALR_steps} reveals that even for $N = 20$, the ALR approach already identifies the region that contributes the most to the estimation of the probability of failure. Moreover, it also depicts that occasionally, points are selected in low-score regions. This behavior results from the clustering applied to the candidate set, which promotes exploration by encouraging diversity among selected points. In this case, the active learning algorithm selects points primarily because they lie in regions of high input density, even if the local variance $\Var{\hat{s}\prt{\ve{x}}}$ is low. As the experimental design grows, the emulator improves locally in the relevant region. At $N = 100$, the contours of the product are remarkably close to the reference ones, indicating that the emulator gained local accuracy.

\begin{figure}[H]
     \centering
     \begin{subfigure}[c]{0.49\textwidth}
         \centering
         \includegraphics[width=0.9\textwidth]{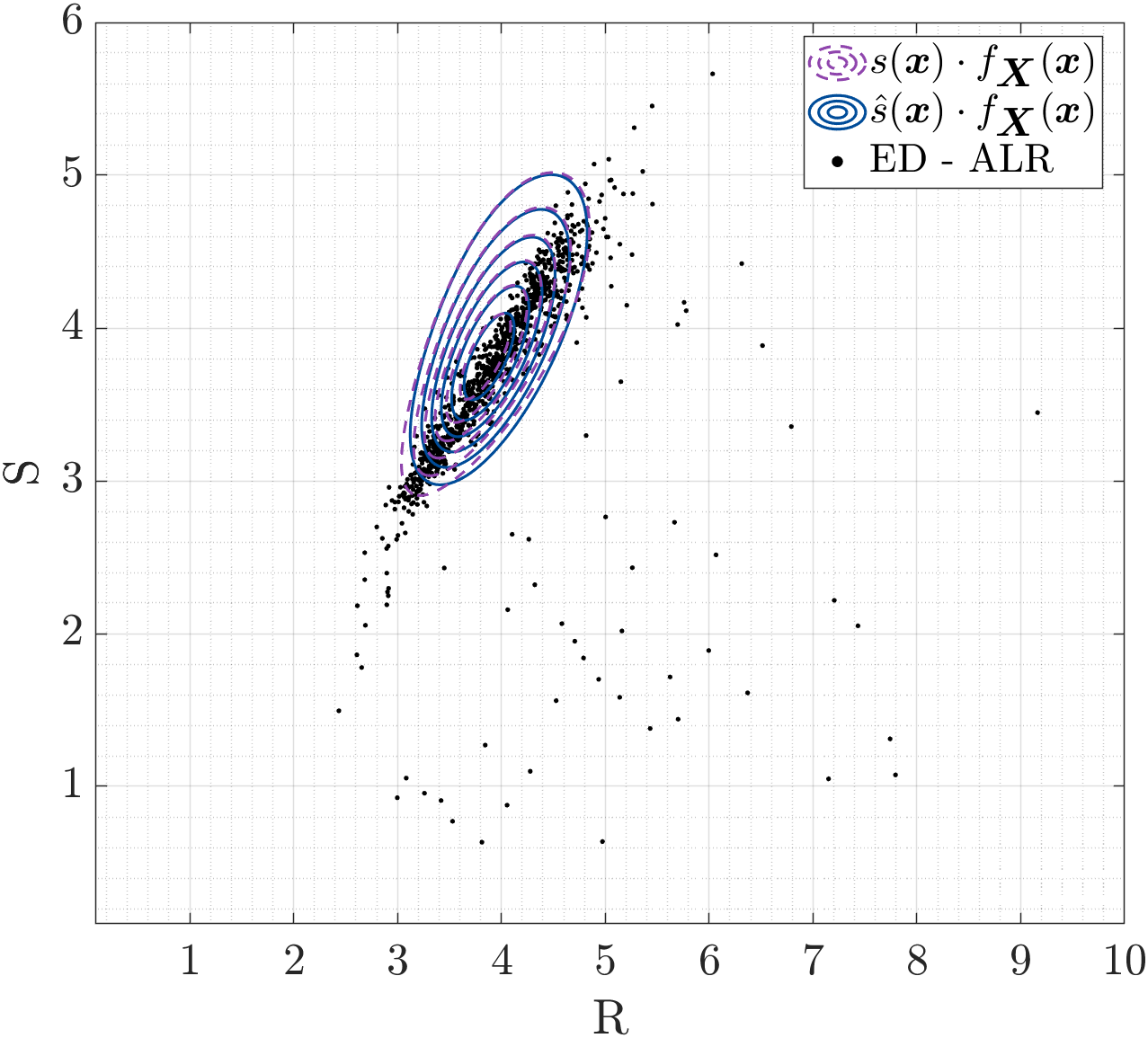}
         \caption{ALR-based estimate $\hat{s}(\ve{x}) \cdot f_{\mathbf{X}}(\ve{x})$ (blue) and reference (dashed black). ED points (black dots).}
         \label{fig_product_ALR_final}
     \end{subfigure}
     \begin{subfigure}[c]{0.49\textwidth}
         \centering
         \includegraphics[width=0.9\textwidth]{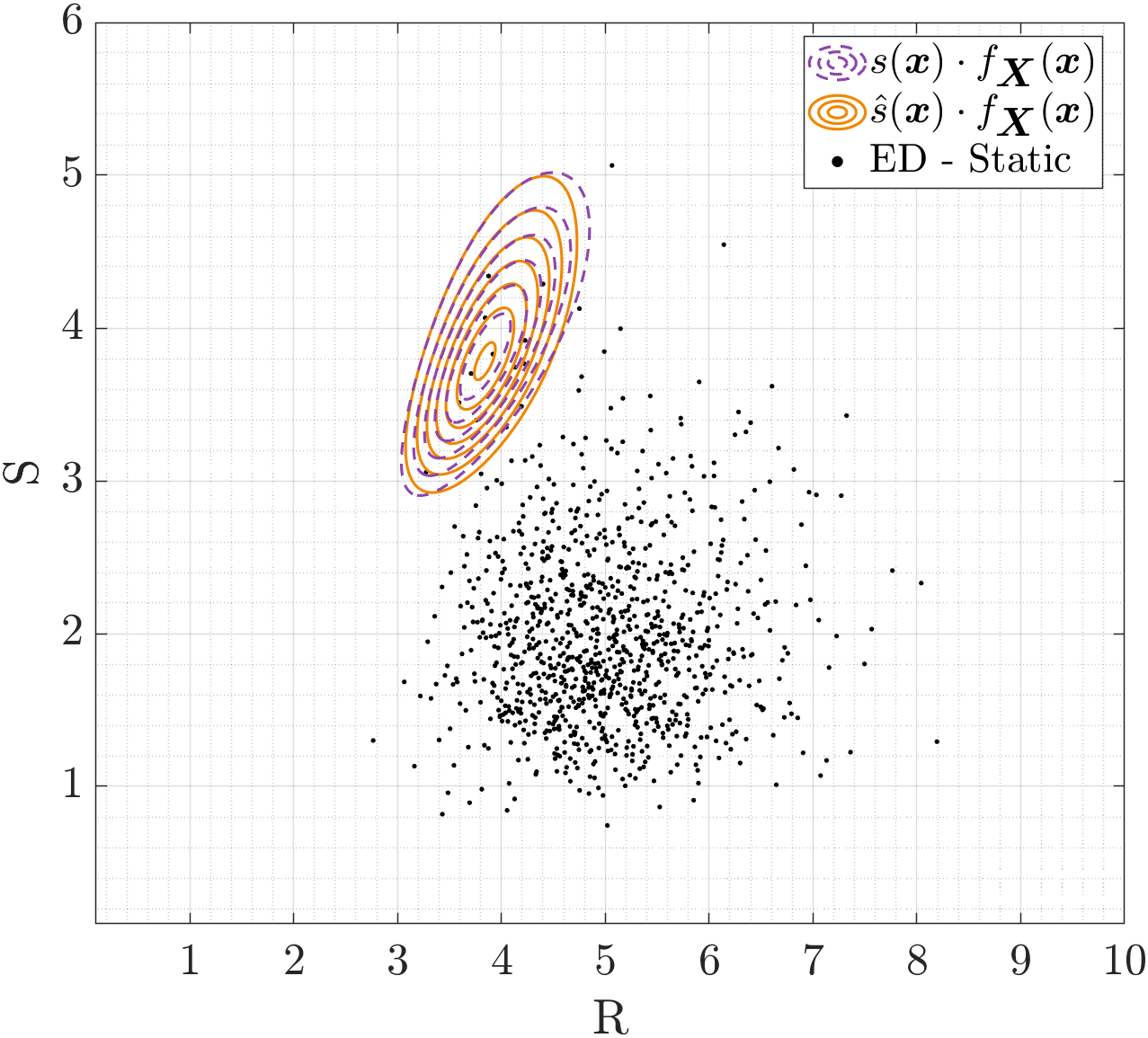}
         \caption{Static ED estimate of $\hat{s}(\ve{x}) \cdot f_{\mathbf{X}}(\ve{x})$ (orange) and reference (dashed black). ED points (black dots).}
         \label{fig_product_Static_final}
     \end{subfigure}
     \caption{R-S example: Comparison of ALR and static ED approaches at $N = 1{,}000$, based on runs yielding the median $\hat{P}_f$.}
     \label{fig_product_ALR_vs_Static}
\end{figure}

The plots for the final iteration ($N = 1{,}000$) are shown in \cref{fig_product_ALR_vs_Static}. The ALR approach concentrates samples along the line $R=S$, corresponding to the ``deterministic'' limit-state function, while the static ED approach covers the input space more uniformly with respect to the input PDF. In this specific case, both methods yield similarly accurate contours since we selected the emulator runs corresponding to the median $\hat{P}_f$. However, because ALR systematically targets the relevant regions, it leads to faster convergence and lower variance across runs, as confirmed by the boxplots in \cref{fig_boxplots_RS_ALR_vs_static}.

\begin{figure}[H]
     \centering
     \begin{subfigure}[t]{0.49\textwidth}
         \centering
         \includegraphics[width=\textwidth]{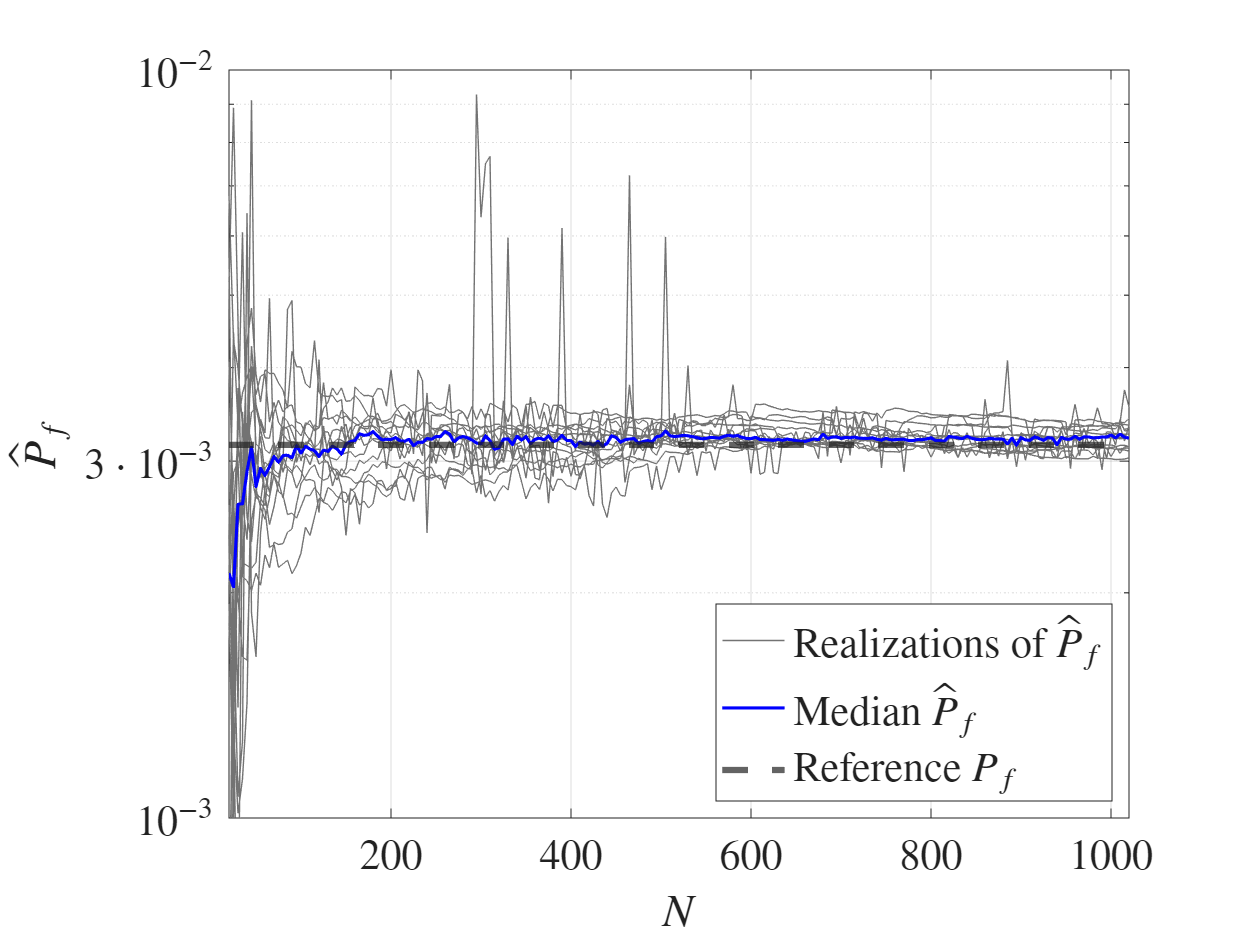}
         \caption{}
         \label{fig_ALR_RS_conv_curve_Pf}
     \end{subfigure}
     \begin{subfigure}[t]{0.49\textwidth}
         \centering
         \includegraphics[width=\textwidth]{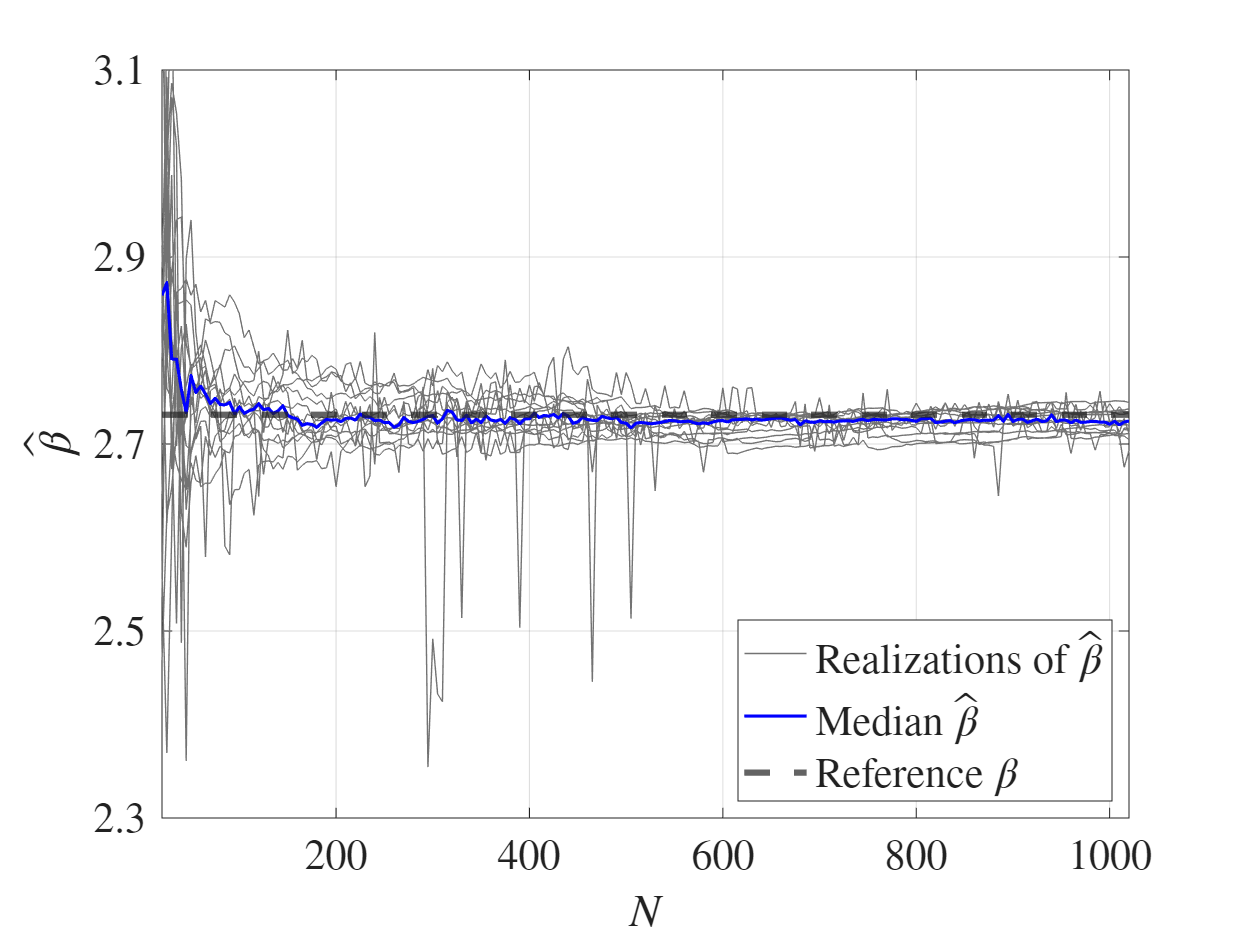}
         \caption{}
         \label{fig_ALR_RS_conv_curve_beta}
     \end{subfigure}
     \caption{R-S example: Convergence of ALR-based failure probability estimates. All 15 ALR runs are shown in gray. The blue curves show the median estimates across the 15 runs at each sample size $N$. (a) Convergence curve of the failure probability $\hat{P}_f$, (b) Convergence curve of the corresponding reliability index $\hat{\beta} = -\Phi^{-1}(\hat{P}_f)$. The dashed lines indicate the reference values.}
     \label{fig_ALR_RS_conv_curve}
\end{figure}

The convergence behavior of the estimated $\hat{P}_f$ across all 15 ALR runs is shown in \cref{fig_ALR_RS_conv_curve}, where the individual trajectories are plotted in gray. The blue curves represent the median estimates across runs at each sample size $N$. Specifically, \cref{fig_ALR_RS_conv_curve_Pf} shows the convergence in terms of the failure probability $\hat{P}_f$, while \cref{fig_ALR_RS_conv_curve_beta} shows the corresponding reliability index $\hat{\beta} = -\Phi^{-1}(\hat{P}_f)$. In both subfigures, the dashed lines indicate the reference values. All individual trajectories have been smoothed out according to the methodology described in \cref{subsec_mitigating_effects}. Although some residual noise remains, most runs stabilize around the true value of $P_f$ after approximately 600 samples. Notably, reasonable estimates can already be achieved with as few as $200$ samples.

\begin{figure}[H]
     \centering
     \begin{subfigure}[c]{0.49\textwidth}
         \centering
         \includegraphics[width=\textwidth]{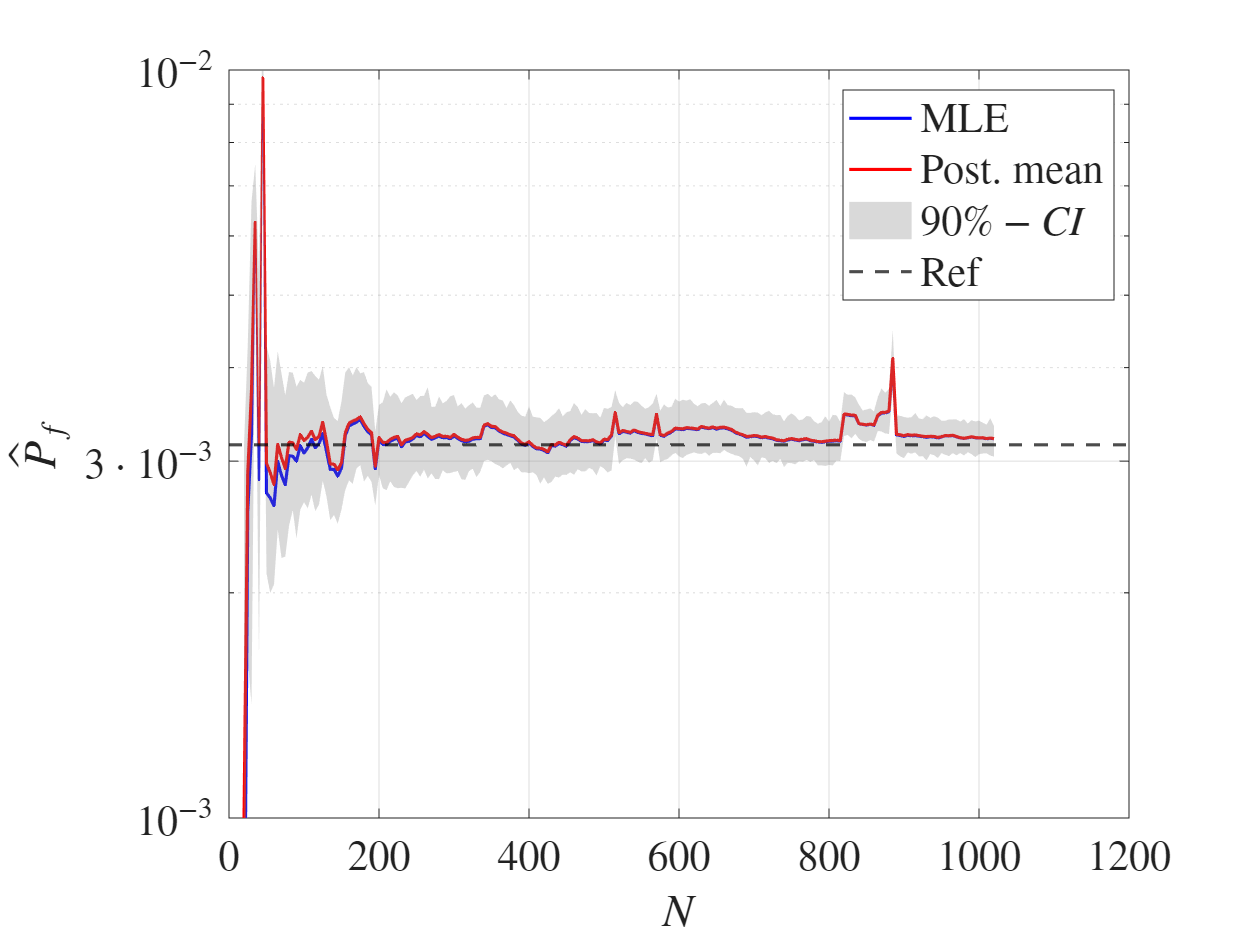}
         \caption{}
         \label{fig_ALR_RS_conv_Pf_CI}
     \end{subfigure}
     \begin{subfigure}[c]{0.49\textwidth}
         \centering
         \includegraphics[width=\textwidth]{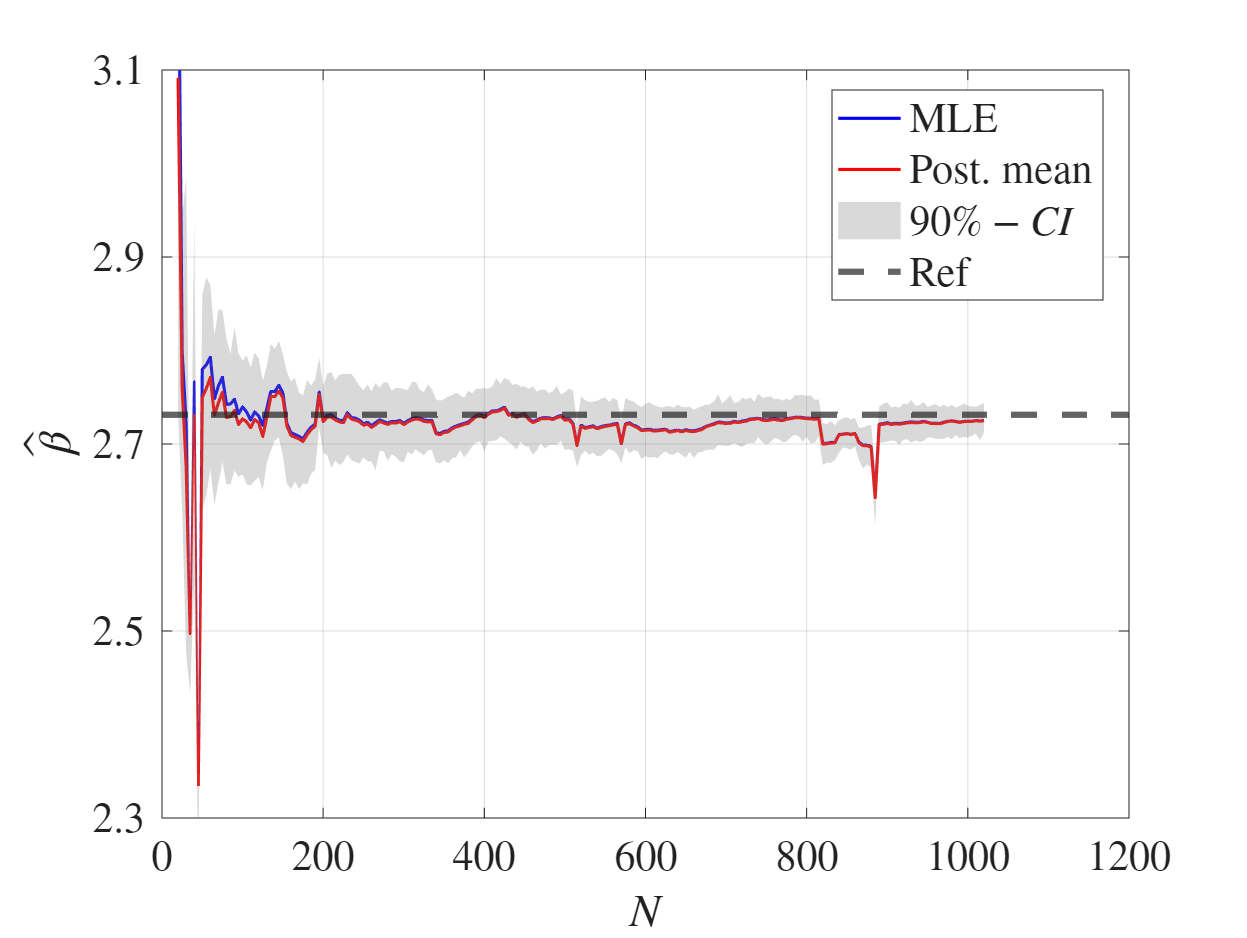}
         \caption{}
         \label{fig_ALR_RS_conv_Beta_CI}
     \end{subfigure}
     \caption{R-S example: Convergence of ALR-based failure probability estimate for the repetition that lead to the median final $\hat{P}_f$ estimate. The blue curve show the MLE-based estimate, while the red one shows the posterior mean. The gray band indicates the $90\%$ confidence interval. (a) Convergence curve of the failure probability $\hat{P}_f$, (b) Convergence curve of the corresponding reliability index $\hat{\beta} = -\Phi^{-1}(\hat{P}_f)$.}
     \label{fig_ALR_RS_conv_CI}
\end{figure}

Figure~\ref{fig_ALR_RS_conv_CI} shows the convergence history for the repetition that lead to the median final $\hat{P}_f$. The blue curve corresponds to the failure probability estimate obtained using the maximum likelihood (MLE) estimate of the SPCE coefficients. The red curve represents the posterior mean estimate obtained by averaging the $\hat{P}_f$ values computed from the resampled SPCE coefficients. The gray band shows the $90\%$ confidence interval defined by the $5$-th and $95$-th percentile of the $\hat{P}_f$ values associated with the resampled coefficients. The MLE-based and posterior estimates have very similar values indicating limited bias due to the uncertainty on the coefficients. Moreover, the confidence interval steadily decreases, showing convergence of the algorithm as the experimental design increases.

\subsection{Stochastic SIR Model}

Our second example is an agent-based simulator: the Susceptible-Infected-Recovered (SIR) model. Widely used in epidemiology, this model simulates the outbreak of an infectious disease spreading through contact between infected and susceptible individuals \citep{Britton_2010}. It assumes a constant population size \( P \), meaning births and deaths are not considered. At any given time \( t \), individuals are divided into three compartments: (1) Susceptible, denoted by $S_t$, are individuals who can contract the disease through contact with infected individuals;  (2) Infected, denoted by $I_t$, are individuals who are currently contagious;  (3) Recovered, denoted by $R_t$, are those no longer susceptible due to immunity after recovery. Since the total population is constant and only these three states are considered, the full state of the system at any time \( t \) is determined by the pair \( (S_t, I_t) \), as \( R_t = P - S_t - I_t \).

\begin{figure}[H]
    \centering
    \includegraphics[width=0.5\linewidth]{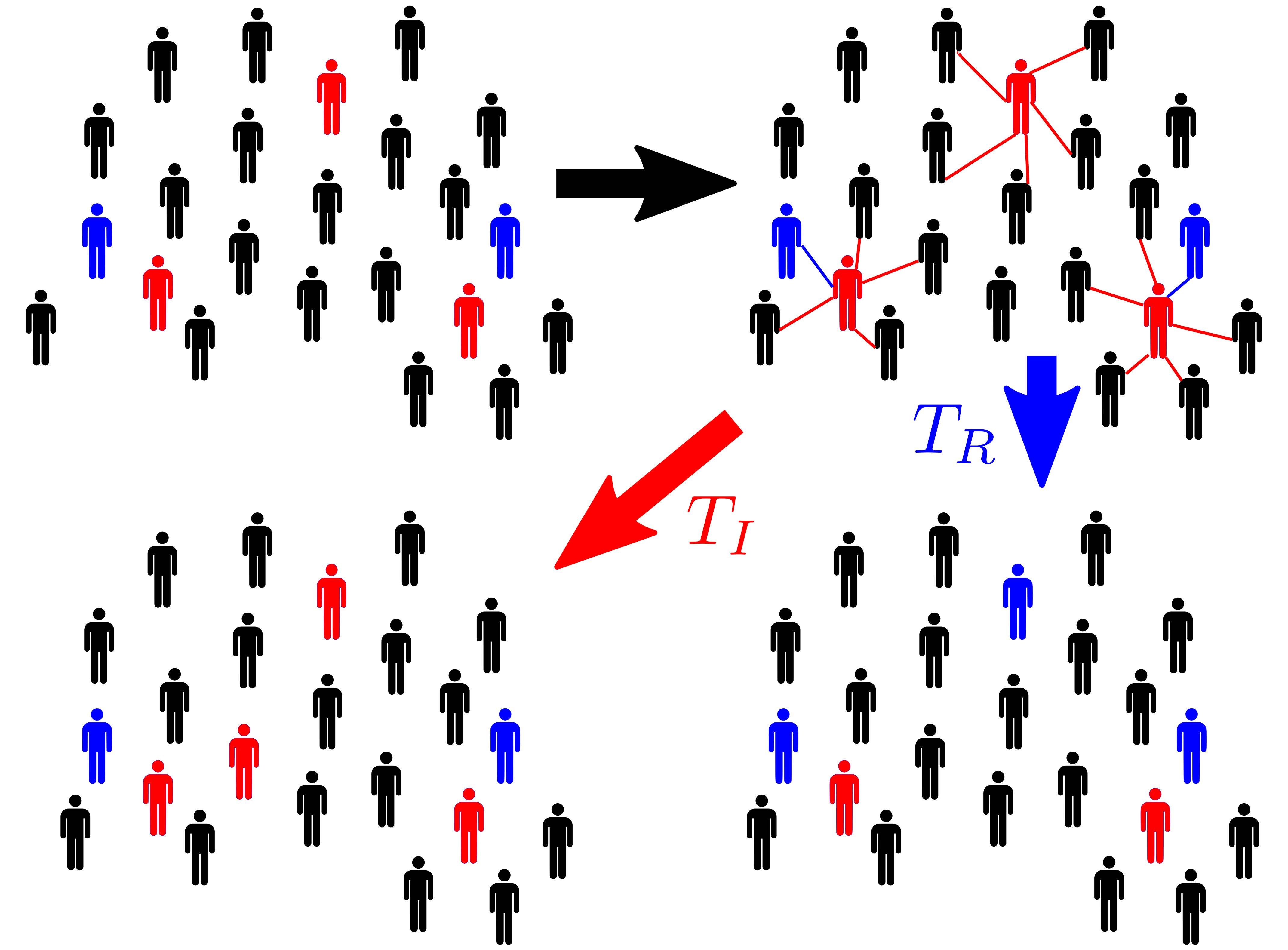}
    \caption{Dynamics of the stochastic SIR model: black icons stand for susceptible individuals, red icons represent infected individuals, and blue icons are the ones who have recovered.}
    \label{fig_SIR_model}
\end{figure}

\cref{fig_SIR_model} illustrates the dynamics of the stochastic SIR model. Susceptible, infected, and recovered individuals are represented by black, red, and blue icons, respectively. The disease propagates through two stochastic transitions: infection and recovery. The infection event leads to a susceptible individual becoming infected, while the recovery event corresponds to an infected individual recovering. These transitions are modeled as random events governed by the exponential distributions:
\begin{equation}
\begin{aligned}
T_I &\sim \operatorname{Exp}(\lambda_I), \quad \lambda_I = \beta \cdot \frac{S_t I_t}{P}, \\
T_R &\sim \operatorname{Exp}(\lambda_R), \quad \lambda_R = \gamma \cdot I_t,
\end{aligned}
\label{eq_update_step_SIR}
\end{equation}
where \( T_I \) and \( T_R \) represent the waiting times until the next infection or recovery event, respectively, and \( \beta \) and \( \gamma \) are the corresponding infection and recovery rates.

The simulation proceeds according to Gillespie's algorithm \citep{Gillespie_1977}, which at each iteration draws a random event (infection or recovery) and updates the state of the system accordingly. The outbreak eventually ends when no infected individuals remain. Given the finite population and the fact that recovered individuals cannot be reinfected, the process always terminates in finite time \( T < + \infty \).

The input vector for this simulator is \( \boldsymbol{X} = \{S_0, I_0, \beta, \gamma\} \), where \( S_0 \) and \( I_0 \) and the infection and recovery rates \( \beta \) and \( \gamma \). The simulator output, denoted by \( \mathrm{SIR}(\boldsymbol{X}, \omega) \), corresponds to the total number of new infections observed over the course of the outbreak, excluding the initially infected individuals. This quantity is an indicator related to the severity of the outbreak and guides the need for mitigation strategies \citep{Binois2018}.

For reliability analysis, we define a failure event as the number of newly infected individuals that exceed a given threshold \( I^\mathrm{lim} \). The associated stochastic limit-state function is given by:
\begin{equation}
g_s(\boldsymbol{X}, \omega) = I^\mathrm{lim} - \mathrm{SIR}(\boldsymbol{X}, \omega).
\end{equation}

The randomness in this model arises from the latent random variables \( T_I \) and \( T_R \), which determine the sequence and outcomes of the simulated transitions. As a consequence, the number of latent variables and the stochastic dimension of the simulator is itself a random quantity, implicitly defined by the progression of the outbreak.

In our experiments, we consider a population of size \( P = 2{,}000 \). All components of the input vector \( \boldsymbol{X} = \{S_0, I_0, \beta, \gamma\} \) are assumed to follow uniform distributions. \tabref{tab_SIR} summarizes the distributions along with their respective bounds. The uncertainties in \( S_0 \) and \( I_0 \) reflect the lack of knowledge of the initial state of the system, while those in \( \beta \) and \( \gamma \) capture variability in disease transmission and recovery, potentially influenced by external interventions such as quarantine measures or medical treatments.
\begin{table}[H]
\centering
\caption{Distributions and support of the input variables in the stochastic SIR model.}
\label{tab_SIR}
\begin{tabular}{@{}ccc@{}}
\toprule
Variable & Distribution & Support \\ \midrule
\( S_0 \)  & Uniform & \( \mathcal{U}(1200,\ 1800) \) \\
\( I_0 \)  & Uniform & \( \mathcal{U}(20,\ 200) \) \\
\( \beta \)  & Uniform & \( \mathcal{U}(0.5,\ 0.75) \) \\
\( \gamma \)  & Uniform & \( \mathcal{U}(0.5,\ 0.75) \) \\ \bottomrule
\end{tabular}
\end{table}

We begin our active learning procedure with an initial experimental design of 100 samples generated via LHS. The failure probability of the emulator $\hat{P}_f$ is estimated using the expression in \cref{eq_Pf_random_variable_POV}, based on a MCS of size $N_{\text{MCS}} = 10^5$. As detailed in \cref{eq_smoothed_Pf}, a moving average with a window of three iterations is employed to smooth the sequence of failure probability estimates. The candidate set for the ALR approach consists of $10^4$ points, and the experimental design is enriched by 5 points per iteration until a total of 2,000 points is reached. During training, the SPCE settings include degree adaptivity in the range $p \in [1, 5]$, and the latent variable is assumed to follow a Gaussian distribution.

\begin{figure}[H]
\centering
\includegraphics[width=0.65\linewidth]{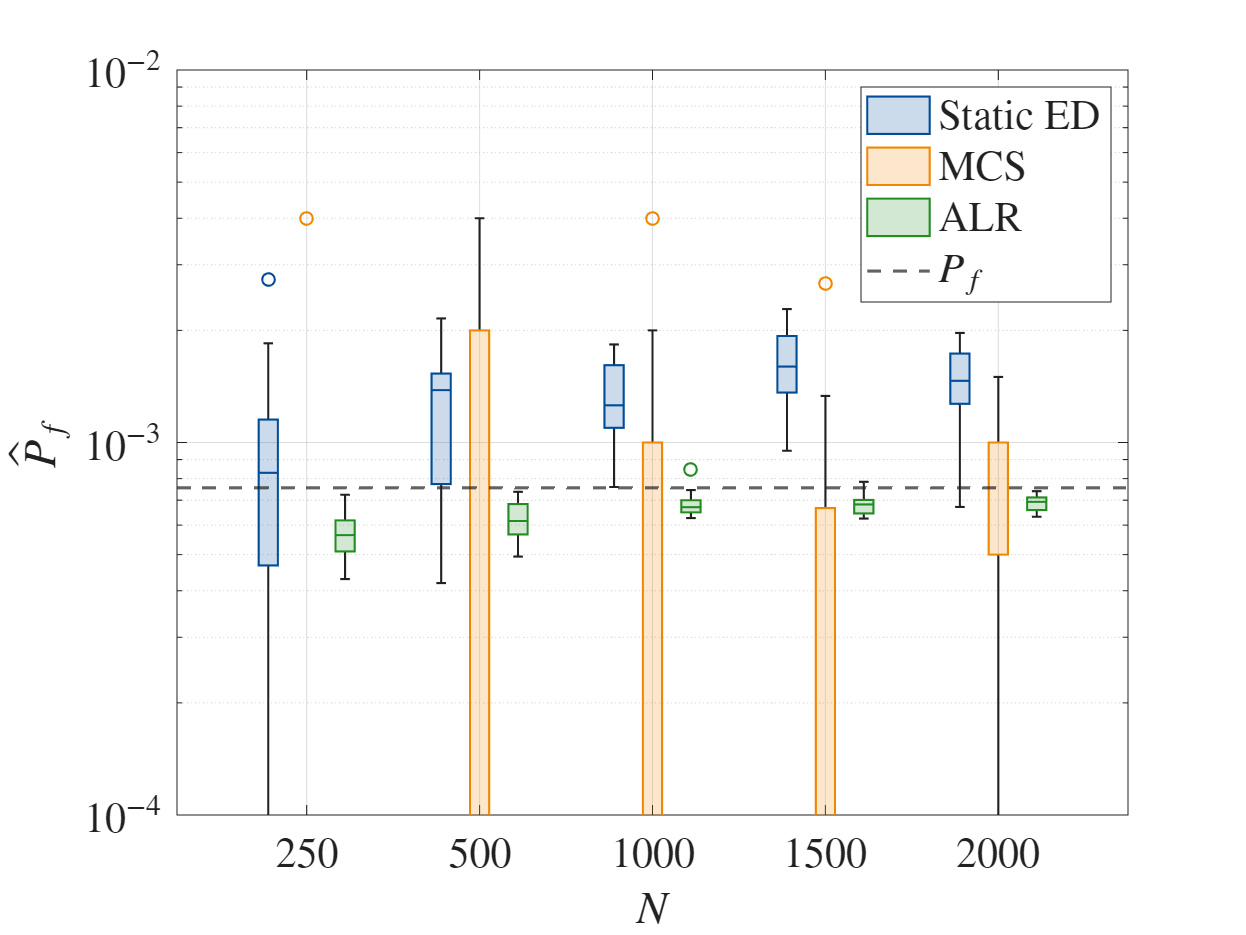}
\caption{ SIR example: Boxplots comparing convergence of the ALR method (in green), SPCE with a static ED approach (in blue), and direct MCS (in orange) across 15 replications. The dashed line indicates the reference $P_f$.}
\label{fig_boxplots_SIR_ALR_vs_static}
\end{figure}

\cref{fig_boxplots_SIR_ALR_vs_static} presents results from 15 independent runs of the experiment. The active learning approach is shown in green, the SPCE approach with a static experimental design in blue, and direct MCS results in orange. In the static approach, the SPCE emulator is trained on an ED of $N$ samples generated via LHS. The emulator settings are identical to those used in the active learning approach, with degree adaptivity in the range $p \in [1, 5]$ and a Gaussian latent variable model. The black dashed line indicates the reference probability of failure $P_f$. Since no analytical solution is available in this case, the reference value is obtained via large-scale MCS with $N_{\text{MCS}} = 10^8$, yielding $P_f = 7.511 \times 10^{-4}$ and a coefficient of variation of approximately $0.5\%$. Despite a slight bias, the ALR approach converges significantly faster than the other two methods, providing reasonable estimates of $P_f$ with as few as 250 samples. The robustness against statistical uncertainty, as measured by the extremely small width of the (blue) box plots is also remarkable.
\begin{table}[H]
\centering
\caption{SIR Model: Comparison of ALR, static ED, and MCS across multiple runs for different sample sizes. Reference $P_f=7.511\times 10^{-4}$.}
\label{tab_additional_results}
\begin{tabular}{ccccccc}
\toprule
\multirow{2}{*}{\( N \)} & \multicolumn{2}{c}{ALR} & \multicolumn{2}{c}{Static ED} & \multicolumn{2}{c}{MCS} \\ \cmidrule{2-7}
 & Median \( \hat{P}_f \) & CoV & Median \( \hat{P}_f \) & CoV & Median \( \hat{P}_f \) & CoV \\
\midrule
250   & $5.642 \times 10^{-4}$ & $14.2\%$  & $8.293 \times 10^{-4}$ & $73.6\%$ & $-$ & $-$ \\
500   & $6.156 \times 10^{-4}$ & $11.4\%$  & $1.382 \times 10^{-3}$ & $40.0\%$ & $-$ & $-$ \\
1000  & $6.704 \times 10^{-4}$ & $8.1\%$  & $1.258 \times 10^{-3}$ & $23.6\%$ & $-$  & $-$  \\
1500  & $6.818 \times 10^{-4}$ & $6.1\%$  & $1.599 \times 10^{-3}$ & $23.7\%$ & $-$  & $-$  \\
2000  & $6.929 \times 10^{-4}$ & $4.5\%$  & $1.464 \times 10^{-3}$ & $26.8\%$ & $5.000 \times 10^{-4}$  & $75.4\%$  \\
\bottomrule
\end{tabular}
\end{table}

\begin{figure}[H]
     \centering
     \begin{subfigure}[c]{0.49\textwidth}
         \centering
         \includegraphics[width=\textwidth]{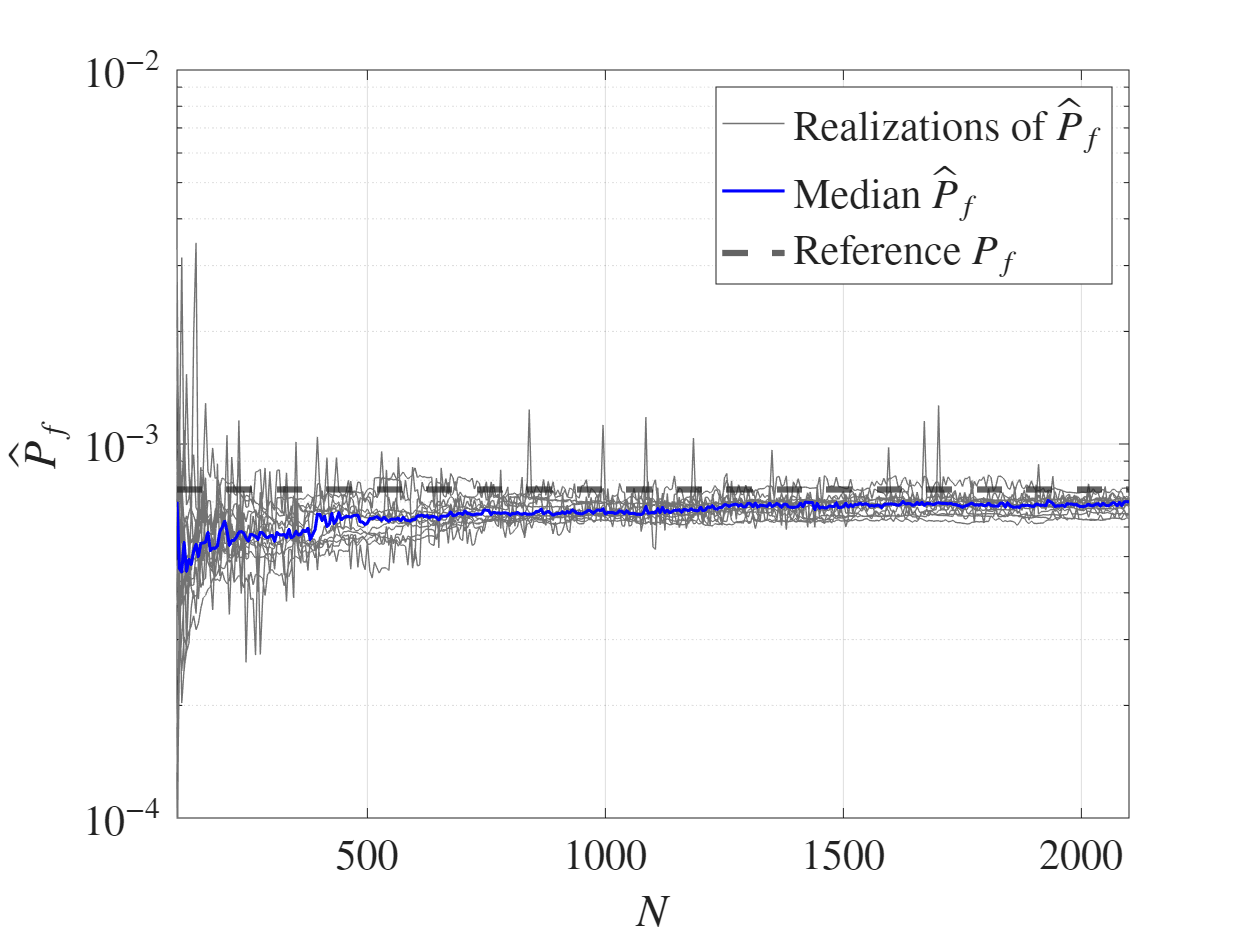}
         \caption{}
         \label{fig_ALR_SIR_conv_curve_Pf}
     \end{subfigure}
     \begin{subfigure}[c]{0.49\textwidth}
         \centering
         \includegraphics[width=\textwidth]{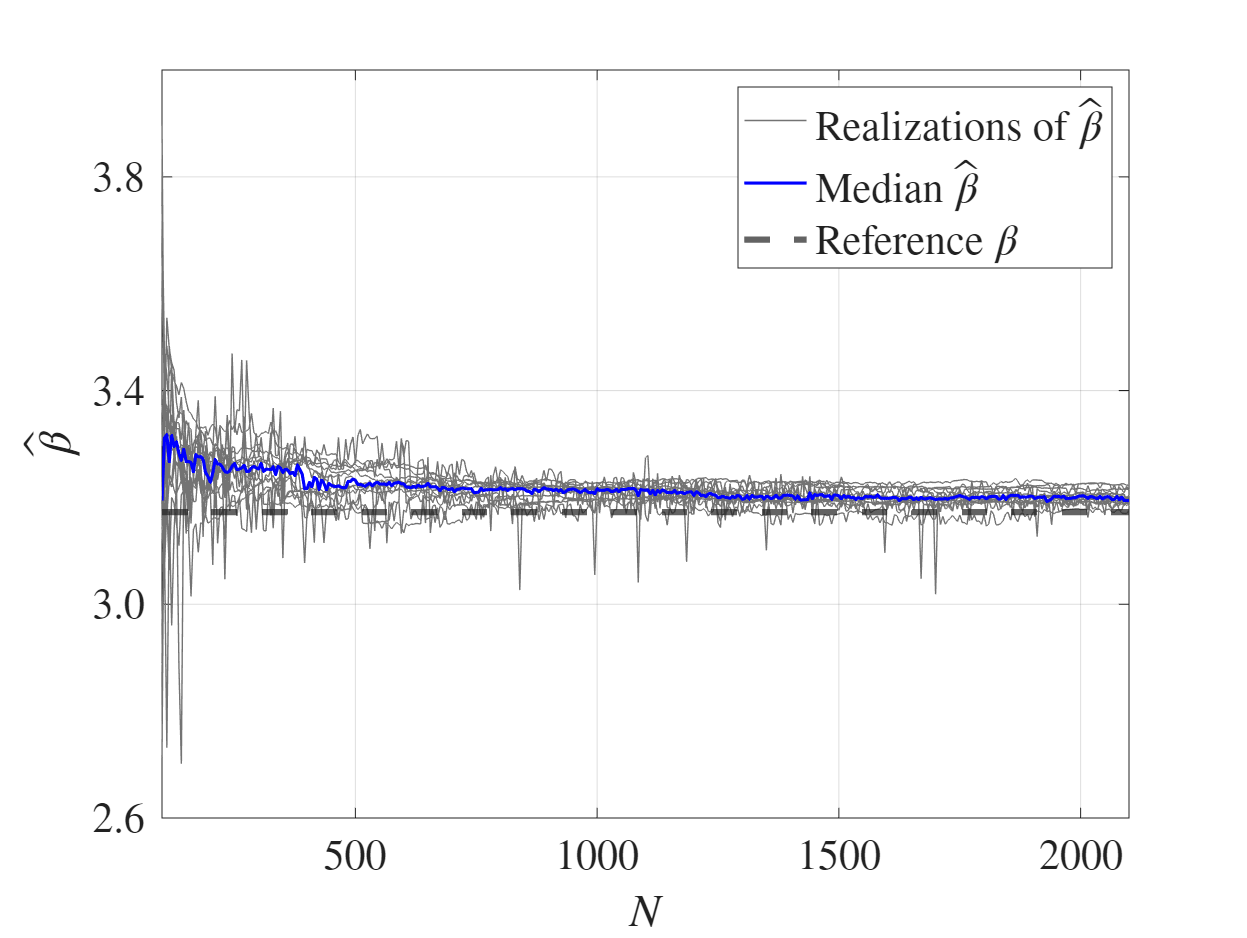}
         \caption{}
         \label{fig_ALR_SIR_conv_curve_beta}
     \end{subfigure}
     \caption{SIR example: Convergence of ALR-based failure probability estimates. All 15 ALR runs are shown in gray. The blue curves show the median estimates across the 15 runs at each sample size $N$. (a) Convergence curve of the failure probability $\hat{P}_f$, (b) Convergence curve of the corresponding reliability index $\hat{\beta} = -\Phi^{-1}(\hat{P}_f)$. The dashed lines indicate the reference values.}
     \label{fig_ALR_SIR_conv_curve}
\end{figure}

\cref{fig_ALR_SIR_conv_curve} shows the convergence curves of all 15 ALR runs in gray, with the median estimates across runs at each sample size $N$ highlighted in blue. The figure is composed of two subfigures: \cref{fig_ALR_SIR_conv_curve_Pf} displays the convergence of the failure probability $\hat{P}_f$, while \cref{fig_ALR_SIR_conv_curve_beta} shows the corresponding reliability index $\hat{\beta} = -\Phi^{-1}(\hat{P}_f)$. Initially, the limited experimental design produces an inaccurate model, which leads to an underestimation of the failure probability. As the algorithm iteratively adds samples to the ED, the estimate rapidly improves. The convergence stabilizes around 800 samples. Similarly to the previous example, minor fluctuations in the curve are observed due to the stochastic nature of the simulator. Most runs exhibit a slight bias at this stage, but are expected to converge fully with additional enrichment steps.

We observe in both  \cref{fig_boxplots_SIR_ALR_vs_static} and \cref{fig_ALR_SIR_conv_curve} that the ALR approach produces slightly biased results and consistently underestimates the failure probability. Interestingly, the static approach also yields biased results, but in the opposite direction, \emph{i.e.}, overestimating the probability of failure. We interpret this behavior as a consequence of a specific characteristic of the problem, namely a strong degree of heteroskedasticity in the conditional model response. More precisely, there are likely small regions of the input space where the tail behavior differs from that of the bulk and which carry non-negligible probability mass in the failure domain. The active learning approach, which is greedy, may then fail to identify those regions when initialized with a small experimental design. In fact, this is the most common cause of bias for classical ALR algorithms in problems featuring multiple failure domains. In contrast, the static ED approach is more likely to cover the entire space, including these small regions of potentially high variance, which may result in an overall higher-variance SPCE model. Overall, this points to a favorable bias–variance trade-off for ALR, which exhibits a small bias while delivering significantly more stable estimates. In both cases, however, we expect this bias to vanish as the size of the training dataset increases.

\begin{figure}[H]
     \centering
     \begin{subfigure}[c]{0.49\textwidth}
         \centering
         \includegraphics[width=\textwidth]{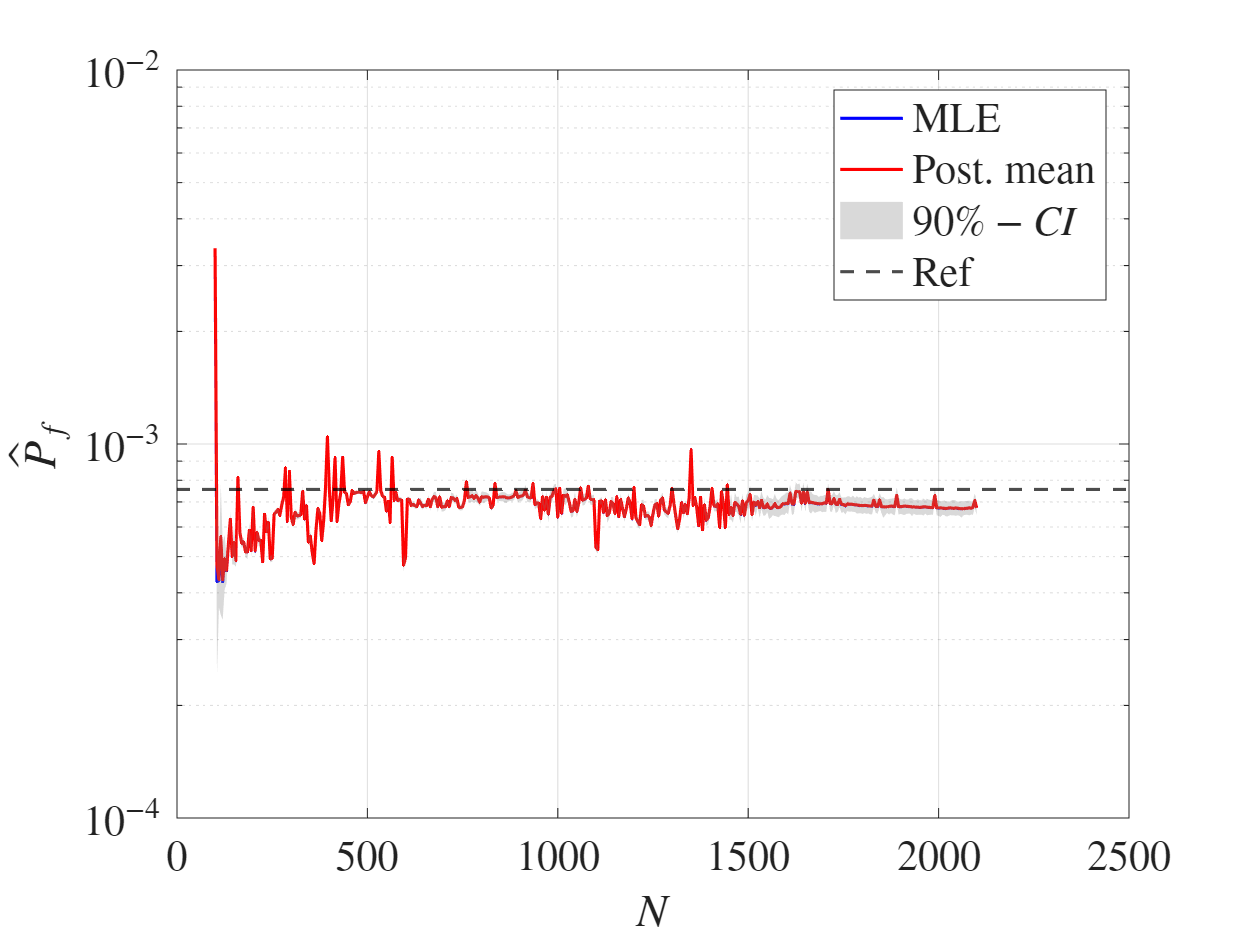}
         \caption{}
         \label{fig_ALR_SIR_conv_CI_Pf}
     \end{subfigure}
     \begin{subfigure}[c]{0.49\textwidth}
         \centering
         \includegraphics[width=\textwidth]{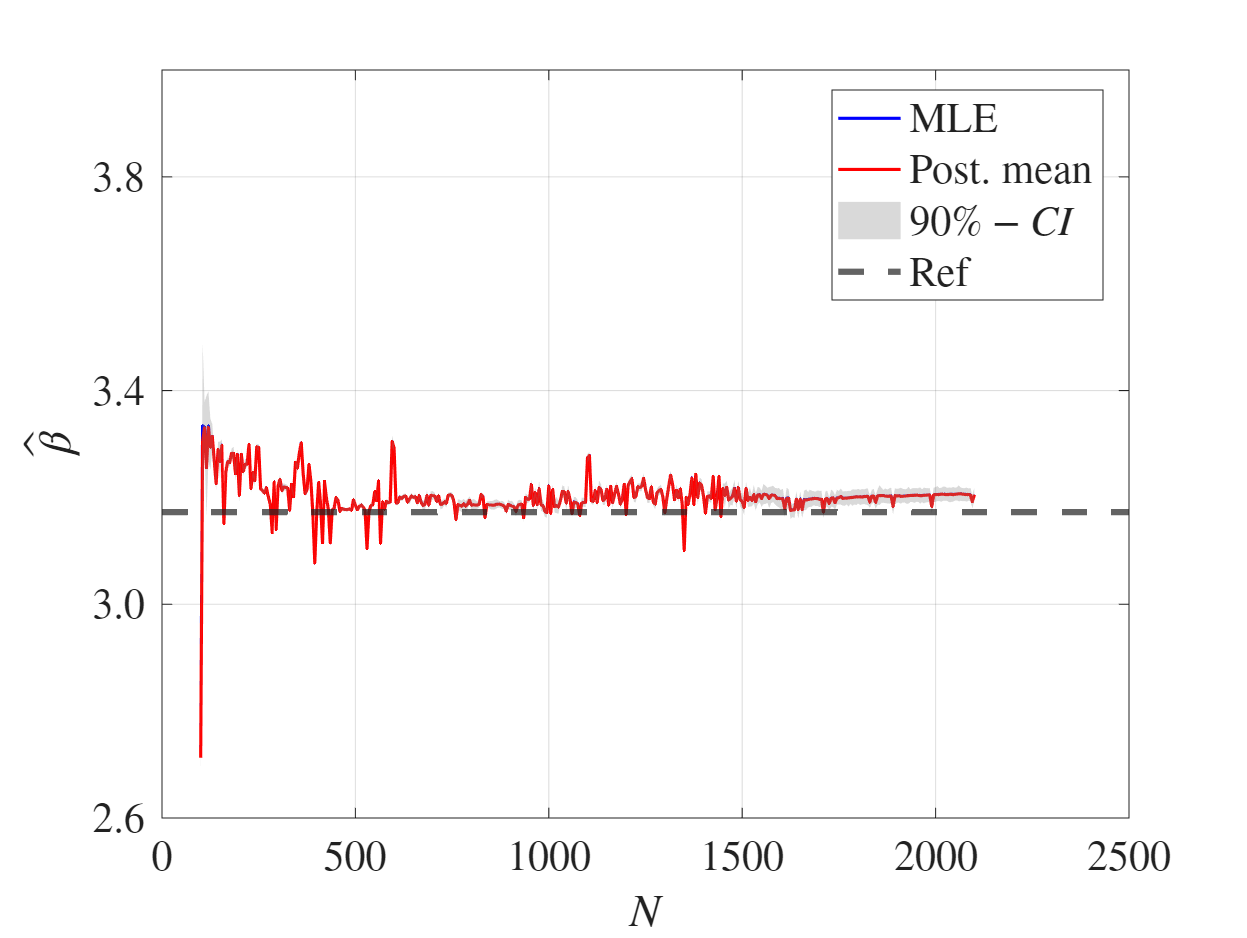}
         \caption{}
         \label{fig_ALR_SIR_conv_CI_beta}
     \end{subfigure}
     \caption{SIR example: Convergence of ALR-based failure probability estimate for the repetition that lead to the median final $\hat{P}_f$ estimate. The blue curve show the MLE-based estimate, while the red one shows the posterior mean. The gray band indicates the $90\%$ confidence interval. (a) Convergence curve of the failure probability $\hat{P}_f$, (b) Convergence curve of the corresponding reliability index $\hat{\beta} = -\Phi^{-1}(\hat{P}_f)$.}
     \label{fig_ALR_SIR_conv_CI}
\end{figure}

Figure~\ref{fig_ALR_SIR_conv_CI} shows the convergence history for the repetition that lead to the median final $\hat{P}_f$. The blue curve corresponds to the failure probability estimate obtained using the maximum likelihood (MLE) estimate of the SPCE coefficients. The red curve represents the posterior mean estimate obtained by averaging the $\hat{P}_f$ values computed from the resampled SPCE coefficients. The gray band shows the $90\%$ confidence interval defined by the $5$-th and $95$-th percentile of the $\hat{P}_f$ values associated with the resampled coefficients. The MLE-based and posterior estimates have very similar values indicating limited bias due to the uncertainty on the coefficients. Moreover, the confidence interval is relatively narrow indicating that the uncertainty on the coefficients has minor impact on the estimated failure probability. This also suggests that the likelihood is peaked around the MLE, leading to small variations in the resampled coefficients.

\subsection{Wind turbine example}\label{sec:WindTurbineExample}

Accurately assessing the performance of wind turbines under realistic conditions is essential for reliable and efficient design. To this end, stochastic reliability analysis has often been applied to wind turbines, as demonstrated in several studies \citep{Choe_2015,Choe_2016,Choe_2017,Cao_2019,Pan_2020,Li_2021}. The typical design process involves two main steps: generation of stochastic wind fields, and aero-elastic simulations that compute the structural response under these conditions. While the aero-elastic simulator is itself deterministic, stochasticity arises from the input wind fields. These are modeled as 10-minute spatio-temporal random fields defined by macroscopic parameters such as average wind speed and turbulence intensity. However, these parameters do not uniquely define the inflow, leading to varying responses even for identical input values. As a result, the simulator behaves as a stochastic model.

We test our ALR methodology in a dataset equivalent to 96 years of simulated operation of the 5 MW NREL reference turbine \citep{Barone_2012,Jonkman_2009_WindTurbine}. Wind inflow fields are generated using NREL’s TurbSim \citep{Jonkman_2009_TurbSim}, with the average wind speed and turbulence intensity as inputs. The wind speed follows a truncated Rayleigh distribution \citep{Moriarty_2008} with a mean of $10$ m/s (for the untruncated distribution) and cut-in/cut-off limits of $3$ m/s and $25$ m/s. Turbulence intensity is modeled deterministically using the IEC normal turbulence model \citep{iec_2005_61400}, leaving wind speed as the only input random variable of our analysis. Structural responses, including the flapwise bending moment at the blade root, are computed using the aero-servo-elastic solver FAST \citep{Jonkman_2005,jonkman2013addendum}.

The focus is on estimating the probability that the maximum flapwise bending moment during a 10-minute simulation \( M_b(U, \omega) \) exceeds a threshold \( \tau \). The associated limit-state function reads:
\begin{equation}
g_s(U, \omega) = \tau - M_b(U, \omega).
\label{eq_limit_state_WT}
\end{equation}

\begin{figure}[H]
     \centering
     \begin{subfigure}[c]{0.49\textwidth}
         \centering
         \includegraphics[width=0.99\textwidth]{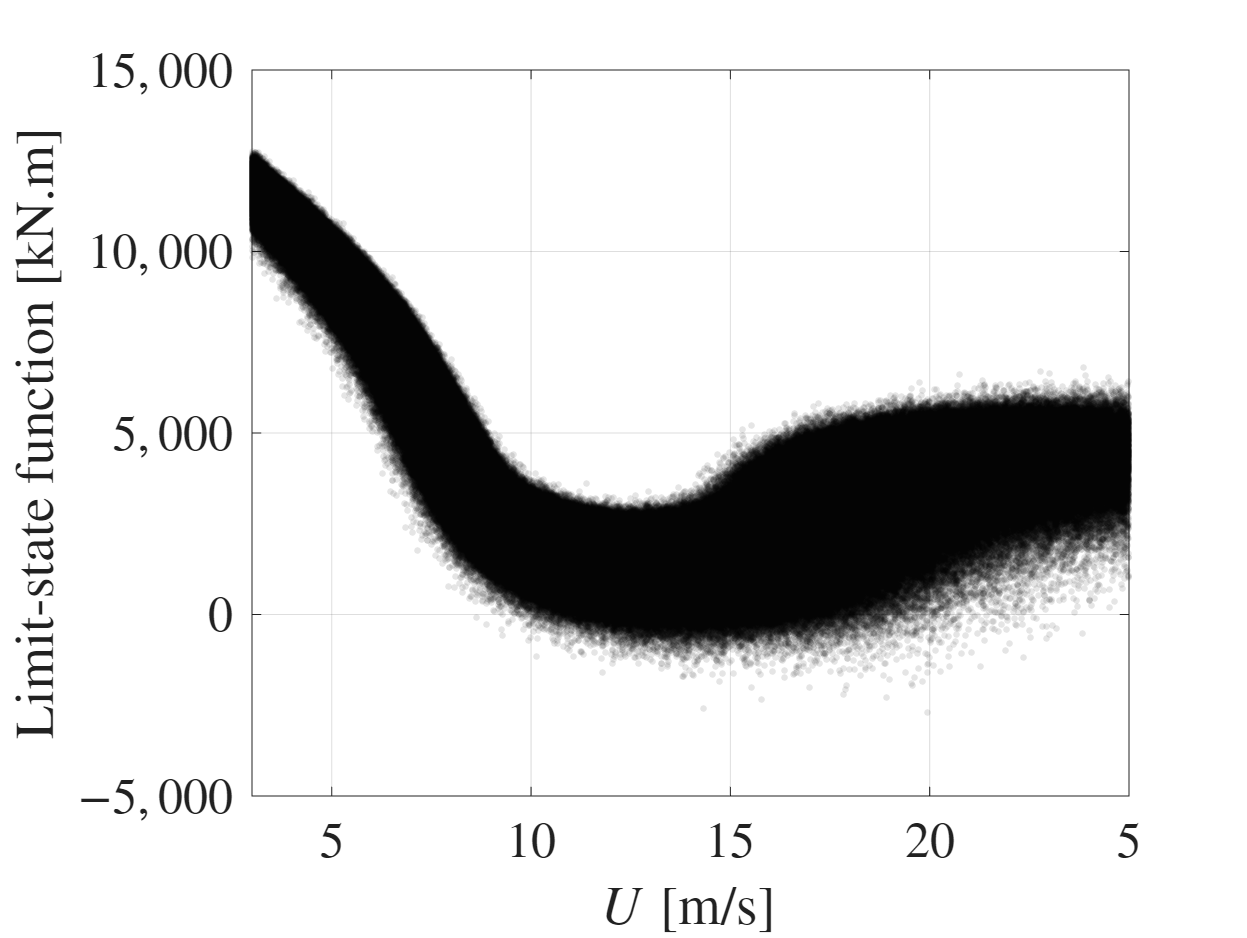}%
         \caption{Scatter plot of the limit-state function versus average wind speed over a 10-minute simulation.}
    \label{fig_dataset_bending_moment}
     \end{subfigure}
     \begin{subfigure}[c]{0.49\textwidth}
         \centering
         \includegraphics[width=0.99\textwidth]{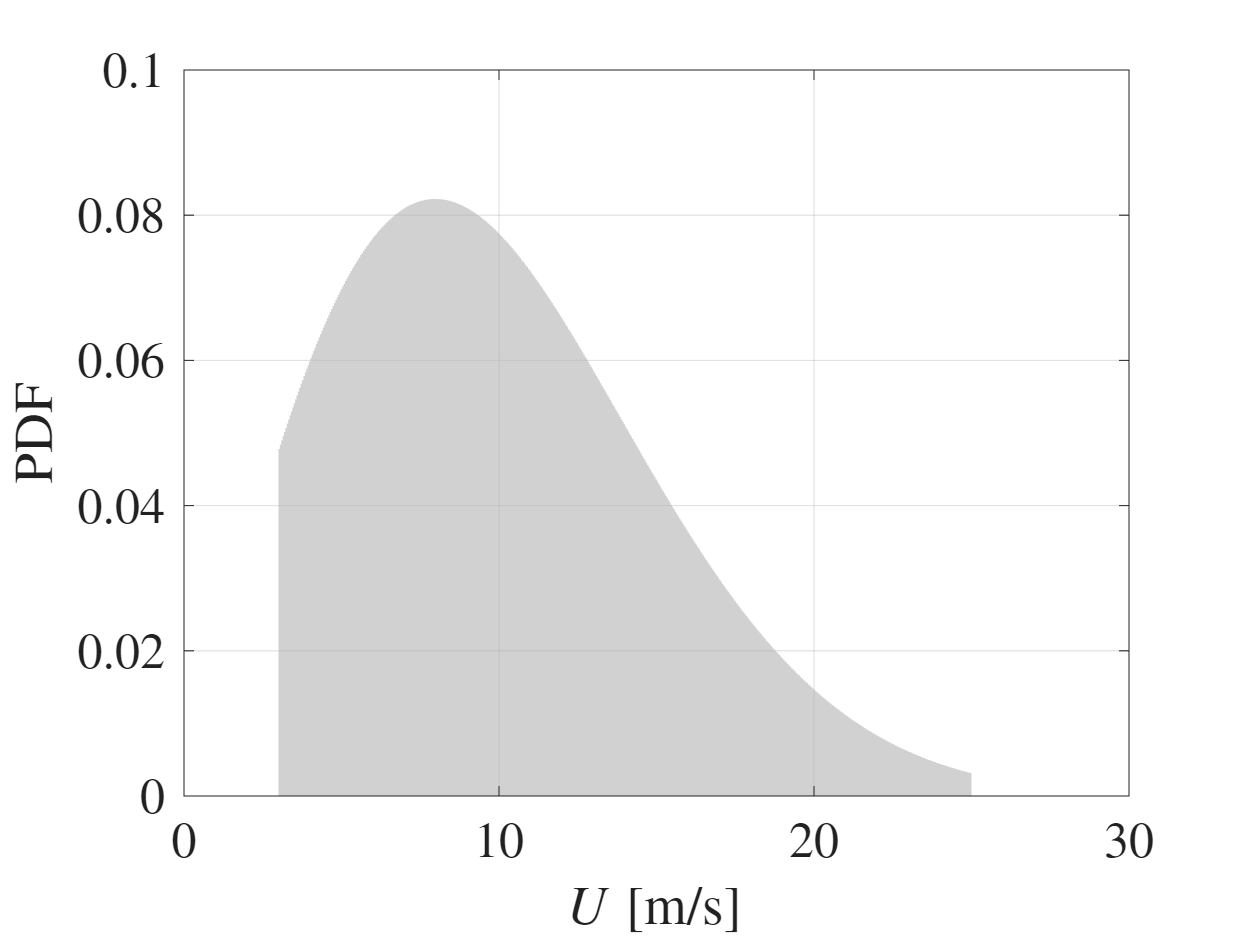}
         \caption{Normalized histogram representing the PDF of the average wind speed for the given dataset.}
    \label{fig_hist_dataset}
     \end{subfigure}
     \caption{Wind turbine application: Available wind turbine dataset and associated input distribution of the 10-minute average wind speed.}
	 \label{fig_dataset_data_and_PDF}
\end{figure}

\cref{fig_dataset_data_and_PDF} depicts the available dataset and its associated PDF. More specifically, \cref{fig_dataset_bending_moment} shows the associated limit-state evaluations, while \cref{fig_hist_dataset} presents the histogram of \( U \). The dataset comprises approximately 5 million simulations, with no replications. We set the failure threshold at \( \tau = 15{,}500 \, \text{kN}\cdot\text{m} \), which yields a reference failure probability \( P_f = 2.031 \times 10^{-3} \) with a coefficient of variation slightly below $0.1\%$, estimated using the entire dataset.

To enable visual validation and a better understanding of the stochastic response, we calculate empirical conditional statistics of \( M_b \) as a function of \( U \), including mean, variance, and quantiles. These are estimated using a moving window approach of width \( \Delta = 0.1 \) m/s. Thus, for a given value \( u \), we collect all observations within the interval \( [u - \Delta, u + \Delta] \). The conditional mean and variance are estimated as:
\begin{align}
\hat{\mu}(u) &= \frac{1}{N_w} \sum_{i: u^{(i)} \in [u - \Delta, u + \Delta]} M_b^{(i)}, \\
\hat{\sigma}^2(u) &= \frac{1}{N_w - 1} \sum_{i: u^{(i)} \in [u - \Delta, u + \Delta]} \left( M_b^{(i)} - \hat{\mu}(u) \right)^2,
\end{align}
where \( N_w \) denotes the number of samples within the interval.

The corresponding conditional quantiles are computed by sorting the values and extracting the desired percentile $\alpha$:
\begin{equation}
Q_{\alpha}(u) = M_{b_{(\lfloor \alpha N_w \rfloor)}},
\end{equation}
where \( \lfloor \cdot \rfloor \) is the floor operator.

\begin{figure}[H]
     \centering
     \begin{subfigure}[c]{0.49\textwidth}
         \centering
         \includegraphics[width=0.99\textwidth]{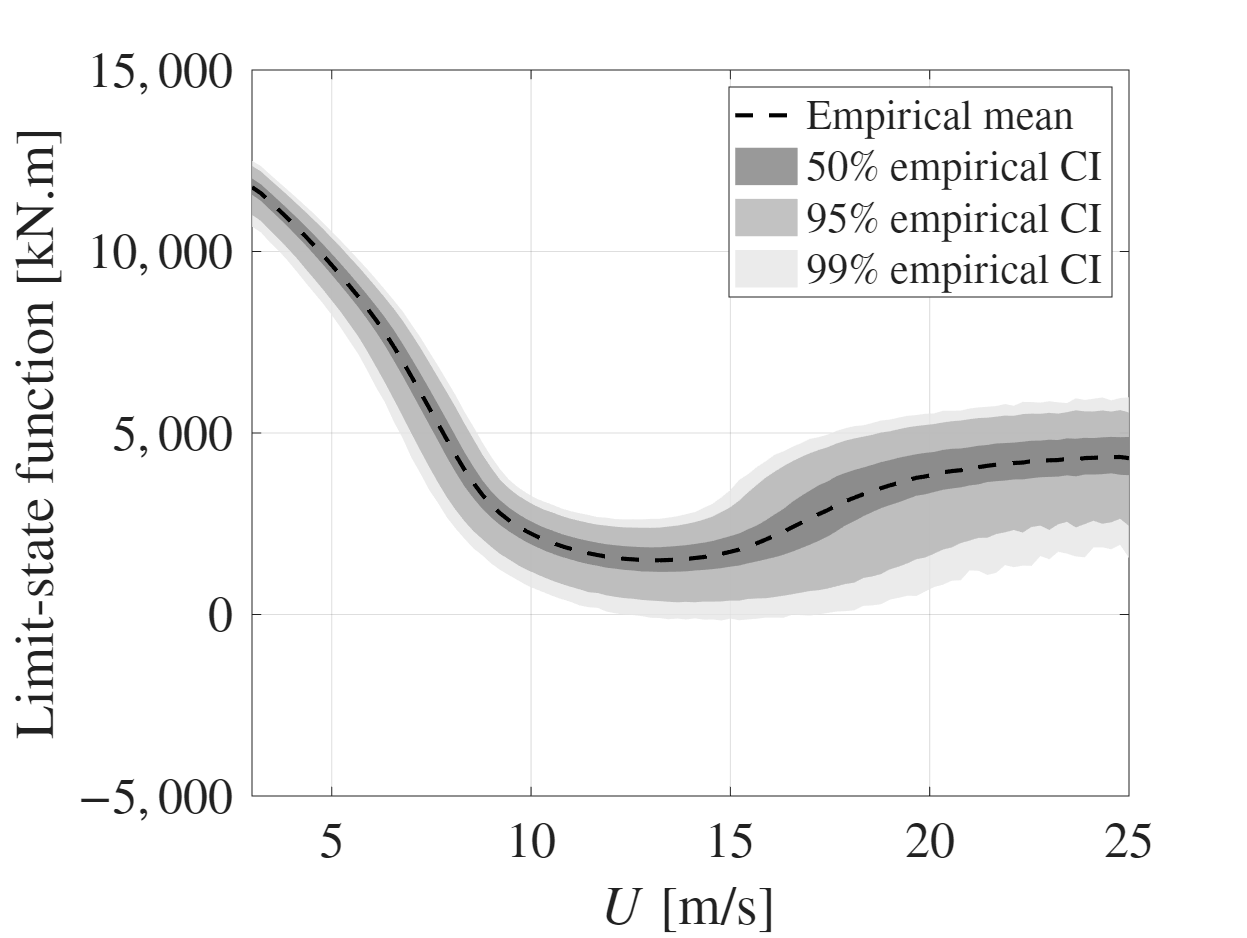}
         \caption{Empirical mean and quantiles.}
         \label{fig_mean_dataset}
     \end{subfigure}
     \begin{subfigure}[c]{0.49\textwidth}
         \centering
         \includegraphics[width=0.99\textwidth]{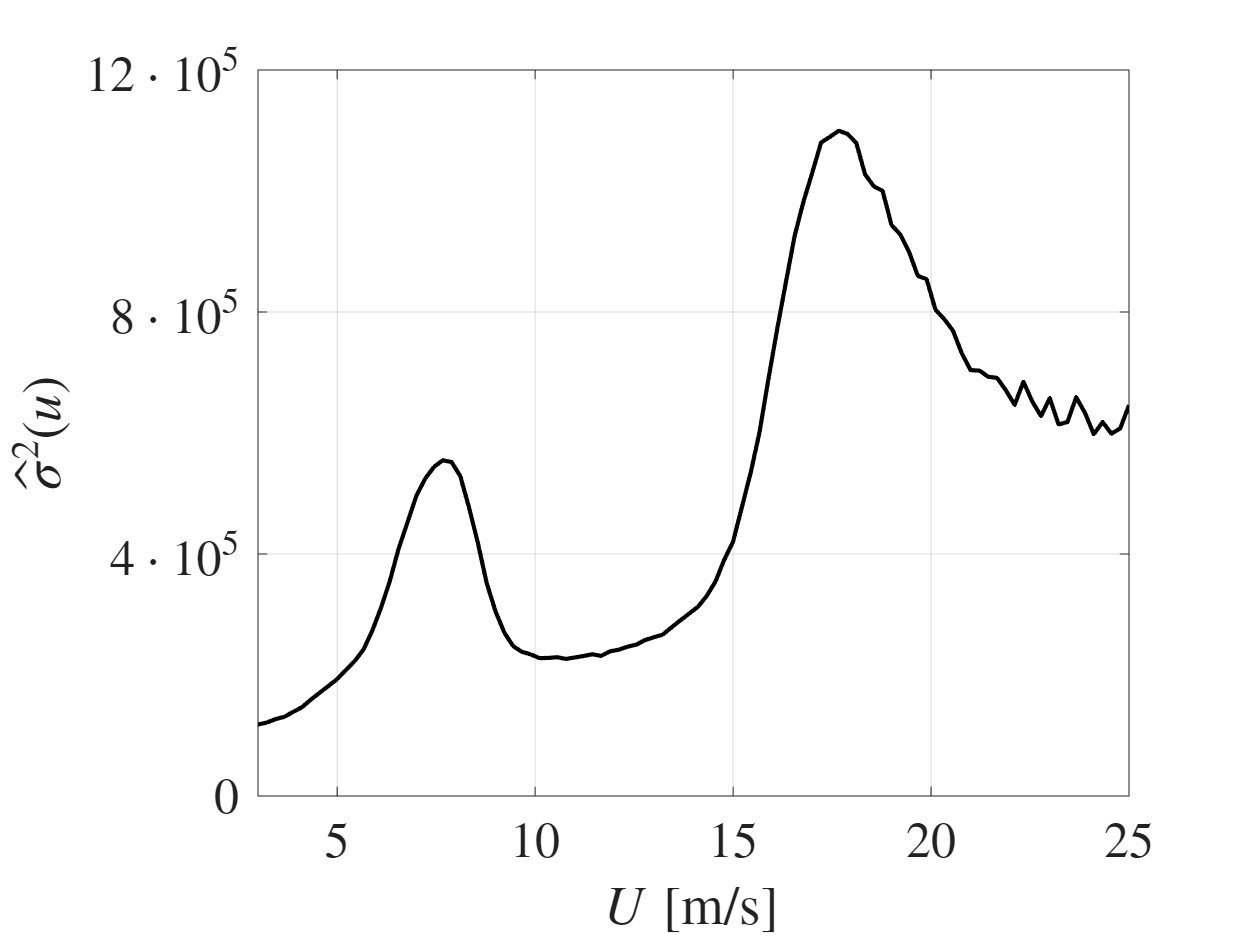}
         \caption{Empirical variance function.}
         \label{fig_variance_dataset}
     \end{subfigure}
     \caption{Wind turbine application: estimated conditional statistics based on the available dataset.}
     \label{fig_dataset_mean_variance}
\end{figure}

\cref{fig_dataset_mean_variance} presents the functions related to conditional statistics. \cref{fig_mean_dataset} shows the conditional mean function and its associated confidence interval bands. \cref{fig_variance_dataset} displays the conditional variance. These plots reveal strong heteroskedasticity and skewed conditional distributions, particularly at high wind speeds. These features pose an additional challenge when constructing accurate surrogates for this problem.

The active learning process begins with an initial experimental design of 20 points generated via LHS. At each iteration, following the procedure described in \cref{sec_local_error_measures_for_SPCE}, 100 SPCE realizations are drawn to compute local error estimates. The failure probability associated with the emulator $\hat{P}_f$ is estimated using \cref{eq_Pf_random_variable_POV} with MCS of size $N_{\text{MCS}} = 10^5$ samples. To reduce the noise effects in the convergence curves, the $\hat{P}_f$  estimates are smoothed using a moving average filter over three iterations (see \cref{eq_smoothed_Pf}). A candidate set of $10^3$ points is used, and 5 points are added to the experimental design at each iteration until a total of $1{,}000$ samples is reached. During training, the SPCE settings include degree adaptivity with polynomial orders in $p \in [1, 4]$, and the latent variable is modeled as Gaussian.

\begin{figure}[H]
\centering
\includegraphics[width=0.65\linewidth]{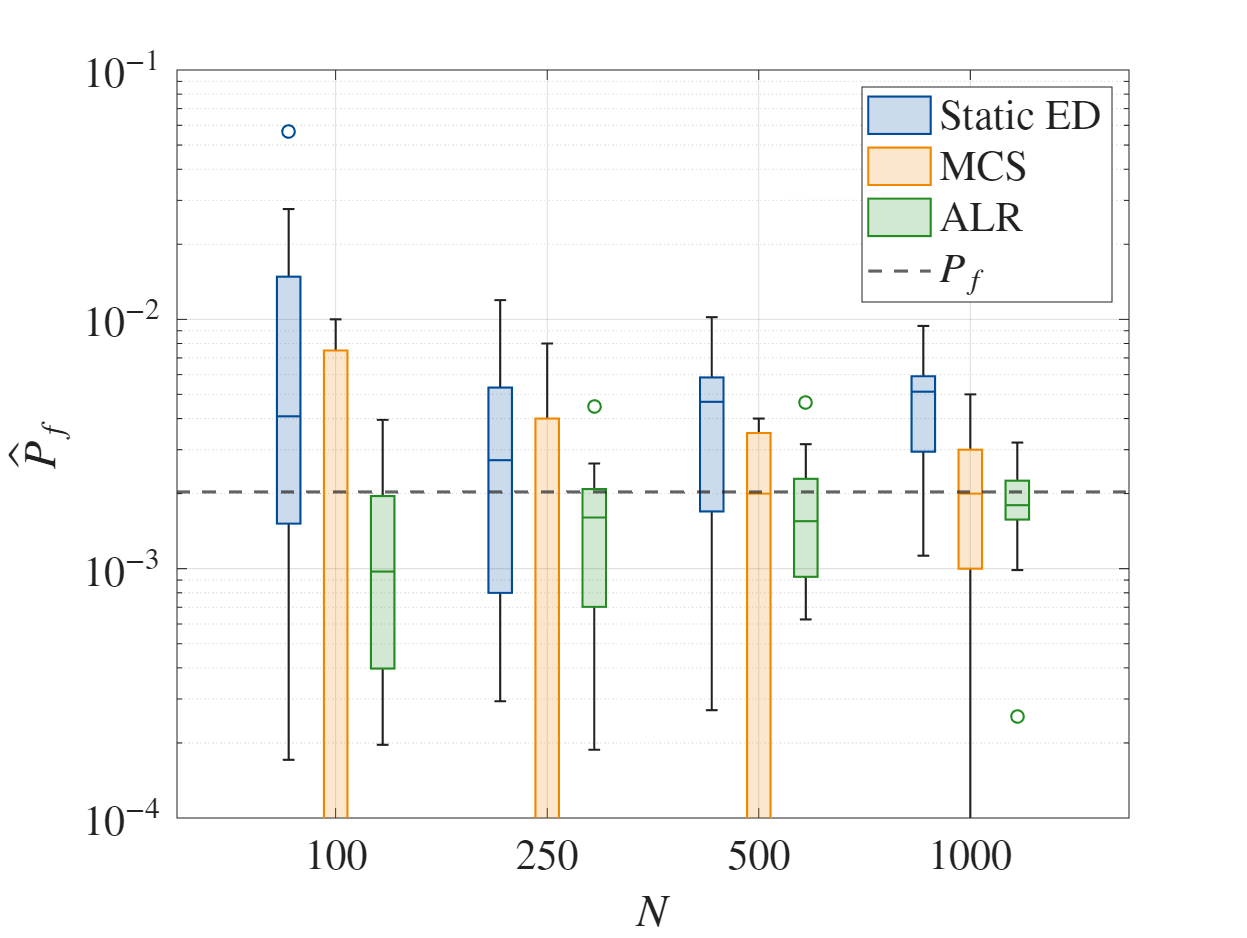}
\caption{Wind turbine example: Boxplots comparing convergence of the ALR method (in green), a static ED approach (in blue), and direct MCS (in orange) across 15 replications. The dashed line indicates the reference $P_f$.}
\label{fig_boxplots_WT_ALR_vs_static}
\end{figure}

\cref{fig_boxplots_WT_ALR_vs_static} compares results across 15 runs of the active learning approach (blue), the SPCE approach with a static experimental design (orange), and direct MCS (green). The static ED model follows the procedure in \citet{Pires_2025b}, where points are uniformly subsampled from the dataset, and the SPCE is trained with \( p \in [1, 10] \). This ensures that the points cover the entire input domain. The dashed black line indicates the reference probability of failure. Despite the pronounced heteroskedasticity, the ALR approach achieves accurate results with as few as 250 samples, significantly outperforming the alternatives.

\begin{table}[H]
\centering
\caption{Wind turbine application: Comparison of ALR, SPCE with a static ED, and MCS across multiple runs for different sample sizes. Reference $P_f=2.031\times 10^{-3}$.}
\label{tab_WT_results}
\begin{tabular}{ccccccc}
\toprule
\multirow{2}{*}{\( N \)} & \multicolumn{2}{c}{ALR} & \multicolumn{2}{c}{Static ED} & \multicolumn{2}{c}{MCS} \\ \cmidrule{2-7}
 & Median \( \hat{P}_f \) & CoV & Median \( \hat{P}_f \) & CoV & Median \( \hat{P}_f \) & CoV \\
\midrule
100   & $9.736 \times 10^{-4}$ & $83.4\%$ & $4.084 \times 10^{-3}$ & $148.2\%$ & $--$ & $--$ \\
250   & $1.604 \times 10^{-3}$ & $75.6\%$ & $2.723 \times 10^{-3}$ & $93.1\%$ & $--$ & $--$ \\
500   & $1.550 \times 10^{-3}$ & $62.3\%$ & $4.668 \times 10^{-3}$ & $67.1\%$ & $2.000 \times 10^{-3}$ & $85.6\%$ \\
1000  & $1.798 \times 10^{-3}$ & $38.6\%$ & $5.127 \times 10^{-3}$ & $46.2\%$ & $2.000 \times 10^{-3}$ & $64.6\%$ \\
\bottomrule
\end{tabular}
\end{table}

%
% \begin{table}[H]
% \centering
% \caption{Wind turbine application: Comparison of ALR, SPCE with a static ED, and MCS across multiple runs for different sample sizes. Reference $P_f=1.022\times 10^{-2}$.}
% \label{tab_WT_results}
% \begin{tabular}{ccccccc}
% \toprule
% \multirow{2}{*}{\( N \)} & \multicolumn{2}{c}{ALR} & \multicolumn{2}{c}{Static ED} & \multicolumn{2}{c}{MCS} \\ \cmidrule{2-7}
%  & Median \( \hat{P}_f \) & CoV & Median \( \hat{P}_f \) & CoV & Median \( \hat{P}_f \) & CoV \\
% \midrule
% 100   & $9.140 \times 10^{-2}$ & $61.0\%$ & $1.527 \times 10^{-2}$ & $99.5\%$ & $ -- $ & $--$ \\
% 250   & $7.173 \times 10^{-3}$ & $40.5\%$ & $8.949 \times 10^{-3}$ & $63.2\%$ & $8.000 \times 10^{-3}$ & $73.3\%$ \\
% 500   & $9.024 \times 10^{-3}$ & $37.2\%$ & $1.448 \times 10^{-2}$ & $52.7\%$ & $8.000 \times 10^{-3}$ & $42.0\%$ \\
% 1000  & $8.940 \times 10^{-3}$ & $21.9\%$ & $1.790 \times 10^{-2}$ & $36.2\%$ & $1.000 \times 10^{-2}$ & $30.3\%$ \\
% \bottomrule
% \end{tabular}
% \end{table}

\begin{figure}[H]
\centering
     \begin{subfigure}[c]{0.49\textwidth}
         \centering
         \includegraphics[width=0.9\textwidth]{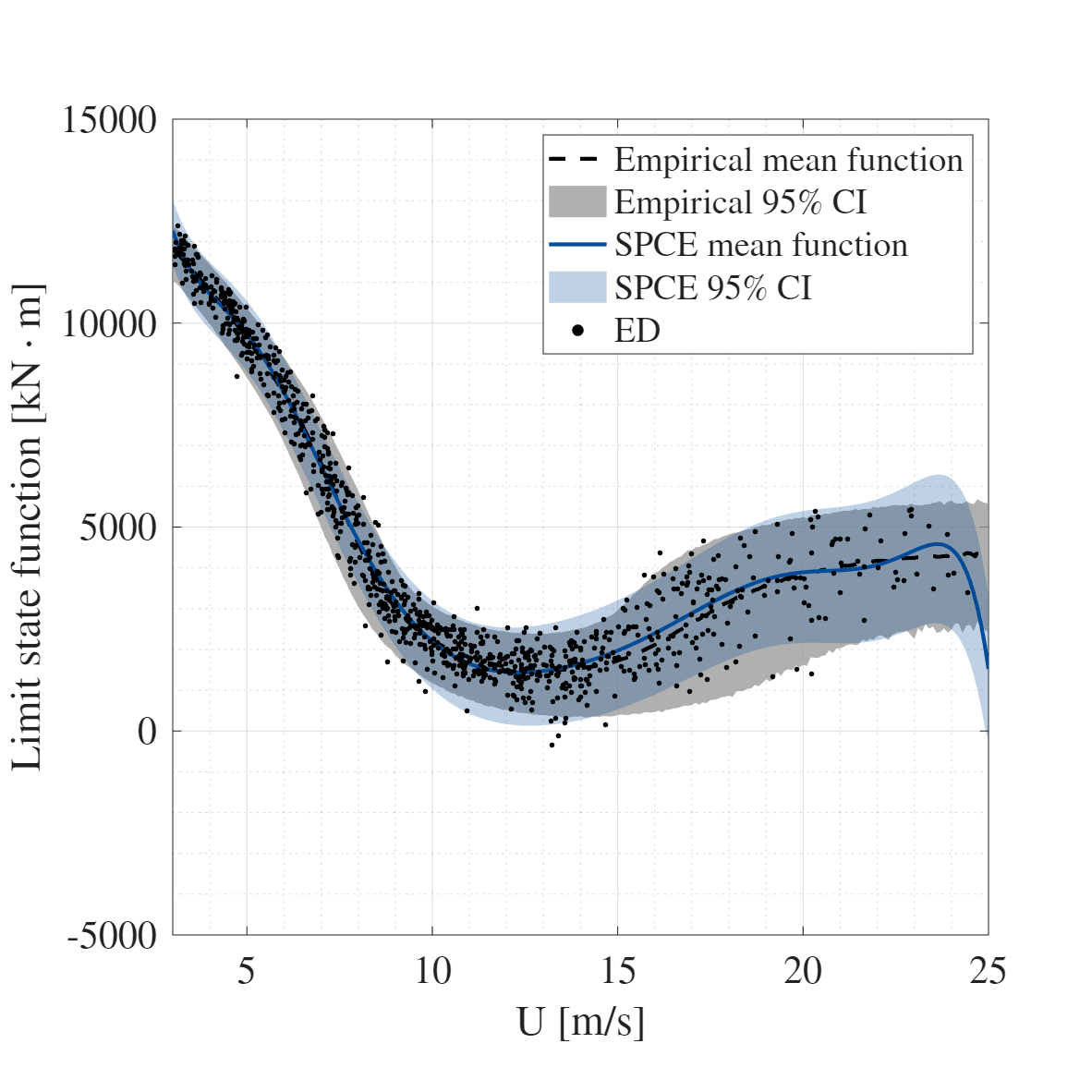}
         \caption{Static ED approach.}
         \label{fig_static_ED_model_WT}
     \end{subfigure}
     \begin{subfigure}[c]{0.49\textwidth}
         \centering
         \includegraphics[width=0.9\textwidth]{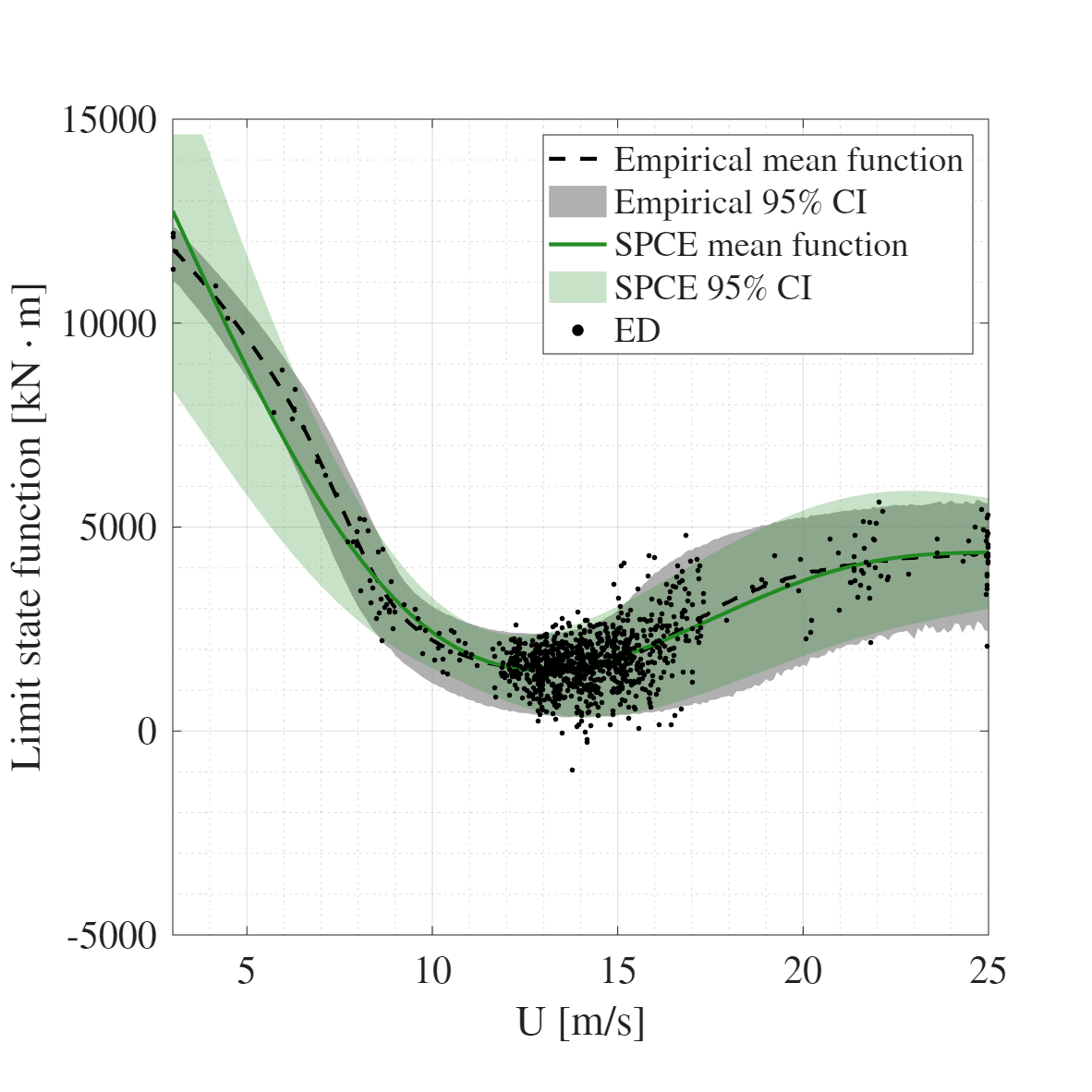}
         \caption{ALR approach.}
         \label{fig_ALR_model_WT}
     \end{subfigure}
\caption{Wind turbine example: comparison of surrogate models for the run yielding the median relative error on $\hat{P}_f$ estimate. The plots show the SPCE mean function and $95\%$ confidence intervals, along with the empirical mean and quantile bands derived from the full dataset.}
\label{fig_boxplots_WT_ALR_vs_static_limit_state}
\end{figure}

\cref{fig_boxplots_WT_ALR_vs_static_limit_state} depicts the surrogate models obtained from the ALR and static ED approaches for the run that produced the median estimate of \( \hat{P}_f \). The plots are shown alongside the empirical mean and $95\%$ confidence interval bands derived from the dataset. The ALR methodology tends to concentrate the samples in regions where the product \( f_{U}(u) \cdot s(u) \) is large, that is, where both the input density and the conditional failure probability are high. In this case, this region corresponds to wind speeds around $10$ m/s, which aligns with the mode of the Rayleigh distribution used for the wind speed, as shown in \cref{fig_hist_dataset}. This behavior reflects the core principle behind the learning function defined in \cref{eq_LF}, which guides sampling toward areas that influence the estimator associated with the probability of failure the most.

An interesting behavior is also observed near \( U = 15 \) m/s. Although the true conditional failure probability in this region is extremely low, the surrogate initially overestimates \( s(u) \) due to limited training data, as evidenced by the static ED approach. As a result, the learning function identifies this region as uncertain and potentially relevant, prompting the addition of several points. Once the emulator learns that \( s(u) \) is in fact small in this region, further enrichment is naturally redirected to more relevant regions. This adaptivity illustrates how ALR balances exploitation, \emph{i.e.} sampling where failure is likely and the input density is high, and exploration, \emph{i.e.} allocating points in uncertain regions to improve local accuracy when needed. This mechanism not only ensures a more reliable estimate of \( \hat{P}_f \) but also avoids unnecessary sampling in regions that have minimal impact on the overall probability of failure.

\begin{figure}[H]
     \centering
     \begin{subfigure}[t]{0.49\textwidth}
         \centering
         \includegraphics[width=\textwidth]{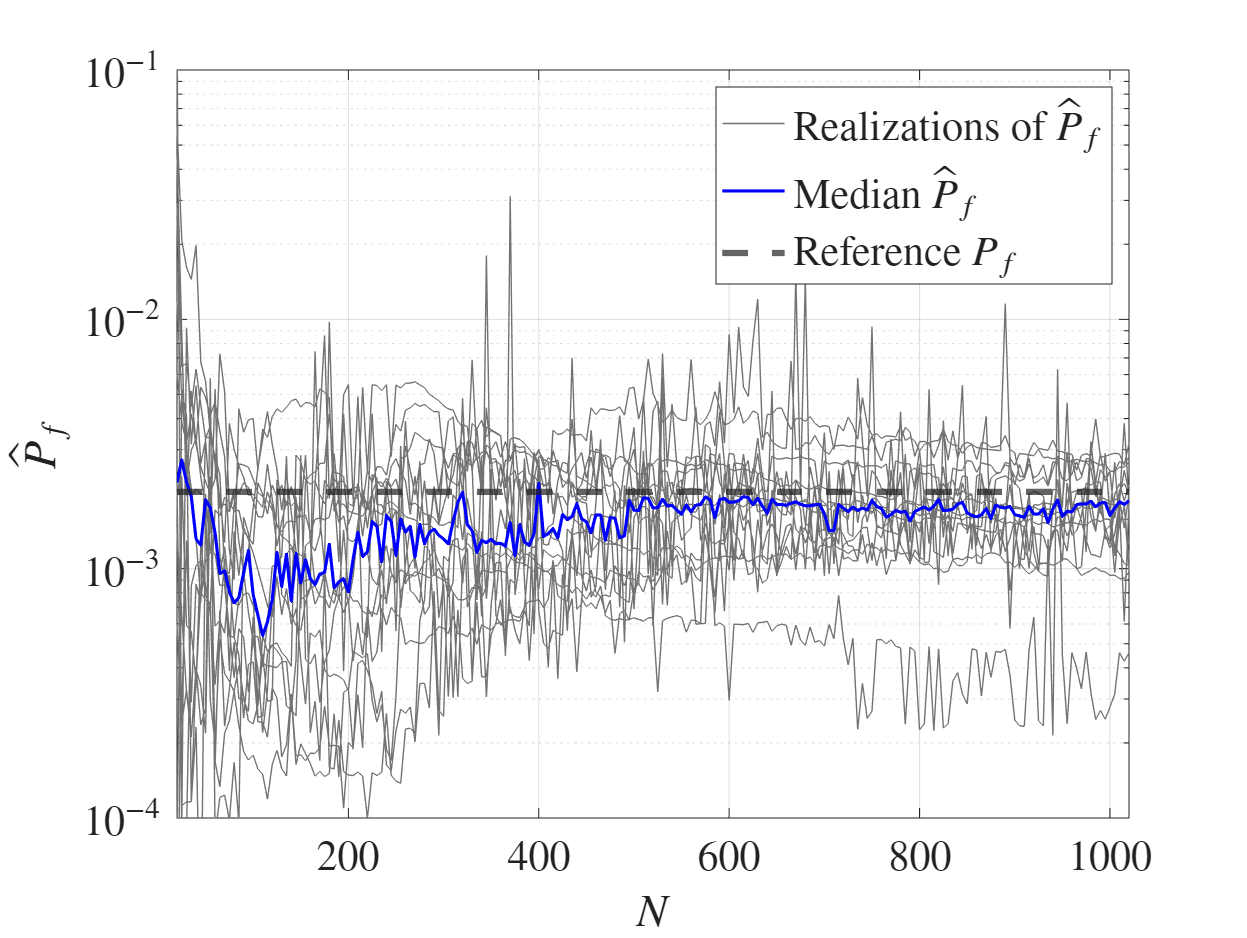}
         \caption{}
         \label{fig_ALR_WT_conv_curve_Pf}
     \end{subfigure}
     \begin{subfigure}[t]{0.49\textwidth}
         \centering
         \includegraphics[width=\textwidth]{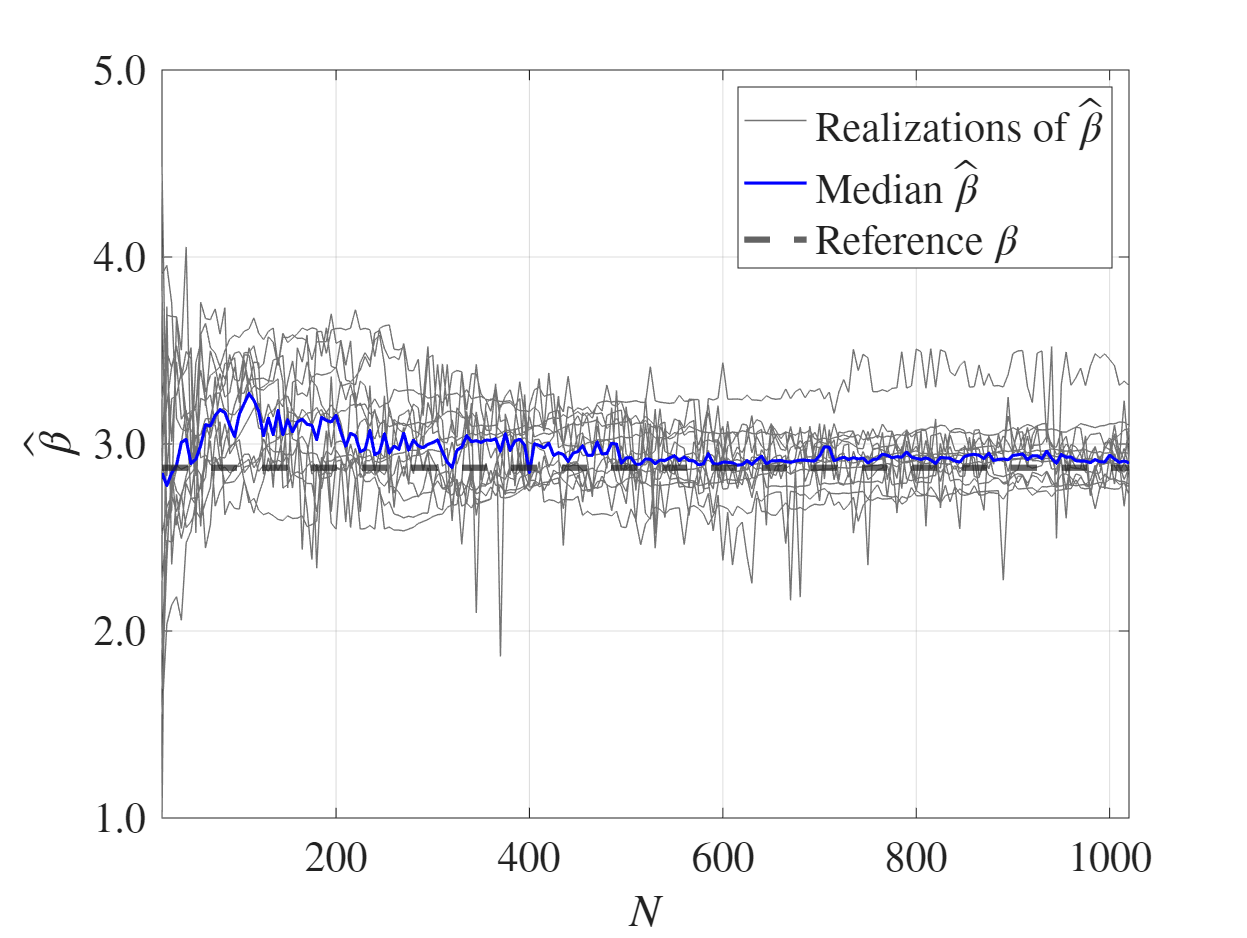}
         \caption{}
         \label{fig_ALR_WT_conv_curve_beta}
     \end{subfigure}
     \caption{Wind turbine example: Convergence of ALR-based failure probability estimates. All 15 ALR runs are shown in gray. The blue curves show the median estimates across the 15 runs at each sample size $N$. (a) Convergence curve of the failure probability $\hat{P}_f$, (b) Convergence curve of the reliability index $\hat{\beta} = -\Phi^{-1}(\hat{P}_f)$. The dashed lines indicate the reference values.}
     \label{fig_ALR_WT_conv_curve}
\end{figure}

\cref{fig_ALR_WT_conv_curve} shows the convergence curves of all 15 ALR runs in gray, with the median estimates across runs at each sample size $N$ highlighted in blue. The figure consists of two subfigures: \cref{fig_ALR_WT_conv_curve_Pf} presents the convergence of the failure probability $\hat{P}_f$, while \cref{fig_ALR_WT_conv_curve_beta} shows the corresponding reliability index $\hat{\beta} = -\Phi^{-1}(\hat{P}_f)$. The estimate stabilizes around the reference value after approximately 600 samples. As in the previous examples, the fluctuations observed in the curve are attributed to the stochasticity of the simulator. Defining an appropriate stopping criterion remains an open challenge but could further reduce the computational cost.

\begin{figure}[H]
     \centering
     \begin{subfigure}[c]{0.49\textwidth}
         \centering
         \includegraphics[width=\textwidth]{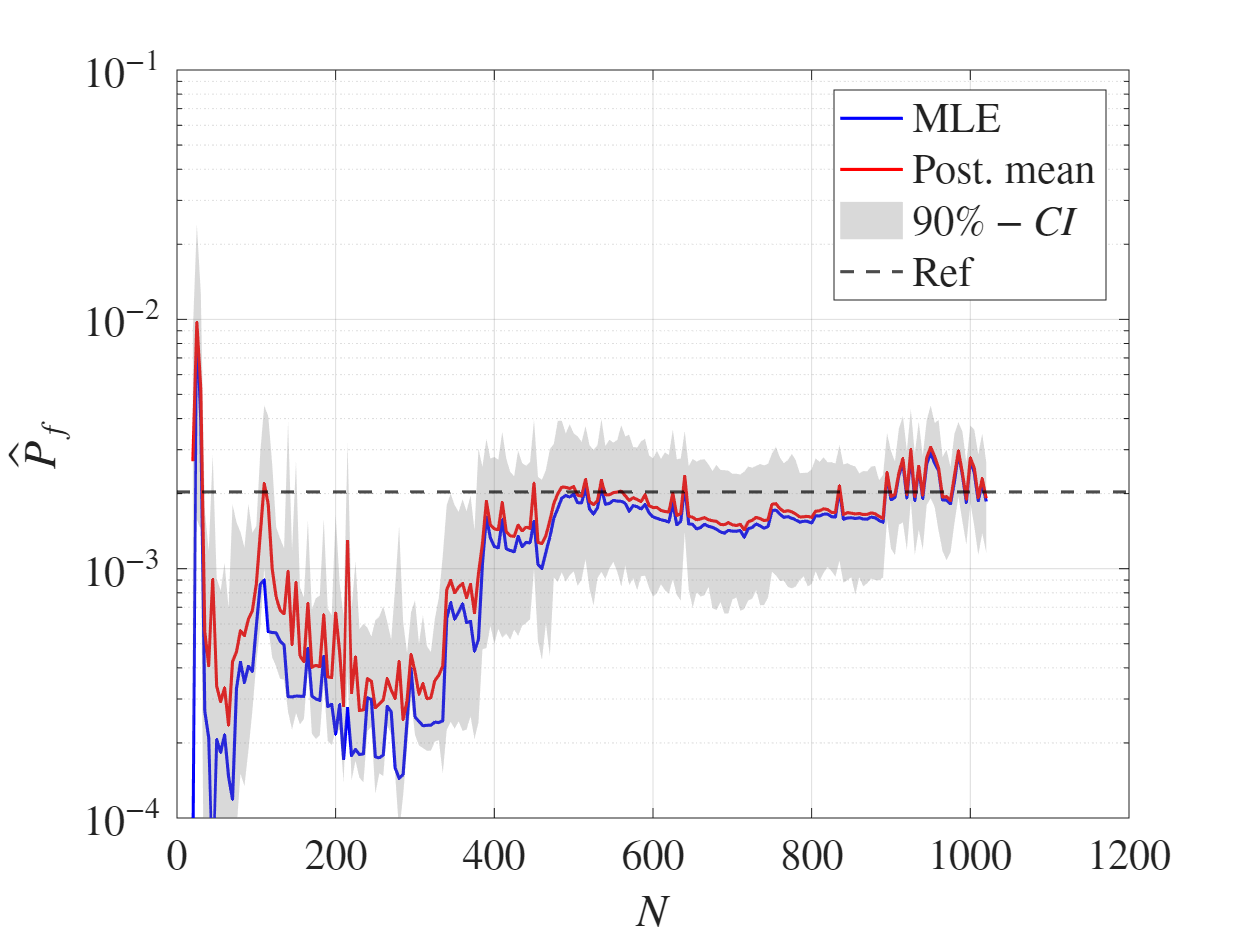}
         \caption{}
         \label{fig_ALR_WT_conv_CI_Pf}
     \end{subfigure}
     \begin{subfigure}[c]{0.49\textwidth}
         \centering
         \includegraphics[width=\textwidth]{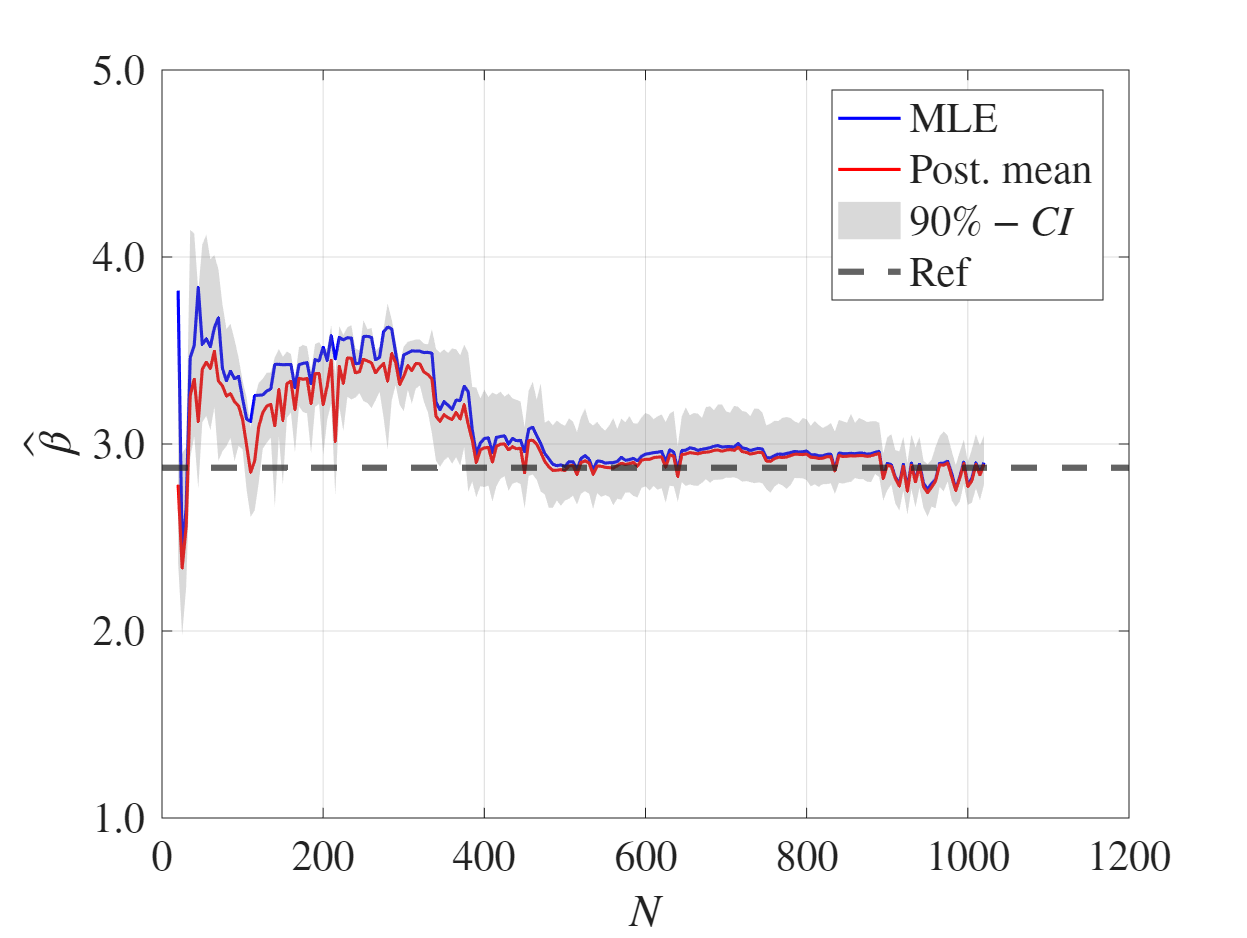}
         \caption{}
         \label{fig_ALR_WT_conv_CI_beta}
     \end{subfigure}
     \caption{Wind turbine example: Convergence of ALR-based failure probability estimate for the repetition that lead to the median relative error on the final $\hat{P}_f$ estimate. The blue curve show the MLE-based estimate, while the red one shows the posterior mean. The gray band indicates the $90\%$ confidence interval. (a) Convergence curve of the failure probability $\hat{P}_f$, (b) Convergence curve of the corresponding reliability index $\hat{\beta} = -\Phi^{-1}(\hat{P}_f)$.}
     \label{fig_ALR_WT_conv_CI}
\end{figure}

Finally, Figure~\ref{fig_ALR_WT_conv_CI} shows the convergence history for the repetition that led to the median error on the final estimate of $\hat{P}_f$. The blue curve corresponds to the failure probability estimate obtained using the maximum likelihood (MLE) estimate of the SPCE coefficients. The red curve represents the posterior mean estimate obtained by averaging the $\hat{P}_f$ values computed from the resampled SPCE coefficients. The gray band indicates the $90\%$ confidence interval defined by the $5$-th and $95$-th percentiles of the $\hat{P}_f$ values associated with the resampled coefficients. There is a small discrepancy between the MLE-based and posterior estimates during the early iterations, but this difference decreases as the experimental design is enriched. Moreover, the width of the confidence interval decreases as the ED size increases, reflecting the progressive reduction of the uncertainty associated with the estimated coefficients.

\section{Discussion}
Beyond the numerical results, a discussion is warranted to contextualize the scope and limitations of the proposed methodology.

The main goal of the proposed algorithm is to demonstrate that the active learning component provides a significant improvement over the original static ED approach introduced in \cite{Pires_2025b}. For this reason, we have not compared our results with other advanced reliability analysis methods for stochastic simulators, such as importance sampling techniques that rely on an approximation of  the conditional failure probability $s(\mathbf{x})$ as importance density \citep{Choe_2015,Choe_2016}. In particular, in the stochastic importance sampling (SIS) framework originally proposed for a wind turbine application similar to the one in Section~\ref{sec:WindTurbineExample}, a failure probability of about $1\%$ required approximately $9{,}000$ samples (see Table~6 in \citet{Choe_2015}). More recent variants, such as those employing cross-entropy optimization \citep{Cao_2019}, achieved similar computational costs, as the sample budget was kept equivalent to the original study despite methodological improvements. In contrast, our proposed ALR method achieves comparable accuracy with around $1{,}000$ samples, while eliminating the need to construct an auxiliary surrogate model to approximate $s(\mathbf{x})$.

Second, the range of application of the proposed method is primarily governed by two factors: the general scalability of SPCE with dimensionality, and the local characteristics of the limit-state function. The first factor depends on the sparsity of the polynomial expansion, which in turn is influenced by the effective dimensionality of the problem. The second factor is more problem-specific, as it mostly depends on the conditional failure probability, which in general can have a dimensionality that is lower, or at most equal, to that of the stochastic limit-state function itself. For instance, the conditional failure probability may be more sensitive to extreme values of a single variable (e.g., loads), even though the limit-state function depends on multiple inputs simultaneously.  Computationally, SPCE can easily handle up to $O(10-20)$ input variables, while higher-dimensional problems would require dimensionality reduction techniques. In general, for a given limit-state complexity, increasing the effective dimensionality entails a need for a larger training set.

The previous paragraph addressed the dimensionality of the explicit input space. The dimensionality of the latent space, however, does not generally affect the performance of SPCE, as already noted in the original work by \cite{Zhu_2023_SPCE}. We expect this observation to hold in the context of reliability analysis as well, particularly given the scarcity of available training data. Similarly, we do not expect that a high variance in the latent space will significantly affect the proposed methodology, compared with challenges arising from nonlinearity or heteroskedasticity in the model response. 

Finally, beyond the direct applications for reliability analysis discussed in this paper, the proposed methodology has potential for applications in other fields of uncertainty quantification involving stochastic simulators. For example, stochastic emulation has recently gained attention in earthquake engineering where inherent aleatoric variability precludes the use of standard deterministic surrogate models \citep{Zhu_2023,Yi_2024,Kim_2024}. In such applications, the objective is often to calculate conditional failure probabilities for the construction of fragility curves. The approach proposed in this paper, which is based on the analytical estimation of the conditional failure probability $s(\mathbf{x})$ and its integration within an active learning framework, could be easily extended to these applications.

\section{Conclusions}
\label{sec_conclusions}

Reliability analysis methods are well-established for deterministic simulators. However, the growing use of stochastic models in engineering, such as agent-based systems or classical simulators with inherent randomness, as found in earthquake or wind turbine modeling, calls for approaches that explicitly account for this stochasticity. In this work, we build upon recent advances in stochastic surrogate modeling to develop an active learning framework tailored to stochastic polynomial chaos expansions (SPCE).

Our contribution is twofold. First, we leverage the asymptotic properties of the maximum likelihood estimator to estimate the uncertainty associated with the coefficients of the SPCE emulator. Although this indicator is not a direct measure of prediction error, it effectively highlights regions where the emulator shows high uncertainty in the conditional failure probability estimation. Second, we propose a learning function that combines this uncertainty with the input PDF of the parameters to drive sampling toward regions that are both uncertain and relevant for estimating the probability of failure. This results in a targeted and efficient active learning strategy for stochastic reliability analysis.

We validate the proposed methodology on three distinct problems. The first one is a stochastic reformulation of the classical R-S problem, chosen for its analytical tractability and for the ability to visualize the behavior of the active learning strategy. The second case considers a stochastic SIR model, where randomness emerges from uncontrollable latent variables in the agent-based dynamics. The third example uses a large dataset of wind turbine simulations for the NREL 5-MW reference turbine. Here, the underlying simulator is not available, and reliability analysis is performed directly on the dataset using a subsampling strategy guided by our active learning approach. Across all examples, the proposed method consistently outperforms both direct Monte Carlo simulation and stochastic polynomial chaos expansion based on a static experimental design, showing significantly faster convergence.

Though our method significantly reduces the number of simulator runs, a robust stopping criterion remains an open issue. This is primarily due to the noisy nature of the convergence curves. Although stabilization is observed, the curves are not smooth enough to allow the use of conventional stopping rules. While some of this noise is a result of the stochastic nature of the simulators, we observed that a considerable share stems from the optimization process carried out when training the emulator. We reduced it by adopting a warm-start strategy for the noise term of the SPCE surrogate across iterations, which helps dampen fluctuations caused by the optimization. Nevertheless, we believe that a more principled solution is the reformulation of SPCE in a Bayesian setting. This would naturally lead both to embedded local error measures, and to include meaningful regularization terms in the emulator.

%%%%%%%%%%%%%%%%%%%%%%%%%%%%%%%%%%%%%%%%%%%%%%%%%%%%%%%%%%%%%%%%%%%%%%%%%%%%
\section*{Acknowledgments}
The support of European Union’s Horizon 2020 research and innovation program under the Marie Skłodowska-Curie grant agreement No 955393 is greatly acknowledged.

%%%%%%%%%%%%%%%%%%%%%%%%%%%%%%%%%%%%%%%%%%%%%%%%%%%%%%%%%%%%%%%%%%%%%%%%%%%%
\newpage
\bibliography{References} 
\end{document}